\documentclass[preprint,3p,fleqn]{elsarticle}

\usepackage{xcolor}
\usepackage{graphicx}
\usepackage{caption}
\usepackage{subcaption}
\usepackage{float}
\usepackage{booktabs}
\usepackage{multirow}
\usepackage{adjustbox}
\usepackage{enumitem}
\usepackage{algorithm}
\usepackage{algpseudocode}
\usepackage{indentfirst}
\usepackage{tabularx} 
\usepackage[colorlinks=true, citecolor=blue, linkcolor=blue, urlcolor=blue]{hyperref}
\algrenewcommand{\algorithmiccomment}[1]{\hfill$\triangleright$ #1}

\biboptions{sort&compress}
\usepackage{amssymb}
\usepackage{amsmath}
\usepackage{textcomp}
\usepackage[noabbrev]{cleveref}

\journal{arXiv}

\begin{document}

\begin{frontmatter}


\title{Hetero-EUCLID: Interpretable model discovery for heterogeneous hyperelastic materials using stress-unsupervised learning\tnoteref{label1}}

\title{} 

\author[iisc_label]{Kanhaiya Lal Chaurasiya} 
\author[iisc_label]{Saurav Dutta}
\author[tudelft_label]{Siddhant Kumar}
\author[iisc_label]{Akshay Joshi\corref{cor1}}
\cortext[cor1]{Corresponding author}
\ead{akshayjoshi@iisc.ac.in}

\affiliation[iisc_label]{organization={Department of Mechanical Engineering, Indian Institute of Science},
            city={Bengaluru},
            country={India}}

\affiliation[tudelft_label]{organization={Department of Materials Science and Engineering, Delft University of Technology},
            city={2628 CD Delft},
            country={The Netherlands}}

\begin{abstract}
    We propose a computational framework, Hetero-EUCLID, for segmentation and parameter identification to characterize the full hyperelastic behavior of all constituents of a heterogeneous material. In this work, we leverage the Bayesian-EUCLID (Efficient Unsupervised Constitutive Law Identification and Discovery) framework to efficiently solve the heterogenized formulation through parsimonious model selection using sparsity-promoting priors and Monte Carlo Markov Chain sampling. We utilize experimentally observable 3D surface displacement and boundary-averaged force data generated from Finite Element simulations of non-equi-biaxial tension tests on heterogeneous specimens. The framework broadly consists of two steps-- residual force-based segmentation and constitutive parameter identification. We validate and demonstrate the ability of the proposed framework to segment the domain and characterize the constituent materials on various types of thin square heterogeneous domains. We validate the framework's ability to segment and characterize materials with multiple levels of displacement noises and non-native mesh discretizations, i.e, using different meshes for the forward FE simulations and the inverse EUCLID problem. This demonstrates the applicability of the Hetero-EUCLID framework in Digital Image/Volume Correlation-based experimental scenarios. Furthermore, the proposed framework performs successful segmentation and material characterizations based on data from a single experiment, thereby making it viable for rapid, interpretable model discovery in domains such as aerospace and defense composites and for characterization of selective tissue stiffening in medical conditions such as \textit{fibroatheroma}, \textit{atherosclerosis}, or cancer.
\end{abstract}


\end{frontmatter}


\section{Introduction}
\label{introduction}

    With the increasing relevance of biomedical engineering, there is a growing focus on understanding the mechanical behavior of biological tissues. Many of these tissues exhibit intrinsic material heterogeneity \citep{guo2022modeling,wei2023bioinspired,lee2024data}, necessitating the development of multi-segmented constitutive models. Such models can significantly enhance our understanding of complex biomechanical processes, including the progression of aortic aneurysms \citep{burris2022vascular,lin2022aortic,vander2023stiffness} and the localized stiffening of tissues due to tumor growth \citep{goriely2017mathematics,micalet20213d}.\par
    
    A more traditional approach to constitutive modeling involves performing phenomenological model calibration of an \textit{a priori} assumed constitutive equation, utilizing full-field displacement data obtained from Digital Image Correlation (DIC) in combination with boundary-averaged force measurements from force transducers \citep{hild2006digital}. The inverse problem of determining the constitutive parameters is solved using Finite Element Model Updating (FEMU) \citep{marwala2010finite} or Virtual Fields Method (VFM) \citep{grediac2006virtual}. However, assuming a form of the constitutive equation for the material limits the materials and constitutive models that can be characterized with this approach. To address the limitations of traditional model calibration, and enable characterization and discovery of materials that are not necessarily well explained by existing models, several works in the past have sought to entirely bypass \citep{kirchdoerfer2016data,ibanez2017data,kirchdoerfer2018data,conti2018data,nguyen2018data,eggersmann2019model,carrara2020data,karapiperis2021data}, or surrogate constitutive models. The approach of surrogating constitutive model identifies a relationship between mechanical stresses and strains using Gaussian Process Regression (GPR) \citep{rocha2020fly,fuhg2022local}, Radial Basis Functions (RBF) \citep{park2021thermal}, piece-wise polynomial interpolation \citep{crespo2017wypiwyg,sussman2009model}, and Artificial Neural Networks (ANN) \citep{ghaboussi1991knowledge,fernandez2021anisotropic,klein2022polyconvex,vlassis2021sobolev,kumar2020inverse,bastek2022inverting,zheng2021data,mozaffar2019deep,bonatti2021one,vlassis2020geometric,kumar2022machine,meng_machine-learning-based_2025}. Several ANN-based constitutive modeling works incorporate novel measures (such as input-convexity) to ensure that the strain-energy density or stress-history predicted by the ANNs are thermodynamically consistent \citep{huang_variational_2022,linka_new_2023,masi_multiscale_2022}, and obey other physical constraints, such as objectivity, material frame indifference and polyconvexity \citep{klein2022polyconvex,geuken_novel_2025}. Furthermore, to address the lack of availability of full stress-tensor data for training supervised training, several frameworks leverage the balance of linear momentum \citep{thakolkaran_nn-euclid_2022} or the virtual fields method \citep{meng_machine-learning-based_2025,lourenco_indirect_2024} to train ANNs or Recurrent Neural Networks (RNNs) to surrogate the stress-strain relationship for the materials using only experimentally available surface displacement and boundary-averaged force data.\par
    
    Although stress-unsupervised ANN/RNN surrogates are capable of learning complicated constitutive models, they do not provide interpretable constitutive models. The EUCLID framework \citep{flaschel_unsupervised_2021} was initially introduced to model complicated homogeneous hyperelastic constitutive behavior in an interpretable manner. This was achieved through sparse symbolic regression using a pre-defined library of known strain energy densities. The EUCLID framework has since been extended to model homogeneous elastoplastic materials \citep{flaschel2022discovering}, viscoelastic materials \citep{marino_automated_2023}, and generalized standard materials \citep{flaschel2023automated}. Most of the aforementioned works focus on modeling homogeneous materials. \par
    
    An earlier work, which used only surface displacement data \citep{guo_physics-driven_2020}, focused on elastography of incompressible linear elastic heterogeneous solids, utilizing novel Physics-Informed Neural Networks (PINNs) to predict the variation of Young's Modulus as a function of position within the material. A later work by the same authors employed multiple PINNs to obtain the spatial variation of Poisson's ratio in elastography \citep{chen_physics-informed_2023}. Another work presented a detailed study on segmenting heterogeneous hyperelastic materials using surface displacement data and associated strain invariants \citep{nguyen_segmenting_2024}. A more recent work used a stress-unsupervised ANN-based framework \citep{shi2025deep} to segment and model a heterogeneous hyperelastic solid, wherein the inputs to the ANN are the strain invariants and the query position, and the output of the ANN is the strain energy density. The novel ANN-based framework was able to segment and reproduce the constitutive models of each segment in the hyperelastic solid. However, the use of ANNs makes the discovered models uninterpretable and provides no uncertainty estimates on the discovered models. Furthermore, since the query position is an essential input to the aforementioned ANN, it would complicate the adoption of the ANN as a constitutive model for forward simulation on other geometries. In this context, we propose the Hetero-EUCLID framework, wherein we extend the Bayesian-EUCLID framework \citep{joshi2022bayesian} to segment a heterogeneous solid and provide interpretable constitutive models for all constituents, using only surface displacement and boundary-averaged force data.
    
\begin{figure}[H]
        \centering
        \includegraphics[width=\linewidth]{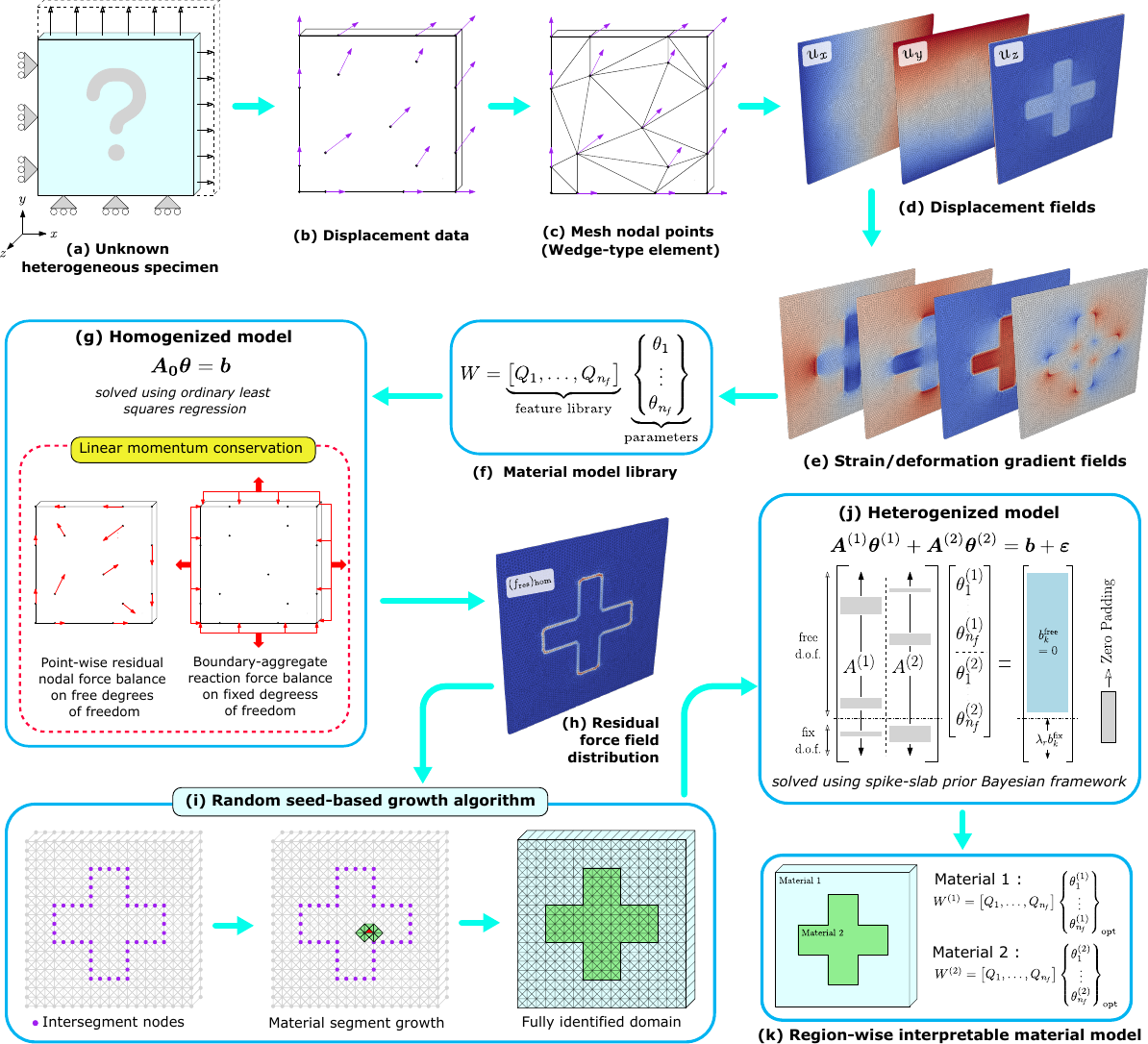}
        \caption{Schematic of the Hetero-EUCLID workflow for unsupervised constitutive model discovery in heterogeneous hyperelastic materials. (a) illustrates the specimen subjected to non-equi-biaxial tensile loading. (b) and (c) show the point-wise displacement observations and corresponding finite element interpolation onto a wedge-type mesh, respectively. (d) depicts the resulting full-field displacement field, while (e) presents the corresponding strain field obtained via spatial differentiation. (f) represents a catalogue of the hyperelastic constitutive models library. (g) outlines the homogenized model formulation incorporating the weak form of the linear momentum balance, which is solved using ordinary least squares regression to yield the residual force norm distribution shown in (h). Material segmentation based on the residual force indicator is performed using a random seed-based island growth algorithm, as illustrated in (i). (j) presents the unified global heterogenized system of equations assembled after material segmentation. The unknown constitutive model coefficients  $\boldsymbol{\theta}^{(i)}$ are solved for using the Bayesian EUCLID framework. (k) highlights the discovered region-wise interpretable constitutive models corresponding to each material segment within the heterogeneous specimen.}
        \label{main_illustration}
    \end{figure}

    The problem of constitutively modeling all the constituents in a heterogeneous geometry can be divided into two subtasks: segmenting the geometry and obtaining the constitutive model for all the materials involved. In \autoref{sec:methods} of the current work we present
    \begin{itemize}
        \item An overview of extending the Bayesian-EUCLID framework \citep{joshi2022bayesian}, which was originally developed for homogeneous materials, to model segmented homogeneous materials.
        \item A survey of existing segmenting techniques that rely only on surface displacement data, and the reason for our choice of the residual force method.
        \item A comment on displacement-noise and the steps taken to mitigate its effect on the proposed framework's results.
    \end{itemize}
    A schematic illustration of the proposed Hetero-EUCLID framework is provided in \autoref{main_illustration}. In \autoref{sec:results}, we present results of validation studies on square geometries containing two or more hyperelastic constituents. The validation studies utilize surface displacement data and boundary-averaged normal reaction force data from biaxial tension tests on thin hyperelastic specimens.

     For the scope of this study, we source displacement and boundary-averaged normal force data from three-dimensional finite element simulations. This is in contrast to previous similar studies, which focused on validating their results in 2D plane strain settings \citep{thakolkaran_nn-euclid_2022,shi2025deep}. In \autoref{sec:results}, we also validate the proposed Hetero-EUCLID framework on ``Non-native'' meshed geometries, wherein the Finite Element mesh used to perform the inverse Hetero-EUCLID analysis was different from the Finite Element mesh used to generate the displacement and boundary-averaged force. The paper concludes in \autoref{conclusion} with comments on the results of the study and potential future work.

\section{Hetero-EUCLID: Methodology}
\label{sec:methods}

\subsection{Incorporating heterogeneity}
    Consider a three-dimensional reference domain \(\mathit{\Omega}\in \mathbb{R}^3\), with boundary \(\partial \mathit{\Omega}\) in a plane stress condition subjected to quasi-static displacement-controlled non-equi-biaxial loading on \( \partial \mathit{\Omega}\). Let \(\chi = \{\boldsymbol{X}^a \in \mathit{\Omega} : a = 1,\dots,n_n \} \) denote \( n_n\) reference points in the given three-dimensional setting with displacement measurements as \( \{ \boldsymbol{u}^{a,t} \in \mathbb{R}^3 : a = 1,\dots,n_n \, ; \, t = 1,\dots,n_t\} \) for \(n_t\) time steps. In addition, the boundary-aggregate reaction force measurements at \(n_b \)(\( \ll n_n\)) load transducer points can be denoted by \( \{R^{k,t}_i : k = 1,\dots,n_b \, ; \, t=1,\dots,n_t \, ; \, i=1,2,3 \}\). \par
    
    \begin{figure}[H]
        \centering
        \includegraphics[width=0.6\linewidth]{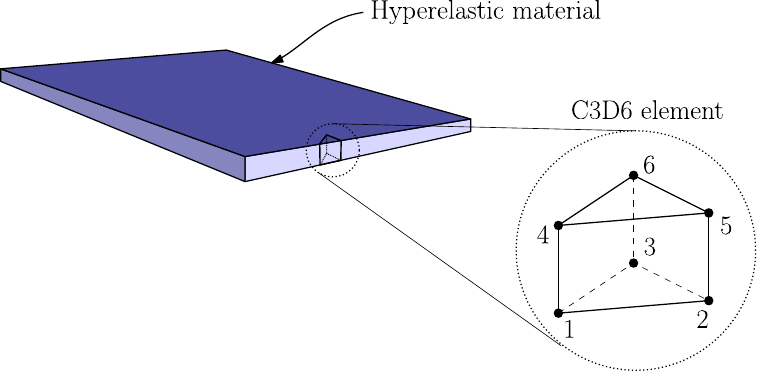}
        \caption{Meshing the given thin plate geometry with C3D6 elements with thickness identical to that of the plate.}
        \label{fig:meshing_example}
    \end{figure}
    
    As shown in \autoref{fig:meshing_example}, the given reference domain, a thin square plate, has been meshed with 6-node linear triangular prism elements (wedge or C3D6) using the nodal set \(\chi\). Given nodal displacements $\boldsymbol{u}^{a,t}$, the value of the displacement field at any point $\boldsymbol{X}$ in the volume is
    \begin{equation} 
        \boldsymbol{u^{t}(X)} = \sum_{a=1}^{n_n} \mathbb{N}^a (\boldsymbol{X}) \boldsymbol{u}^{a,t} 
        \label{eq1}
    \end{equation}
    
    where \( \mathbb{N}^a : \mathit{\Omega} \rightarrow \mathbb{R} \) denotes the interpolation shape function associated with the \(a^{th}\) node in the reference domain. The corresponding deformation gradient can thus be computed as
    \begin{equation} 
        \boldsymbol{F^{t}(X)} = \boldsymbol{I} + \sum_{a=1}^{n_n} \boldsymbol{u}^{a,t} \otimes \nabla \mathbb{N}^a (\boldsymbol{X}) 
        \label{eq2}
    \end{equation}
    
    where \( \nabla (\cdot) = \partial (\cdot) /\partial \boldsymbol{X} \) is defined as the gradient operator in the reference coordinates. Due to the use of linear shape functions in the wedge mesh type, $\nabla \mathbb{N}^a (X)$ remains constant within each element. For the scope of this work, a single-element-thick mesh is generated over the reference domain. This would imply that all nodes are located on the surface and their displacements $\boldsymbol{u}^{a,t}$ could be obtained through 3D-Digital Image Correlation (DIC) experiments \citep{luo_accurate_1993}. Although single-surface 3D-DIC experiments are somewhat routinely performed, it is challenging to simultaneously perform 3D-DIC to obtain the 3D displacement field on both-- the top and bottom surfaces of a thin deforming sample. Such dual-surface displacement field acquisition can be performed using dual-camera stereo-DIC, dual-surface DIC, or mirror-assisted multi-view DIC systems \citep{takada2023biaxial, genovese2021multi, sun2023study, chen2022mirror}. Furthermore, for brevity and without loss of generality, we employ only the data 
    corresponding to a single large deformation state, \( t = 1 \). We will henceforth be dropping the use of $t$ in the superscript. \par
    
    We now consider a reference domain composed of multiple hyperelastic material segments. Since each segment belongs to the hyperelastic material class, they can be assigned a strain energy density \( W(\boldsymbol{F}) \) and the corresponding first Piola–Kirchhoff stress tensor would be \( \boldsymbol{P}(\boldsymbol{F}) = \partial W / \partial \boldsymbol{F} \). In order to satisfy the material objectivity constraint, the strain energy density \(W(\boldsymbol{F})\), will be a function of the right Cauchy-Green deformation tensor \(\boldsymbol{C} = \boldsymbol{F}^T\boldsymbol{F}\) and can be defined as
    \begin{equation}
        W(\boldsymbol{C}) = W(I_1,I_2,I_3,I_a)
        \label{eq3}
    \end{equation}
    \begin{equation}
        I_1(\boldsymbol{C}) = \mathrm{tr}(\boldsymbol{C}), \quad 
        I_2(\boldsymbol{C}) = \frac{1}{2}[\mathrm{tr}^{2}(\boldsymbol{C}) - \mathrm{tr}(\boldsymbol{C}^2)], \quad
        I_3(\boldsymbol{C}) = \mathrm{det}(\boldsymbol{C}), \quad I_{a}(\boldsymbol{C},\boldsymbol{a}) = \boldsymbol{a} \cdot\boldsymbol{Ca}
        \label{eq4}    
    \end{equation}
    
    where \(I_1,I_2,I_3\) are isotropic invariants of right Cauchy-Green deformation tensor \(\boldsymbol{C}\) and \(I_a\) represents anisotropic invariant of \(\boldsymbol{C}\) with fiber orientation angle \(\alpha \) with \(\boldsymbol{a}  = (\cos{\alpha}, \sin{\alpha}, 0)^{T}\). \par
    
    We address the inverse problem of identifying the strain energy density function for a specimen subjected to a non-equi-biaxial loading test by introducing a material feature library \( \boldsymbol{Q} : \mathbb{R}^4 \rightarrow \mathbb{R}^{\mathit{n_f}} \), where \( \mathit{n_f} \) denotes the number of nonlinear feature functions. Accordingly, the constitutive relation in \autoref{eq3} can be reformulated using this feature library \( \boldsymbol{Q} \) and a vector of unknown material parameters \( \boldsymbol{\theta} \in \mathbb{R}^{\mathit{n_f}}_{+} \), consisting of non-negative scalar coefficients \citep{joshi2022bayesian}. 
    \begin{equation}
        W(I_1,I_2,I_3,I_a;\boldsymbol{\theta}) = \boldsymbol{Q}^{T}(I_1,I_2,I_3,I_a) \: \boldsymbol{\theta}
        \label{eq5}
    \end{equation}
    
    For the scope of this work, we choose a feature library which can be used to represent the compressible hyperelastic material models such as the Neo-Hookean \citep{treloar1943elasticity}, Isihara \citep{isihara1951statistical}, Biderman \citep{biderman1958calculation}, Haines-Wilson \citep{haines1979strain}, Ogden \citep{ogden1972large} and Holzapfel-Gasser-Ogden \cite{holzapfel2000new} models:
    
    \begin{align}
        \boldsymbol{Q}\left(I_{1}, I_{2}, I_{3}, I_{a}\right) &=\underbrace{\left[\left(\tilde{I}_{1}-3\right)^{i}\left(\tilde{I}_{2}-3\right)^{j-i}: j \in\{1, \dots, N_{\text{poly}}\}, i \in\{0, \dots, j\}\right]^{T}}_{\text{Generalized Mooney-Rivlin features }} \nonumber \\ & \quad  \oplus \quad \underbrace{\left[(J-1)^{2 k}: k \in\{1, \dots, N_{\text{vol}}\}\right]^{T}}_{\text{Volumetric energy features }} \quad \oplus \quad  \underbrace{\left[(\tilde{I}_{a}-1)^{l}: l \in\{2, \dots, N_{\text{aniso}}\}\right]^{T}}_{\text{Anisotropy features }} 
        \label{eq6}
    \end{align}
    
    where \(J = \mathrm{det}(\boldsymbol{F}) = I_3^{1/2}\) and \( \tilde{I}_1 = J^{-2/3}I_1, \: \tilde{I}_2 = J^{-4/3}I_2\) are the isotropic deviatoric invariants of the unimodular Cauchy-Green deformation tensor \((\boldsymbol{\tilde{C}} = J^{-2/3}\boldsymbol{C})\), \(\tilde{I}_{a} = J^{-2/3} (\boldsymbol{a} \cdot\boldsymbol{Ca}) \) is the anisotropic deviatoric invariant of \(\boldsymbol{\tilde{C}}\) and the symbol \(\oplus\) represents the concatenation of the features of the chosen library. The physical constraints such as objectivity, material symmetry, and a stress-free reference configuration (\( \boldsymbol{P} (\boldsymbol{F} = \boldsymbol{I}) = 0\)) are automatically satisfied \citep{joshi2022bayesian} by \autoref{eq5} and \ref{eq6}. It should be noted that any terms involving $(\tilde{I}_{2}-3)$ in the feature library $\boldsymbol{Q}$ are not polyconvex, and should ideally be replaced with $(\tilde{I}_{2}^{3/2}-3^{3/2})$ \citep{hartmann_polyconvexity_2003}. However, we choose to retain the more traditional $(\tilde{I}_{2}-3)$ term, which appears in the Isihara model \citep{isihara1951statistical}. The addition of this term to the feature library makes it convenient to benchmark the Hetero-EUCLID framework using synthetic surface displacement data from commercial finite element software that utilizes the Isihara model. If the $(\tilde{I}_{2}^{3/2}-3^{3/2})$ term were to be included in the feature library instead of the $(\tilde{I}_{2}-3)$ term, then the non-negativity of the entries of $\boldsymbol{\theta}$ automatically ensures that the strain energy density $W(\boldsymbol{F})$ will be polyconvex.\par
    
    Our approach is based on the stress-unsupervised EUCLID framework, where the learning of material models is performed without relying on experimentally inaccessible stress tensors. To achieve this, we incorporate additional physics-based constraints in the form of conservation of linear momentum. Assuming negligible body forces, the weak form of the linear momentum balance equation under quasi-static loading conditions is applied to the three-dimensional reference domain \( \mathit{\Omega} \) as follows
    \begin{equation}
        \int_{\Omega} \boldsymbol{P}: \nabla \boldsymbol{v} \mathrm{~d} V-\int_{\partial \Omega} \hat{\boldsymbol{t}} \cdot \boldsymbol{v} \mathrm{~d} S=\boldsymbol{0} \quad \forall \quad \boldsymbol{v} \text{ differentiable in } \Omega
        \label{eq7}
    \end{equation}
    where \( \hat{\boldsymbol{t}} \) represents the traction acting on \( \partial \Omega \). Using \autoref{eq5}, the first Piola–Kirchhoff stress tensor in indicial notation can be written as
    \begin{equation}
        P_{i\mathrm{J}} = \frac{\partial W(\boldsymbol{F})}{\partial F_{i\mathrm{J}}} =  \frac{\partial \boldsymbol{Q}^{T}(I_1,I_2,I_3)}{\partial F_{i\mathrm{J}}} \boldsymbol{\theta}
        \label{eq8}
    \end{equation}
    
    To compute the partial derivatives of the various features in the material library, \( \partial \boldsymbol{Q}^{T} / \partial F_{i\mathrm{J}} \), we use the automatic differentiation \citep{baydin2018automatic} feature in the torch library of Python. An advantage of using the weak form of the linear momentum balance is that it inherently reduces the sensitivity issues associated with the second-order spatial derivatives present in the strong form \citep{flaschel2021unsupervised}.\par
    
    Let \(\mathcal{D} = \{(a,i) : a=1,\dots,n_n \, ; \, i = 1,2,3 \}\) represents the set of all nodal d.o.f. (degrees of freedom) for the three-dimensional discretized reference domain. The set \(\mathcal{D}\) can be further divided into two subsets, \(\mathcal{D}^{\text{free}} \subset \mathcal{D}\) consisting of free d.o.f while \(\mathcal{D}^{\text{fix}} = \mathcal{D} \backslash \mathcal{D}^{\text{free}} \) features the fixed d.o.f (via Dirichlet constraints). Using the Bubnov-Galerkin approximation, the test functions are computed as follows
    \begin{equation}
        v_i(\boldsymbol{X}) = \sum_{a=1}^{n_n} \mathbb{N}^a(\boldsymbol{X})v_i^a
        \label{eq9}
    \end{equation}
    
    Substituting  \(\boldsymbol{v}\) from \autoref{eq9} into \autoref{eq7}, the weak form then can be written as
    \begin{equation}
        \sum_{a=1}^{n_{n}} \left[\int_{\Omega} P_{i\mathrm{J}} \mathbb{N}_{,\mathrm{J}}^{a}(\boldsymbol{X}) v_{i}^{a} \mathrm{~d} V - \int_{\partial \Omega} \hat{t_{i}} \mathbb{N}^{a}(\boldsymbol{X}) v_{i}^{a} \mathrm{~d} S\right] = 0
        \label{eq10}
    \end{equation}
    
    Now, we choose various nodal values of the test function \( \boldsymbol{v}^a \) to generate equations to solve for the constitutive properties of the material
    \begin{enumerate}[label=(\alph*)]
        \item \( v_i^a = 0\)  \(\forall\)  \((a,i) \in \mathcal{D}^{\mathrm{fix}} \)  -- \textit{Homogeneous Dirichlet boundary condition}
        
        Substituting the value of \( v_i^a\) in \autoref{eq10}, the modified equation is obtained as
        \begin{equation}
             \sum_{a=1}^{n_{n}} \left[ v_{i}^{a}\int_{\Omega} P_{i\mathrm{J}} \mathbb{N}_{,\mathrm{J}}^{a}(\boldsymbol{X}) \mathrm{~d} V \right] = 0 
            \label{eq11}
        \end{equation}
        It is easy to show that \autoref{eq11} holds for all values of \( v_i^a \) if and only if (substituting \(P_{i\mathrm{J}}\) using \autoref{eq8})
        \begin{equation}
            \int_{\Omega} \left(\frac{\partial \boldsymbol{Q}^{T}}{\partial F_{i\mathrm{J}}} \boldsymbol{\theta}\left(\boldsymbol{X}\right)\right)  \mathbb{N}_{,\mathrm{J}}^{a}(\boldsymbol{X}) \mathrm{~d} V = 0 \quad \forall \quad (a,i) \in \mathcal{D}^{\text{free}} (= \mathcal{D} \backslash \mathcal{D}^{\text{fix}} )
            \label{eq12}
        \end{equation}
        \autoref{eq12} consists of an integral over the entire domain $\Omega$. For a domain consisting of $n_c$ material segments, $\Omega = \cup_{k=1}^{n_c}\Omega^{(k)}$. We can thus rewrite \autoref{eq12} as follows
    
        \begin{equation}
            \int_{\Omega^{(1)}} \left(\frac{\partial \boldsymbol{Q}^{T}}{\partial F_{i\mathrm{J}}} \boldsymbol{\theta^{(1)}}\right)  \mathbb{N}_{,\mathrm{J}}^{a}(\boldsymbol{X}) \mathrm{~d} V + \dots + \int_{\Omega^{(n_c)}} \left(\frac{\partial \boldsymbol{Q}^{T}}{\partial F_{i\mathrm{J}}} \boldsymbol{\theta^{(n_c)}}\right)  \mathbb{N}_{,\mathrm{J}}^{a}(\boldsymbol{X}) \mathrm{~d} V= 0 \quad \forall \quad (a,i) \in \mathcal{D}^{\text{free}}
            \label{eq:splitintegrals}
        \end{equation}
    
        Within each integral, the material model  $\boldsymbol{\theta}\left(\boldsymbol{X}\right) = \boldsymbol{\theta^{(k)}}, \text{for } \boldsymbol{X}\in \Omega^{(k)};k\in\{1,\dots,n_c\}$. Since $ \boldsymbol{\theta^{(k)}}$ is constant in each material segment, it can be taken out of the integral. \autoref{eq:splitintegrals} can thus be rewritten as
        \begin{equation}
                \boldsymbol{A}^{\text{free}(1)} \boldsymbol{\theta}^{(1)} \; + \; \boldsymbol{A}^{\text{free}(2)} \boldsymbol{\theta}^{(2)} \; + \: \cdots \: + \; \boldsymbol{A}^{\text{free}(j)} \boldsymbol{\theta}^{(j)} \; + \: \cdots \: + \; \boldsymbol{A}^{\text{free}(n_c)} \boldsymbol{\theta}^{(n_c)} = \boldsymbol{0}
                \label{eq:MultipleAsbs}
        \end{equation}
        
        where $\boldsymbol{A}^{\text{free}(k)} \in \mathbb{R}^{|\mathcal{D}^{\text{free}}| \times n_{f}}\quad \forall \quad k \in \{1\dots n_c\} \quad \text{and} \quad \boldsymbol{b}^{\text{free}} \in \mathbb{R}^{|\mathcal{D}^{\text{free}}|}$. \autoref{eq:MultipleAsbs} can be further written as
        \begin{equation}
            \boldsymbol{A}^{\text{free}} \boldsymbol{\theta}=\boldsymbol{b}^{\text{free}} (= \boldsymbol{0})
            \label{eq13}
        \end{equation}
        where $\boldsymbol{A}^{\text{free}} = \left[\boldsymbol{A}^{\text{free}(1)} \quad \boldsymbol{A}^{\text{free}(2)} \dots \, \boldsymbol{A}^{\text{free}(n_c)}\right]$ and $\boldsymbol{\theta} = \left[\boldsymbol{\theta}^{(1)}\quad\boldsymbol{\theta}^{(2)}\dots \, \boldsymbol{\theta}^{(n_c)}\right]^T$
        
        \item \((v_i^a)_{k} = \begin{cases} 
                          1 & \forall \: (a,i) \in \mathcal{D}^{\mathrm{fix}}_{k} \\
                          0 & \text{otherwise}
                       \end{cases} \quad \text{for } k = \{1,\dots,n_b\}\)

            where, $\mathcal{D}_k^{\text{fix}}\subset\mathcal{D}^{\text{fix}}$ denotes the set of degree of freedoms belonging to the $k^{\text{th}}$ boundary $\partial\Omega_k$ and \(n_b\) denotes the total number of boundaries where force is measured. Substituting the value of chosen test function \( (v_i^a)_{k}\), \autoref{eq10} gets modified and results in \(n_b\) set of equations (one at every boundary) as follows
            \begin{equation}
                \int_{\Omega^{e}_{k}}  P_{i\mathrm{J}} \mathbb{N}_{,\mathrm{J}}^{a}(\boldsymbol{X})  \mathrm{~d} V - \int_{\partial \Omega_k} \hat{t_{i}} \mathbb{N}^{a}(\boldsymbol{X}) \mathrm{~d} S = 0
                \label{eq14}
            \end{equation}
    
            where the term $\int_{\partial \Omega_k} \hat{t_{i}} \mathbb{N}^{a}(\boldsymbol{X}) \mathrm{~d} S$ can be interpreted as the total force experimentally recorded on boundary $k$ along the $i$ direction, denoted by $R_i^k$. \(\Omega^{e}_{k}\) represents volume enclosed by the set of mesh elements \(e\) for which at least one node lies on the external boundary $\partial \Omega_k$ of the reference domain \(\Omega\). It is important to note that \(\Omega^{e}_{k}\) is different from \(\Omega^{(k)}\) defined earlier, wherein the latter corresponds to the volume enclosed by the $k^{\text{th}}$ material segment. Further, \autoref{eq14} can be recast as
            \begin{equation}
                \int_{\Omega^{e}_{k}}  P_{i\mathrm{J}} \mathbb{N}_{,\mathrm{J}}^{a}(\boldsymbol{X})  \mathrm{~d} V = R_i^k
                \label{eq:Rik}
            \end{equation}

 Using \autoref{eq8} and the methodology adopted in \autoref{eq:splitintegrals}--\ref{eq13}, \autoref{eq:Rik} can be written as
        \begin{equation}
        \begin{pmatrix}
            \int_{\Omega^{e(1)}_{1}}  \frac{\partial \boldsymbol{Q}^{T}}{\partial F_{i\mathrm{J}}} \mathbb{N}_{,\mathrm{J}}^{a}(\boldsymbol{X})  \mathrm{~d} V  \quad \dots \quad \int_{\Omega^{e(n_c)}_{1}}  \frac{\partial \boldsymbol{Q}^{T}}{\partial F_{i\mathrm{J}}} \mathbb{N}_{,\mathrm{J}}^{a}(\boldsymbol{X})  \mathrm{~d} V\\
            \vdots \\
            \int_{\Omega^{e(1)}_{k}}  \frac{\partial \boldsymbol{Q}^{T}}{\partial F_{i\mathrm{J}}} \mathbb{N}_{,\mathrm{J}}^{a}(\boldsymbol{X})  \mathrm{~d} V   \quad \dots \quad \int_{\Omega^{e(n_c)}_{k}}  \frac{\partial \boldsymbol{Q}^{T}}{\partial F_{i\mathrm{J}}} \mathbb{N}_{,\mathrm{J}}^{a}(\boldsymbol{X})  \mathrm{~d} V\\
            \vdots \\
            \int_{\Omega^{e(1)}_{n_b}}  \frac{\partial \boldsymbol{Q}^{T}}{\partial F_{i\mathrm{J}}} \mathbb{N}_{,\mathrm{J}}^{a}(\boldsymbol{X})  \mathrm{~d} V   \quad \dots \quad \int_{\Omega^{e(n_c)}_{n_b}}  \frac{\partial \boldsymbol{Q}^{T}}{\partial F_{i\mathrm{J}}} \mathbb{N}_{,\mathrm{J}}^{a}(\boldsymbol{X})  \mathrm{~d} V
        \end{pmatrix}\boldsymbol{\theta} =
        \begin{pmatrix}
            R^{1}_{i} \\
            \vdots \\
            R^{k}_{i} \\
            \vdots \\
            R^{n_b}_{i}
        \end{pmatrix}
        \quad \text{where} \quad 
            \begin{split}
              i = 1, 2, 3
            \end{split}
        \label{eq15}
    \end{equation}
    \end{enumerate}
    
    where $\Omega_k^{e(j)}$ denotes the volume enclosed by elements of material segment $j$ in the volume $\Omega_k^e$, and $\boldsymbol{\theta}$ retains its definition from \autoref{eq13}. However, in routine material testing experiments that do not use specialized shear beam load cells, the load cells measure only one component of force, i.e, normal force. Designating the $i^{\text{th}}$ component of the normal vector and magnitude of the normal force at the $k^{\text{th}}$ boundary as $n^k_i$ and $F^{k}$, respectively, we have,
    
    \begin{equation}
        F^k = \sum_{i=1}^3 n^k_i R_i^k \qquad k=1,2,\dots,n_b
    \end{equation}
    Thus, \autoref{eq15} gets modified to:
    \begin{equation}
        \begin{pmatrix}
            \sum_{i=1}^3n_i^k\int_{\Omega^{e(1)}_{1}}  \frac{\partial \boldsymbol{Q}^{T}}{\partial F_{i\mathrm{J}}} \mathbb{N}_{,\mathrm{J}}^{a}(\boldsymbol{X})  \mathrm{~d} V  \quad \dots \quad \sum_{i=1}^3n_i^k\int_{\Omega^{e(n_c)}_{1}}  \frac{\partial \boldsymbol{Q}^{T}}{\partial F_{i\mathrm{J}}} \mathbb{N}_{,\mathrm{J}}^{a}(\boldsymbol{X})  \mathrm{~d} V\\
            \vdots \\
            \sum_{i=1}^3n_i^k\int_{\Omega^{e(1)}_{k}}  \frac{\partial \boldsymbol{Q}^{T}}{\partial F_{i\mathrm{J}}} \mathbb{N}_{,\mathrm{J}}^{a}(\boldsymbol{X})  \mathrm{~d} V   \quad \dots \quad \sum_{i=1}^3n_i^k\int_{\Omega^{e(n_c)}_{k}}  \frac{\partial \boldsymbol{Q}^{T}}{\partial F_{i\mathrm{J}}} \mathbb{N}_{,\mathrm{J}}^{a}(\boldsymbol{X})  \mathrm{~d} V\\
            \vdots \\
            \sum_{i=1}^3n_i^k\int_{\Omega^{e(1)}_{n_b}}  \frac{\partial \boldsymbol{Q}^{T}}{\partial F_{i\mathrm{J}}} \mathbb{N}_{,\mathrm{J}}^{a}(\boldsymbol{X})  \mathrm{~d} V   \quad \dots \quad \sum_{i=1}^3n_i^k\int_{\Omega^{e(n_c)}_{n_b}}  \frac{\partial \boldsymbol{Q}^{T}}{\partial F_{i\mathrm{J}}} \mathbb{N}_{,\mathrm{J}}^{a}(\boldsymbol{X})  \mathrm{~d} V
        \end{pmatrix}\boldsymbol{\theta} =
        \begin{pmatrix}
            F^{1} \\
            \vdots \\
            F^{k} \\
            \vdots \\
            F^{n_b}
        \end{pmatrix}
        \label{eq:revised_fix}
    \end{equation}
    
 \autoref{eq:revised_fix} results in a system of \(n_b\) linear equations (one corresponding to each load-cell reading) following numerical integration and thus can be assembled as follows
    \begin{equation}
        \boldsymbol{A}^{\text{fix}} \boldsymbol{\theta}=\boldsymbol{b}^{\text{fix}} \quad \text{with} \quad \boldsymbol{A}^{\text{fix}} \in \mathbb{R}^{n_{b} \times n_{f}n_c} \quad \text{and} \quad \boldsymbol{b}^{\text{fix}} \in \mathbb{R}^{n_{b}}
        \label{eq16}
    \end{equation}  
    
    By combining \autoref{eq13} and \ref{eq16}, we formulate a unified model of the form \( \boldsymbol{A \theta} = \boldsymbol{b} \), where a hyperparameter \( \lambda_r \) is introduced to control the relative weights between the free and fixed degrees of freedom in the reference domain within the force balance equation.
    \begin{equation}
        \boldsymbol{A \theta} = \boldsymbol{b} \quad \text{with} \quad \boldsymbol{A} = \begin{bmatrix}
                \boldsymbol{A}^{\text{free}}  \\
                \lambda_r \boldsymbol{A}^{\text{fix}}
                \end{bmatrix}
        \quad \text{and} \quad \boldsymbol{b} = \begin{bmatrix}
                \boldsymbol{b}^{\text{free}}  \\
                \lambda_r \boldsymbol{b}^{\text{fix}}
                \end{bmatrix}
        \label{eq17}
    \end{equation}
    
    \begin{figure}[t]
        \centering
        \includegraphics[width=0.85\linewidth]{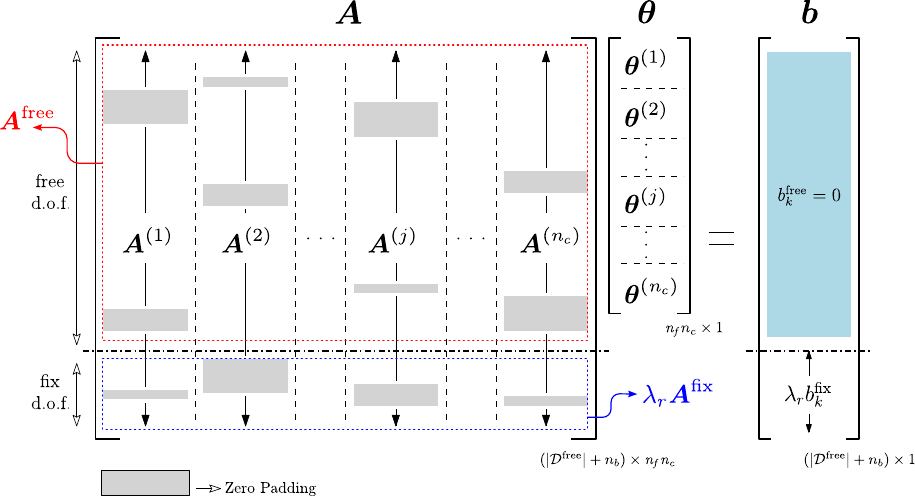}
        \caption{Graphical illustration of the unified global heterogenized model \( \boldsymbol{A\theta} = \boldsymbol{b} \). The matrix \( \boldsymbol{A} \) comprises \( \mathcal{D}^{\text{free}} \) rows corresponding to the free degrees of freedom, along with additional rows for the Dirichlet boundary nodes \( \mathcal{D}^{\text{fix}} \). Each segment-specific block \( \boldsymbol{A}^{(j)} \), containing \( \mathit{n_f} \) columns associated with the constitutive feature library \( \boldsymbol{Q} \), is zero-padded (gray regions) to ensure dimensional compatibility. It is assembled via column-wise concatenation over \( n_c \) identified material segments. The parameter vector \( \boldsymbol{\theta} \in \mathbb{R}^{\mathit{n_f n_c} \times 1} \) encodes the model coefficients for each material segment. The right-hand side vector \( \boldsymbol{b} \) has two components: \( \boldsymbol{b}^{\text{free}} = \boldsymbol{0} \) representing equilibrium at internal nodes, and \( \boldsymbol{b}^{\text{fix}} \) containing the boundary-aggregated reaction force.}
        \label{Atheta_b}
    \end{figure}
    
    \autoref{Atheta_b} provides a schematic to aid in better understanding the compact \autoref{eq17}. The global matrix \( \boldsymbol{A} \) is assembled such that its rows contain non-zero entries corresponding to \( \boldsymbol{A}^{(j)} \) only if the associated degrees of freedom belong to material segment \( j \), as illustrated in \autoref{Atheta_b}.\par
    
    To ensure dimensional compatibility, zero-padding is applied to the rows (d.o.f.) of \( \boldsymbol{A}^{(j)} \) not belonging to the material segment \( j \). The resulting right-hand side vector \( \boldsymbol{b} \) consists of two components: \( \boldsymbol{b}^{\text{free}} \), representing the zero internal force condition at nodes in \( \mathcal{D}^{\text{free}} \), and \( \boldsymbol{b}^{\text{fix}} \), denoting the boundary-aggregated reaction forces \( R^{k}_{i} \) $\left(k\in\{1,\dots, n_b\}\, ; \, i(k) = 1 \text{ or }2\right)$ for the subset \( \mathcal{D}^{\text{fix}} \).\par
    
    Thus, as shown in \autoref{Atheta_b}, the problem of obtaining each segment's material parameters $\boldsymbol{\theta}^{(j)}$, can be recast into the form $\boldsymbol{A \theta} = \boldsymbol{b}$. From the shape of $\boldsymbol{A}$, it can be noted that $\left|\mathcal{D}^{\text{free}}\right|+n_b>> n_fn_c$. This would imply $\boldsymbol{A \theta} = \boldsymbol{b}$ is an overdetermined system, and the solution for $\boldsymbol{\theta}$ can be obtained in an ordinary least-square sense as    
    \begin{equation}
        \label{eqn:ols}
        \boldsymbol{\theta}^{\text{ols}} = (\boldsymbol{A}^T\boldsymbol{A})^{-1}\boldsymbol{A}^T\boldsymbol{b}
    \end{equation}
    
    However, such a solution would suffer from two main drawbacks -- (i) the solution would be susceptible to overfitting and displacement noise \citep{flaschel_unsupervised_2021} (see supplementary subsection S1.2), (ii) we would be unable to ensure that each component of $\boldsymbol{\theta}$ would be non-negative, which is essential to ensure polyconvexity of the strain energy density \citep{joshi2022bayesian}. Thus, following \citep{joshi2022bayesian} and \citep{nayek2021spike}, we formulate the equilibrium problem as a sparse-prior Bayesian regression for $\boldsymbol{\theta}$. The spike-slab prior \citep{nayek2021spike} is used for $\boldsymbol{\theta}$ to enforce parsimony, avoid overfitting, and provide noise-robustness.
    
    The equilibrium \autoref{eq17} can be rewritten as
    \begin{equation*}
        \boldsymbol{b} = \boldsymbol{A \theta} + \boldsymbol{\epsilon}
    \end{equation*}
    where $\epsilon \thicksim \mathcal{N}(\boldsymbol{0},\sigma^2\boldsymbol{I})$ is sampled from an independent and identical normal distribution with mean as $\boldsymbol{0}$ and variance as $\sigma^2$.
    The above equation can be rewritten as
    \begin{equation}
        \boldsymbol{b} \: | \: \boldsymbol{\theta}, \sigma^2, \boldsymbol{A} \sim \mathcal{N}(\boldsymbol{A \theta},\sigma^2 \boldsymbol{I})
        \label{eq20}
    \end{equation}
    
    \autoref{eq20} forms a likelihood probability, and using Bayes' rule, we can extract the posterior distribution
    $p(\boldsymbol{\theta}, \boldsymbol{z}, p_0, \nu_s, \sigma^2 \: | \: \boldsymbol{A}, \boldsymbol{b})$ using the approach outlined in \citet{joshi2022bayesian}. For brevity, we do not reiterate the Bayesian-EUCLID approach in the main text. A synopsis of the approach is provided in \ref{BayesianEUCLID}.
    The solution for the material models $\boldsymbol{\theta}_{(k)}\left(k\in\{1\dots N_G\}\right)$ would be a collection of $N_G$ number of values corresponding to the retained elements of the Markov chain, which approximate the posterior distribution for $\boldsymbol{\theta}$. For practical purposes, such as use in forward finite element simulations, it would be convenient to use the approximated mean of the distribution \(\boldsymbol{\theta}_{\text{mean}} = \dfrac{\sum_{k\in \{1\dots N_{G}\}}\boldsymbol{\theta}_{(k)}}{N_G}\) as material model parameters. The standard deviation $\boldsymbol{\theta}_{\text{std}} = \sqrt{\dfrac{\sum_{k\in \{1\dots N_{G}\}}\left(\boldsymbol{\theta}_{(k)}-\boldsymbol{\theta}_{\text{mean}}\right)^2}{N_G}}$ provides an approximate measure of the uncertainty in the prediction. The distribution for $\boldsymbol{\theta}$ is multivariate, and $\boldsymbol{\theta}_{\text{std}}$ provides only a practical, but approximate measure of the uncertainty associated with the material models. In this study, we provide the results as $\boldsymbol{\theta}_{\text{mean}}\pm \boldsymbol{\theta}_{\text{std}}$, instead of the violin plots in \citet{joshi2022bayesian}.
    
    \subsection{Comment on \autoref{eq20}}\label{subsec:corr_noise}
    
    An important assumption that is made in writing \autoref{eq20} is that the ``noise'' term $\boldsymbol{\epsilon}$ can be sampled from a normal distribution with a diagonal covariance matrix. Since \autoref{eq20} forms a likelihood of satisfying the weak form of the linear momentum balance, the term $\boldsymbol{A\theta-b}$ can be interpreted as a nodal residual force $\boldsymbol{f}_{\mathrm{res}}$. Thus, $\boldsymbol{\epsilon}$ serves as a ``noise'' for the nodal residual force. However, as shown by \citet{avril_general_2007} for the deformation of a linear elastic solid, a spatially uncorrelated displacement noise induces a spatially correlated noise in the nodal residual force $\boldsymbol{f}_{\mathrm{res}}$($=\boldsymbol{A\theta-b}$). This implies that $\boldsymbol{\epsilon}$ cannot be sampled from a normal distribution with a diagonal covariance matrix. In \ref{MoransI_study} we perform spatial correlation studies on the residual force vector field $\boldsymbol{f}_{\mathrm{res}}$ for a homogeneous hyperelastic material to verify that for a given displacement field, the nodal residual forces are indeed spatially correlated, i.e, possess a non-diagonal covariance matrix. Further, in \ref{LikelihoodMSEOLS} we demonstrate that a similar incorrect assumption regarding the lack of spatial correlation in nodal residual forces is made by many important contemporary stress-unsupervised constitutive modeling frameworks that use the mean-squared error loss function \citep{thakolkaran_nn-euclid_2022,shi2025deep,meng_machine-learning-based_2025} or use a variation of the ordinary least square solution to determine the material parameters \citep{haustrate2024mechanical,flaschel_unsupervised_2021}. However, as will be seen in \autoref{sec:results}, the mean value of the material parameters ($\boldsymbol{\theta}_{\text{mean}}$) obtained from sampling the posterior probability distribution are predicted with good accuracy, despite the incorrect assumption made in \autoref{eq20}. By accounting for spatial correlations in the residual force vector in \autoref{eq20}, we anticipate more robust uncertainty quantification of the predicted material parameters. In the current work, we proceed with the use of \autoref{eq20} as the likelihood distribution to obtain the posterior material parameter distribution, without accounting for spatial correlations in the nodal residual force.
    \begin{figure}[H]
        \centering
        \includegraphics[width=0.98\linewidth]{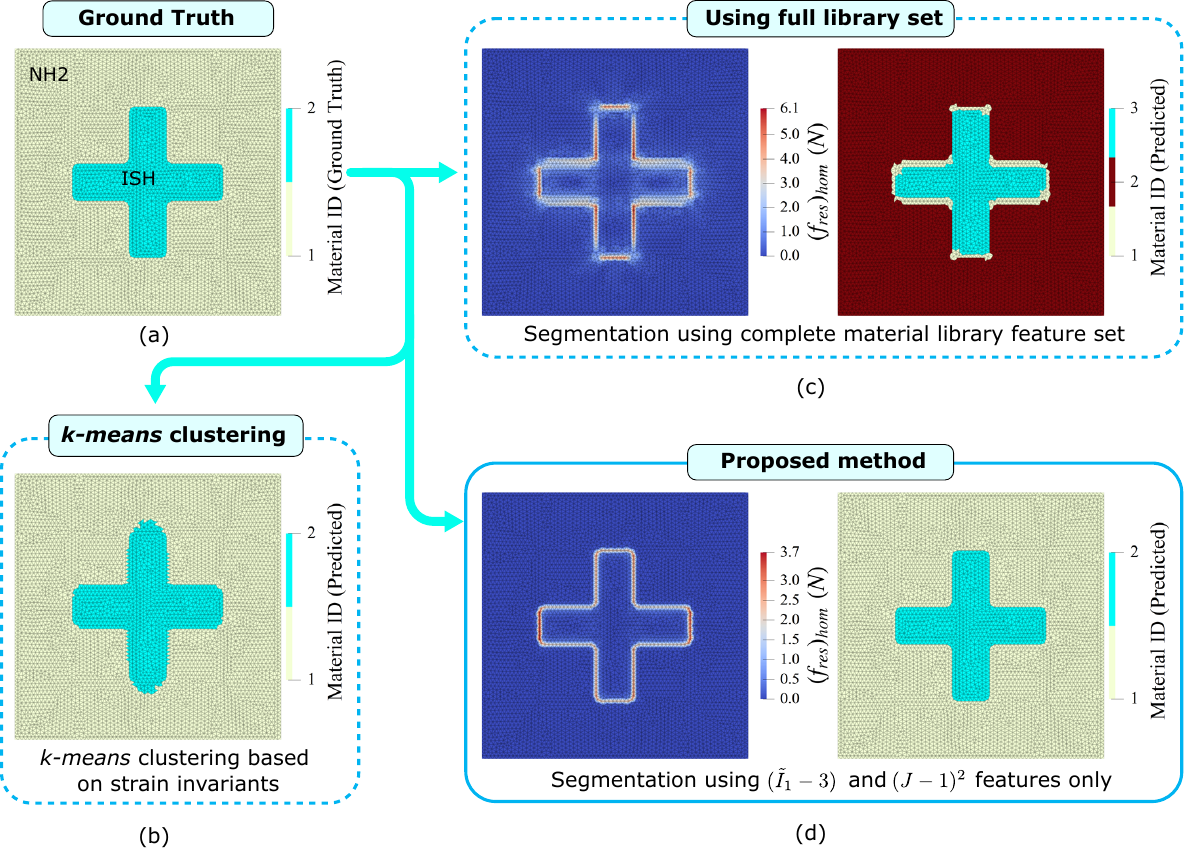}
        \caption{Comparative study of different segmentation techniques. (a) shows the ground truth for a heterogeneous bi-material specimen consisting of a cross-shaped inclusion governed by the Isihara (ISH) constitutive model, embedded in a base matrix described by the Neo-Hookean (NH2) model. (b) presents the segmentation result using \( k \)-means clustering based on strain invariants \citep{nguyen2024segmenting}. (c) utilizes the complete material feature library (\autoref{eq6}) to compute the 2D residual force norm distribution \((f^a_{\text{res}})\), and segments the material region by flagging nodes with residual force values above a specified threshold. (d) illustrates the result from our proposed method, where segmentation is performed using a reduced Neo-Hookean feature set  \( (\tilde{I}_1 - 3) \) and \( (J - 1)^2 \), yielding clean and accurate material cluster identification in strong agreement with the ground truth.}
        \label{clsuter_compare}
    \end{figure}

\subsection{Segmenting based on displacement data}
\label{segmentation}

    \citet{nguyen2024segmenting} provide a study on segmenting mechanically heterogeneous hyperelastic materials using surface displacement data. The isotropic invariants of the Cauchy-Green tensor ($I_1, I_2$ and $I_3$), alongside the displacement vector at the centroid of each finite element, were used as features to cluster each element using either \textit{k}-means clustering or spectral clustering. This approach was validated against various heterogeneity patterns, including circular inclusions, cross-shaped inclusions, Cahn-Hillard patterns \citep{kobeissi2022enhancing}, and split domains. In most cases, their results found that the segmented regions approximated the underlying true segments and contained non-smooth boundaries. Furthermore, their results did not account for correlated or uncorrelated noise in the displacement field, which could be expected to deteriorate the segmentation further. Subpart (b) of \autoref{clsuter_compare} provides results of \textit{k}-means clustering of strain invariants and displacements from a non-equi-biaxial tension test on a square geometry with a cross-shaped inclusion. In \autoref{clsuter_compare}b, it can be seen that a significant number of elements are mis-clustered. \ref{flagged-subsample} demonstrates that even 25 mis-clustered elements can deteriorate material property prediction accuracy by around 30\%. Thus, any adopted segmentation approach should provide accurate boundaries in order to achieve reasonably accurate material property predictions.\par

    \begin{figure}[H]
        \centering
        \includegraphics[width=0.9\linewidth]{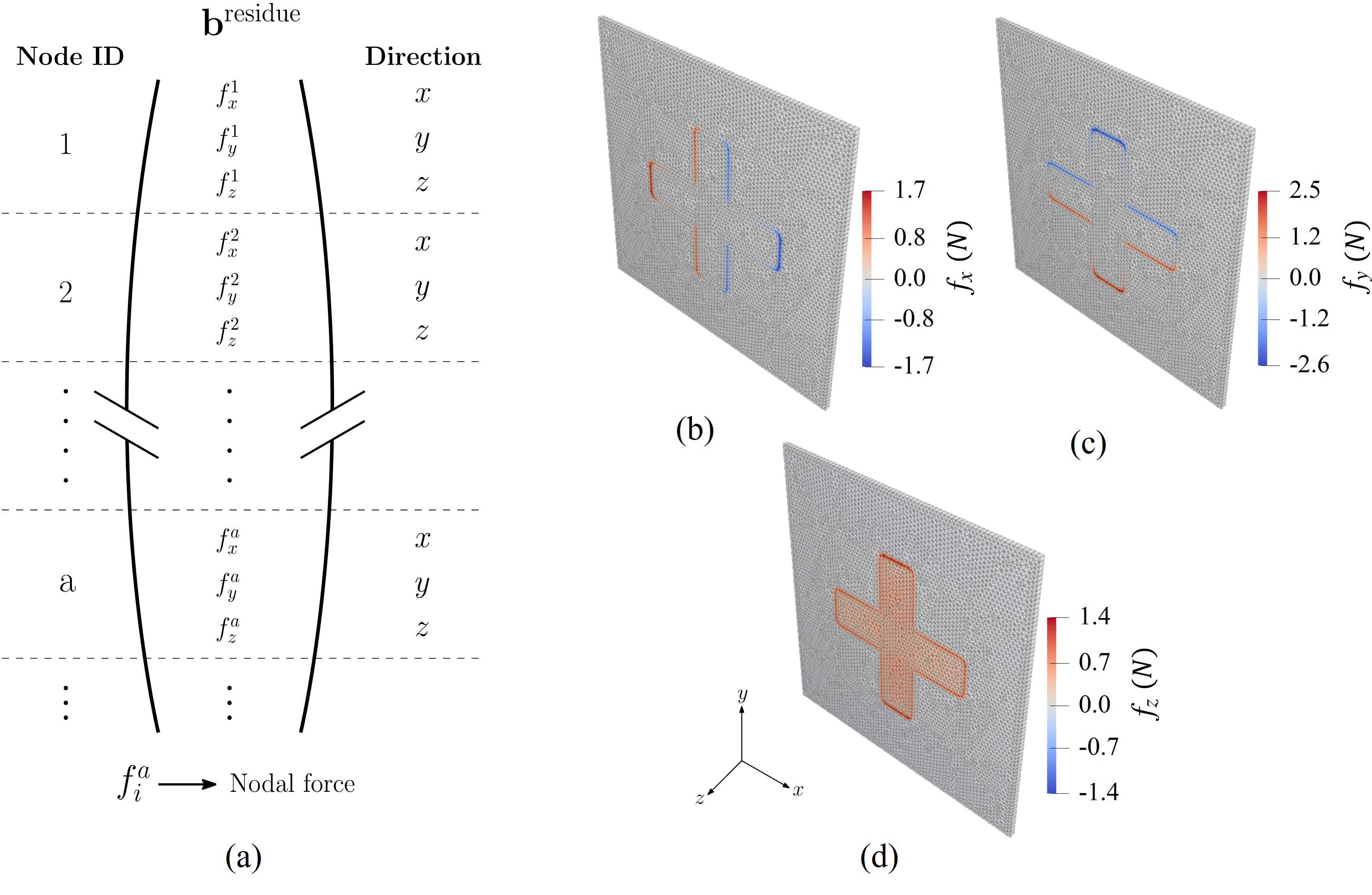}
        \caption{(a) Schematic of extracting nodal forces $\boldsymbol{f}^a$ from $\boldsymbol{b}^{\text{residue}}$, (b)-(d) shows the three components of residual force.}
        \label{fig:ResForce_xyz}
    \end{figure}
    
    In this work, we adopt a two-step approach to achieve segmentation. The first step involves identifying the inter-segment boundaries in the unknown material, and the second step utilizes these identified boundaries to register all elements with their respective segments. The identification of inter-segment boundaries is based on using ``residual forces'' within the geometry, assuming a homogeneous model. The $\boldsymbol{A}$ (\autoref{eq17}) matrix is initially constructed assuming that the geometry contains only one segment (homogeneous) and only the NH2 features,  \( (\tilde{I}_1 - 3) \) and \( (J - 1)^2 \) in the library. This is done only for segmenting the heterogeneous material to get a sharper inter-material interface (see \autoref{clsuter_compare}d), and in the subsequent Bayesian-EUCLID solution step, we keep all features in the library active. The homogeneous matrix $\boldsymbol{A}$ constructed here is denoted by $\boldsymbol{A}_0 (\in \mathbb{R}^{(|\mathcal{D}^{\text{free}}|+ n_b)\times 2})$ to disambiguate it from the fully segmented $\boldsymbol{A}$ matrix that will be used in the Bayesian-EUCLID step. As seen in the discussion preceding \autoref{eq13}, $\boldsymbol{b}^{\text{free}}$ indicates the net force acting on each node within the volume of the material. For an appropriate guess of non-trivial material parameters ($\boldsymbol{\theta^{*}}$), we would have $\boldsymbol{A^{\text{free}}}\boldsymbol{\theta}^{*} = \boldsymbol{b}^{\text{free}} = \boldsymbol{0}$. However, no non-trivial solution exists for $\boldsymbol{A^{\text{free}}}\boldsymbol{\theta} = \boldsymbol{0}$. The Ordinary-Least Square (OLS) approximation $\boldsymbol{\theta^{\text{ols}}}$ (\autoref{eqn:ols}) leaves the residue $\boldsymbol{A^{\text{free}}}\boldsymbol{\theta^{\text{ols}}} = \boldsymbol{b^{\text{residue}}} \neq \boldsymbol{b^{\text{free}}}$. Thus, we compute $\boldsymbol{b^{\text{residue}}}$ for the homogeneous material system $\boldsymbol{A}_0$ as follows
    \begin{equation}
        \label{eqn:bresidue}
        \boldsymbol{b^{\text{residue}}} = \boldsymbol{A_0^{\text{free}}}\left((\boldsymbol{A}_0^T\boldsymbol{A}_0)^{-1}\boldsymbol{A}_0^T\boldsymbol{b}\right)
    \end{equation}
    
    For a node $a : (a,i) \in \mathcal{D}^{\text{free}}$, the nodal force ($\boldsymbol{f^{a}}$) along each direction is obtained using $\boldsymbol{b^{\text{residue}}}$ as shown in \autoref{fig:ResForce_xyz}. Thus, based on the homogenized model, we compute $\boldsymbol{b}^{\text{residue}}$ and then extract the residual nodal forces $\boldsymbol{f}^a$ (see \autoref{fig:ResForce_xyz}a). We then make use of the residual nodal forces to demarcate the boundaries between two material segments by computing the 2D magnitude (ignoring $f_{z}^{a}$) as $f_{\text{res}}^a$
    \begin{equation}
        f_{\mathrm{res}}^{a} = \sqrt{(f_{x}^{a})^2 + (f_{y}^{a})^2} \quad \forall \quad a \in \boldsymbol{N}_{\text{free}}
        \label{eq19}
    \end{equation}
    where,
    \begin{equation}
        \boldsymbol{N}_{\text{free}} = \left\{a: (a,i) \in \mathcal{D}^{\text{free}} \quad \forall \quad i \in \{1,2,3\}\right\}
        \label{eq:Nfree}
    \end{equation}

    \begin{figure}[H]
        \centering
        \includegraphics[width=0.9\textwidth]{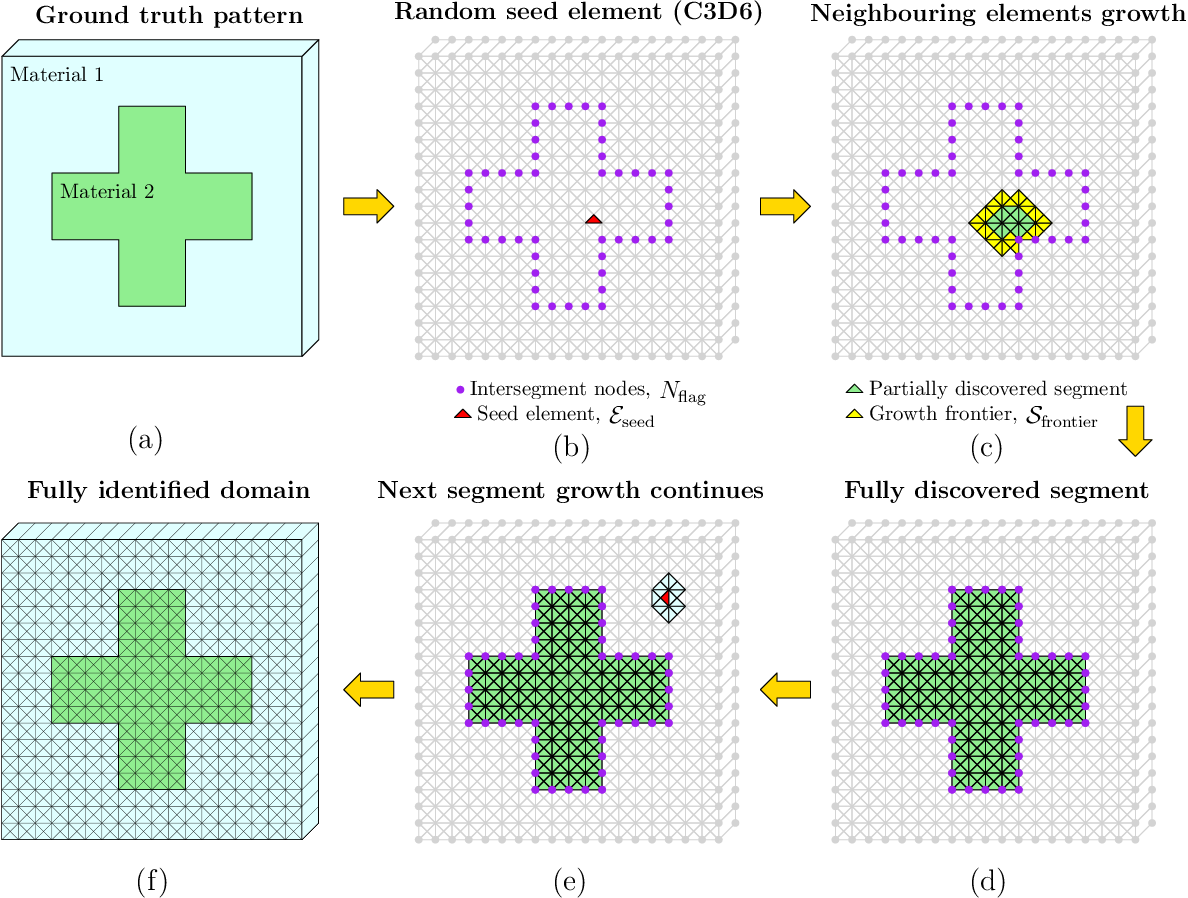}
        \caption{(a)-(f) Schematic illustration of the random seed-based growth algorithm for segmenting material sub-domains. The algorithm initiates from a randomly selected seed element (wedge type) and iteratively grows by adding neighboring elements until it reaches flagged interface boundaries. A new seed element is then chosen from the remaining unvisited elements, and the growth process is repeated until all distinct regions are identified, terminating at either an interface or an external boundary.}
        \label{GrowthAlgo}
    \end{figure}
    
    Since, we consider heterogeneity in a thin plate geometry, the material properties vary along the \textit{x} and \textit{y} directions (2D). The mismatch in material properties across inter-segment boundaries induces higher \textit{x} and \textit{y} components of residual forces on the inter-segment boundary nodes compared to nodes elsewhere in the domain. In contrast, the \textit{z} components of the residual force result from material misidentification and are high throughout each segment. Therefore, it was observed that the use of 2D magnitude of the nodal forces provides a more precise inter-segment boundary as opposed to the 3D magnitude for the thin plate geometry considered in this study $\left(\text{See \autoref{fig:ResForce_xyz}b-d}\right)$.
    
    The nodes with the 2D force magnitude  $f^a_{\text{res}}$, higher than a certain threshold, are ``flagged'' as inter-segment boundary nodes. Generally, the magnitude of residual force is commensurate with the extent of mismatch between the ``guessed'' material property (homogeneous assumption) and the underlying true material property. At the nodes that lie on the interface of two (or more) material segments, a homogeneous material assumption would induce significantly larger residual force due to the imbalance between elements with different material properties. Thus, these interface nodes would be characterized by a considerably higher than average (across all nodes) residual force. Therefore, we employ the following scheme to ``flag'' inter-segment nodes.
    \begin{equation}
        a \in \boldsymbol{{N}_{\text{flag}}}, \quad \iff \quad f_{\text{res}}^a > \lambda \, \mathit{\sigma_{f_{\text{res}}}}
        \label{eq21}
    \end{equation}
    
    where $a$ is an integer, $\boldsymbol{{N}_{\text{flag}}}$ denotes the set of all flagged nodes, $\lambda$ is a tunable scalar (usually ranging between 1.5 and 2.5) and the standard deviation $\mathit{\sigma_{f_{\text{res}}}}$ is given by

    \begin{algorithm}[H]
    \caption{Random seed-based linear element growth for material segmentation}
    \label{alg:segm}
    \begin{algorithmic}[1]
    \Statex \textbf{Inputs:} Flagged nodes {$\boldsymbol{N}_\text{flag}$}, Nodal positions, Connectivity
    \Statex Definitions: $N_a \rightarrow$ Node $a$, with  $a \in \{1\dots n_{n}\}$
    \Statex Definitions: $\mathcal{E}_i \rightarrow$ Element with ID $i$, with  $i \in \{1\dots n_{\text{els}}\}$ \Comment{An element is a set of nodes}
    \Statex Definitions: $\mathcal{S}_i \rightarrow$ Material segment with ID $i$ \Comment{A segment is a set of elements}
    \Statex Definitions: $\mathcal{S}_{\text{domain}} \rightarrow$ Set of all elements in domain
    \Function{$\mathtt{getMaterialSegments}$}{\textbf{Inputs}}
        \State $\boldsymbol{N}_\text{flag} = \{N_a : (a,i) \in \mathcal{D}^{\text{free}} \quad \forall \quad i \in \{1,2,3\} \quad \& \quad f_{\mathrm{res}}^{a} > \lambda  \sigma_{f_{\mathrm{res}}} \}$
        \Comment{Refer \autoref{eq21} for $\lambda$ and $\sigma_{f_{\mathrm{res}}}$}
             
            \State Initialize $i \gets 1$
            \State Initialize $\mathcal{S}_{i} \gets \emptyset$
            \While{$\left(\cup_{k=1}^i\mathcal{S}_{k}\right) \neq \mathcal{S}_{\text{domain}}$}
        
                \State Select random seed element: $\mathcal{E}_{\mathrm{seed}} \in \mathcal{S}_{\mathrm{domain}} \setminus \left(\cup_{k=1}^i\mathcal{S}_{k}\right)$ \Statex \quad \qquad\text{With additional requirement:} $(\mathcal{E}_{\mathrm{seed}}) \cap (N_{\mathrm{flag}}) = \emptyset $
                
                \If{$\mathcal{E}_{\text{seed}}==\emptyset$}
                \State \textbf{break loop}
                \EndIf
                \State Initialize growth frontier: $\mathcal{S}_{\text{frontier}} \gets \{\mathcal{E}_{\text{seed}}\}$
                \State $\mathcal{S}_i \gets \mathcal{S}_{\text{frontier}}$
                \While{$\mathcal{S}_{\text{frontier}}\neq \emptyset$}
                \State $\mathcal{S}_{\text{neighbors}}\gets\emptyset$
                \For{$\mathcal{E}_k$ in $\mathcal{S}_{\text{frontier}}$}
                \If{$\mathcal{E}_k\cap (N_{\mathrm{flag}})=\emptyset$}
                \State $\mathcal{S}_{\text{neighbors}} \gets \mathcal{S}_{\text{neighbors}} \cup \text{Neighbors}\left(\mathcal{E}_k\right)$
                \Statex\Comment{Neighbors($\mathcal{E}$) denotes the set of elements having at least one node in common with element $\mathcal{E}$}
                \EndIf
                \EndFor
                \State $\mathcal{S}_{\text{neighbors}} \gets \mathcal{S}_{\text{neighbors}} \setminus \left(\mathcal{S}_{\text{neighbors}} \cap \mathcal{S}_i\right)$
                \State $\mathcal{S}_{i} \gets \mathcal{S}_i \cup \mathcal{S}_{\text{neighbors}}$
                \State $\mathcal{S}_{\text{frontier}}\gets \mathcal{S}_{\text{neighbors}}$
                \EndWhile
                \State $i \gets i + 1$
                \State Initialize $\mathcal{S}_{i} \gets \emptyset$
            \EndWhile
            \State Initialize $\mathcal{S}_0 \gets \mathcal{S}_{\mathrm{domain}} \setminus \left(\cup_{k=1}^i\mathcal{S}_{k}\right)$
            \Statex \Comment{Elements that have all their nodes belonging to $\boldsymbol{N}_\text{flag}$ will not be covered by the growth algorithm.} 
            \Statex \qquad Step 26 ensures that all such elements are assigned to the segment $\mathcal{S}_0$.
            \Statex \qquad $i-1$ indicates the number of discovered material segments (excluding $\mathcal{S}_0$). 
            \If{$\mathcal{S}_0 \neq \emptyset$}
            \For{$k=i, (i-1), \dots, 1$}
            \State $\mathcal{S}_{k} \gets \mathcal{S}_{k-1}$
            \EndFor
            \State $i \gets i+1$
            \EndIf
            \Statex \textbf{Output:} \{$\mathcal{S}_k: k \in {1 \dots (i-1)}$\} \Comment{$\mathcal{S}_k$ is the set of elements contained by the $k^{\text{th}}$ material segment}
        \EndFunction
    \end{algorithmic}
    \end{algorithm}

    \begin{equation*}
        \mathit{\sigma_{f_{\text{res}}}} =  \sqrt{\frac{\sum_{a \in \boldsymbol{{N}_{\text{flag}}}}\left(f^a_{\text{res}}-\mu\right)^2}{\sum_{a \in \boldsymbol{{N}_{\text{flag}}}}1}} \quad ;\quad 
        \mathit{\mu_{f_{\text{res}}}} = \frac{\sum_{a \in \boldsymbol{{N}_{\text{flag}}}}\left(f^a_{\text{res}}\right)}{\sum_{a \in \boldsymbol{{N}_{\text{flag}}}}1}
    \end{equation*}

        \begin{figure}[H]
        \centering
        \begin{subfigure}[b]{0.30\textwidth}
            \includegraphics[width=\textwidth]{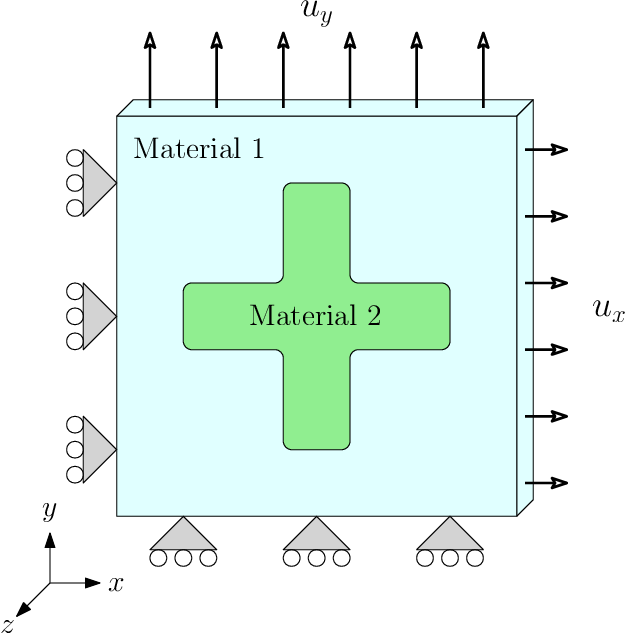}
            \caption{Type 1: Cross-shaped inclusion}
            \label{hetero_pattern:type1}
        \end{subfigure}
        \begin{subfigure}[b]{0.30\textwidth}
            \includegraphics[width=\textwidth]{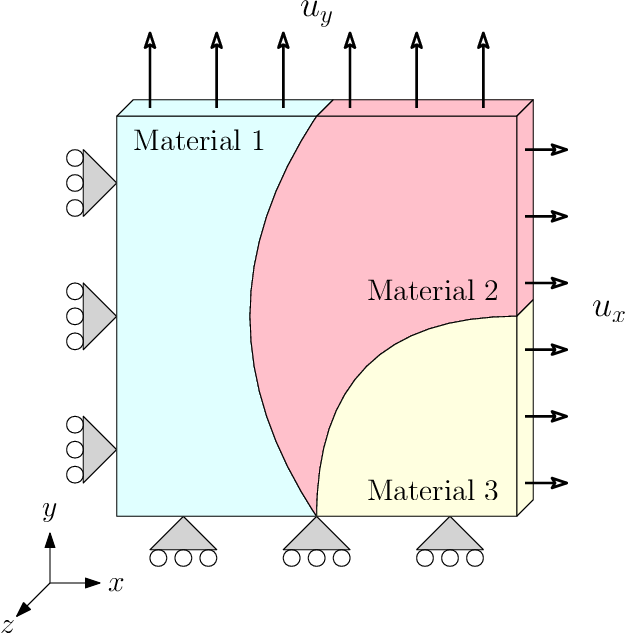}
            \caption{Type 2: Split-domain pattern}
            \label{hetero_pattern:type2}
        \end{subfigure}
        \begin{subfigure}[b]{0.30\textwidth}
            \includegraphics[width=\textwidth]{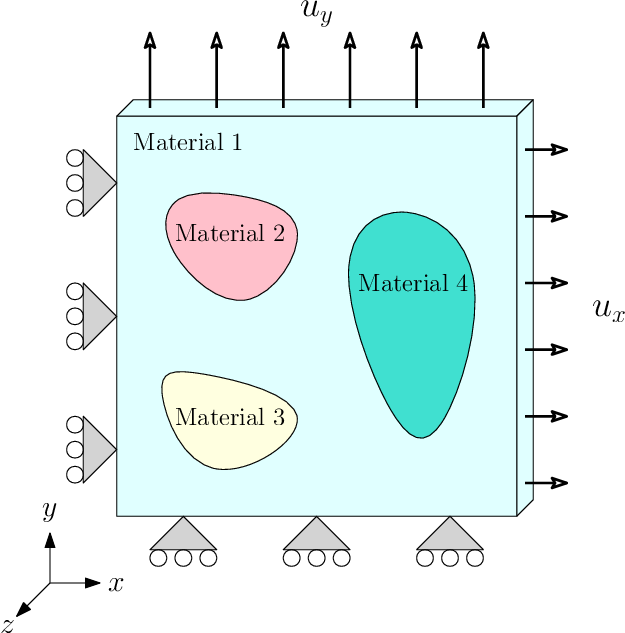}
            \caption{Type 3: Multiple inner inclusions}
            \label{hetero_pattern:type3}
        \end{subfigure}
        \vskip\baselineskip
        \begin{subfigure}[b]{0.30\textwidth}
            \includegraphics[width=\textwidth]{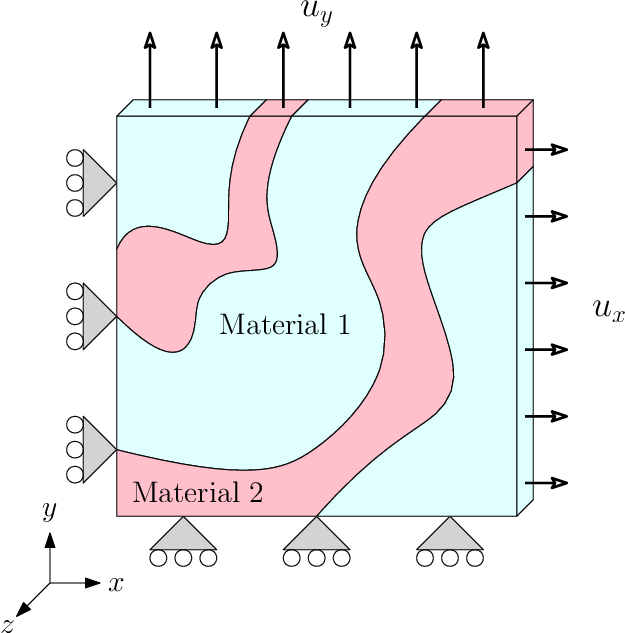}
            \caption{Type 4: Cahn-Hilliard pattern}
            \label{hetero_pattern:type4}
        \end{subfigure}
        \begin{subfigure}[b]{0.30\textwidth}
            \includegraphics[width=\textwidth]{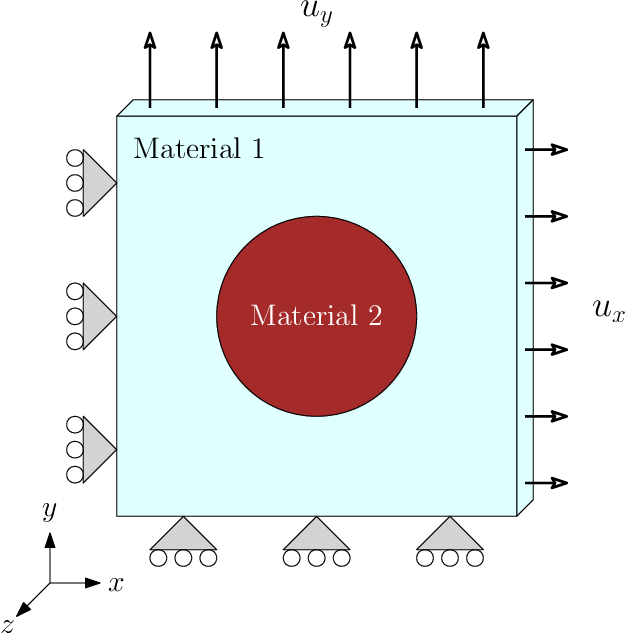}
            \caption{Type 5: Circular inclusion}
            \label{hetero_pattern:type5}
        \end{subfigure}
        \begin{subfigure}[b]{0.30\textwidth}
            \includegraphics[width=\textwidth]{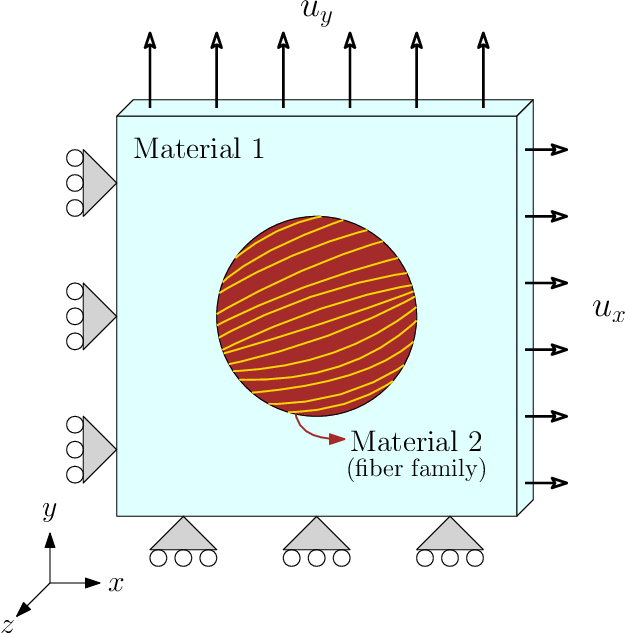}
            \caption{Type 6: Circular inclusion with fibers}
            \label{hetero_pattern:type6}
        \end{subfigure}
   
        \caption{Graphical illustration of heterogeneous test specimen geometries subjected to non-equi-biaxial tensile loading under three-dimensional plane stress conditions. The heterogeneity patterns include: (a) cross-shaped inclusion, (b) split-domain configuration, (c) multiple non-circular inclusions, (d) Cahn-Hilliard pattern, (e) circular inclusion, and (f) circular inclusion with anisotropic fiber reinforcement.}
        \label{hetero_pattern}
    \end{figure}
    
    For zero and moderate noise, it was observed that the average of $f^a_{\text{res}}$ (i.e, $\mathit{\mu_{f_{\text{res}}}}$) was significantly smaller than its standard deviation $\mathit{\sigma_{f_{\text{res}}}}$. However, $\mathit{\mu_{f_{\text{res}}}}$ was observed to increase with an increase in displacement noise, while $\mathit{\sigma_{f_{\text{res}}}}$ was nearly constant. This indicates that $\mathit{\mu_{f_{\text{res}}}}$ is a measure of the noise prevalent in the displacement measurement. Further, this implies that $f^a_{\text{res}}$ could be high in nodes that are either on the inter-segment boundary or have high displacement noise. In order to disambiguate between the inter-segment nodes and those with high displacement noise, we employ a scalar multiple of $\mathit{\sigma_{f_{\text{res}}}}$ as a lower threshold. In experiments where the displacement noise is sufficiently low (measured by the ratio: $\mathit{\mu_{f_{\text{res}}}}/\mathit{\sigma_{f_{\text{res}}}}$), the inter-segment nodes can be expected to mostly have $f^a_{\text{res}}$ above the lower threshold, while the ``noisy'' nodes would mostly have $f^a_{\text{res}}$ below the threshold. This distinction blurs with increasing displacement noise, rendering residual force-based segmentation inaccurate for data with high displacement noise. In \autoref{subsec:noise} we provide further discussion on the observed statistical trends for nodal residual force $f^a_{\text{res}}$. Notably, \citet{shi2025deep} makes similar use of residual forces from a homogenized assumption of their NN surrogate model to highlight the inter-subdomain boundaries, with user-defined lower thresholds.
    \par
    
    After identifying the inter-segment boundaries, we employ an approach adapted and modified from \citet{shi2025deep} to register each element in the domain to a material segment. The approach is essentially a single-threaded (non-parallelizable) recursive growth algorithm. It involves choosing a seed element $\mathcal{E}_{\text{seed}}$ from anywhere within the domain ($\mathcal{S}_{\text{domain}}$), with a condition that none of the nodes constituting the element are ``flagged'' as inter-segment nodes $\left(\boldsymbol{N}_{\text{flag}}\right)$. Next, the seed element and all its neighboring elements $\left(\mathcal{S}_{\text{neighbors}}\right)$ are added into the material segment being discovered $\mathcal{S}_i$, with $i$ denoting the index of the material segment. Next, those elements in the neighboring elements set $\mathcal{S}_{\text{neighbors}}$ that were previously absent in $\mathcal{S}_i$ and do not contain flagged inter-segment nodes are designated as ``growth-frontiers'' $\left(\mathcal{S}_{\text{frontier}}\right)$. The neighbors of each element belonging to $\mathcal{S}_{\text{frontier}}$ are then added to the material segment set $\mathcal{S}_i$, and some of these neighboring elements are then designated as growth frontiers and the growth continues till a single material segment demarcated by a closed loop of flagged inter-segment nodes is completely discovered. The process then continues with the selection of another seed element from the set $\mathcal{S}_{\text{domain}}\setminus\mathcal{S}_i$ until all elements are registered to a material segment set $\mathcal{S}_i$. A more formal description of the approach is provided in \autoref{alg:segm}, and is depicted in \autoref{GrowthAlgo}.\par
    The important outcome of segmentation is the $\boldsymbol{A}$ matrix as shown in \autoref{Atheta_b}, with all elements and nodes mapped to appropriate material segments.

\section{Results and discussion}
\label{sec:results}

\subsection{Computational data generation}
    To evaluate the proposed method, we perform finite element simulations using Abaqus/CAE 2025 software to generate a synthetic computational dataset corresponding to various heterogeneity patterns, as illustrated in \autoref{hetero_pattern}. The model consists of a thin square plate measuring \(50 \, \text{mm} \times 50 \, \text{mm} \times 1 \, \text{mm}\), composed of multiple hyperelastic material segments and subjected to quasi-static, displacement-controlled non-equi-biaxial tension. The specimen is analyzed under the assumption of plane stress conditions. The domain is discretized using 6-node linear triangular prism elements (wedge or C3D6 type) as shown in \autoref{fig:meshing_example}. Simulations are conducted under prescribed boundary conditions, and both nodal displacements and boundary-aggregate normal reaction forces (not shear forces) are recorded over a single load step. To emulate realistic DIC measurements, artificial Gaussian white noise is added to the displacement field, as described in \autoref{eq26}, with subsequent noise handling procedures detailed in \autoref{noise-robustness}. \par
    
    For each heterogeneity pattern shown in \autoref{hetero_pattern}, we assign different combinations of hyperelastic material models that include a quadratic volumetric strain energy term. These configurations are used to benchmark the performance of the Hetero-EUCLID framework. For the scope of the current work, we use the material library  \(\boldsymbol{Q}\) consisting of nineteen features (\autoref{table_feature-list}) across two categories of simulation studies. In the first set of studies (Type 1--4), we employ only \(\mathit{n_f} = 6\) features \((\theta_{1} - \theta_{5}, \theta_{19})\) from the library \(\boldsymbol{Q}\) for the pattern-based case studies corresponding to Type~1--4 configurations (\autoref{hetero_pattern}a--d), as detailed in \autoref{case_study} and \ref{non_native_study}. This is because, for a domain with 5 material segments ($n_c=5$), the number of columns of the $\boldsymbol{A}$ matrix would be $\mathit{n_{f} n_c}=30$. Incorporating a larger number of features (\(\mathit{n_f}\)) would substantially increase the size of the $\boldsymbol{A}$ matrix, and thereby increase the computational cost of the Markov Chain solution. We have therefore avoided including more than $n_f=6$ features in these classes of demonstrative studies, although adding more number of features would not compromise the accuracy of the results \citep{joshi2022bayesian}. As detailed in \autoref{benchmark_study}, the second set of simulations evaluates the performance of the proposed framework for the Type~5 and Type~6 configurations (\autoref{hetero_pattern:type5}, \ref{hetero_pattern:type6}), corresponding to the Ogden and Holzapfel--Gasser--Ogden (HGO) models, respectively. In this study, all features listed in \autoref{table_feature-list} are employed while intentionally omitting the Ogden- and HGO-specific terms (refer to \autoref{eq_ogden} and \ref{eq_hgo}) to examine the framework’s robustness in predicting complex hyperelastic material behavior under feature exclusion. The various hyperelastic material models considered in this work are \par
    \begin{enumerate}[label=\Alph*.]
        \item Neo-Hookean model (NH2) \citep{treloar1943elasticity}
             \begin{equation}
                 W_{\text{NH2}}(\boldsymbol{C};\boldsymbol{\theta}) = \theta_{1}(\tilde{I}_{1}-3) + \theta_{19}\left(J-1\right)^{2}
                 \label{eq28}
             \end{equation}
        \item Isihara model (ISH) \citep{isihara1951statistical}
             \begin{equation}
                 W_{\text{ISH}}(\boldsymbol{C};\boldsymbol{\theta}) = \theta_{1}(\tilde{I}_{1}-3) + \theta_{2}(\tilde{I}_{2}-3) + \theta_{3}(\tilde{I}_{1}-3)^{2} + \theta_{19}\left(J-1\right)^{2}
                 \label{eq29}
             \end{equation}
        \item Haines-Wilson model (HW) \citep{haines1979strain}
             \begin{equation}
                 W_{\text{HW}}(\boldsymbol{C};\boldsymbol{\theta}) = \theta_{1}(\tilde{I}_{1}-3) + \theta_{2}(\tilde{I}_{2}-3) + \theta_{4}(\tilde{I}_{1}-3)(\tilde{I}_{2}-3) + \theta_{5}(\tilde{I}_{1}-3)^{3} + \theta_{19}\left(J-1\right)^{2}
                 \label{eq30}
             \end{equation}
        \item Ogden model (OG)  \citep{ogden1972large}
             \begin{equation}
                 W_{\text{OG}}(\boldsymbol{C};\boldsymbol{\theta}) = \frac{2\mu}{\eta}(\tilde{\lambda}_{1}^{\eta} + \tilde{\lambda}_{2}^{\eta} +\tilde{\lambda}_{3}^{\eta} - 3 ) + \theta_{19}\left(J-1\right)^{2} 
                 \label{eq_ogden}
             \end{equation}
             where \(\tilde{\lambda}_{1}, \tilde{\lambda}_{2}, \tilde{\lambda}_{3}\) are principal stretches with \( \tilde{\lambda}_{k} = J^{-1/3} \lambda_{k}\) and \(\tilde{\lambda}_{1}\tilde{\lambda}_{2}\tilde{\lambda}_{3} = 1\).
        \item Holzapfel-Gasser-Ogden model (HGO) \cite{holzapfel2000new}
             \begin{equation}
                 W_{\text{HGO}}(\boldsymbol{C};\boldsymbol{\theta}) = \theta_{1}(\tilde{I}_{1}-3)  + \theta_{18}\left[\frac{(J^2 -1)}{2} -\ln{(J)}\right] + \frac{k_1}{2k_2}\left[e^{k_2(\tilde{I}_a - 1)^2} -1 \right] 
                 \label{eq_hgo} 
             \end{equation}
             where \(\tilde{I}_{a} = J^{-2/3} (\boldsymbol{a} \cdot\boldsymbol{Ca}) \) denotes anisotropic invariant with one fiber family having orientation angle \(\alpha \) with \(\boldsymbol{a}  = (\cos{\alpha}, \sin{\alpha}, 0)^{T}\).
    \end{enumerate}
        
    \begin{table}[H]
        \centering
        \begin{tabular}{c l|c l|c l}
            \toprule
            \textbf{Model} & \textbf{Library} & \textbf{Model} & \textbf{Library} & \textbf{Model} & \textbf{Library}\\
            \textbf{coefficient} & \textbf{feature} & \textbf{coefficient} & \textbf{feature} & \textbf{coefficient} & \textbf{feature}\\
            \midrule
            $\theta_{1}$ & $(\tilde{I}_1 - 3)$ & $\theta_{8}$ & $(\tilde{I}_1 - 3)(\tilde{I}_2 - 3)^2$ & $\theta_{15}$ & $(\tilde{I}_a - 1)^2$ \\
            $\theta_{2}$ & $(\tilde{I}_2 - 3)$ & $\theta_{9}$ & $(\tilde{I}_2 - 3)^3$ & $\theta_{16}$ & $(\tilde{I}_a - 1)^3$ \\
            $\theta_{3}$ & $(\tilde{I}_1 - 3)^2$ & $\theta_{10}$ & $(\tilde{I}_1 - 3)^4$ & $\theta_{17}$ & $(\tilde{I}_a - 1)^4$\\
            $\theta_{4}$ & $(\tilde{I}_1 - 3)(\tilde{I}_2 - 3)$ & $\theta_{11}$ & $(\tilde{I}_1 - 3)^3(\tilde{I}_2 - 3)$ & $\theta_{18}$ & $0.5(J^2 -1) -\ln{(J)}$ \\
            $\theta_{5}$ & $(\tilde{I}_2 - 3)^2$ & $\theta_{12}$ & $(\tilde{I}_1 - 3)^2(\tilde{I}_2 - 3)^2$ & $\theta_{19}$ & $(J - 1)^2$ \\
            $\theta_{6}$ & $(\tilde{I}_1 - 3)^3$ & $\theta_{13}$ & $(\tilde{I}_1 - 3)(\tilde{I}_2 - 3)^3$ & $-$ & $-$ \\
            $\theta_{7}$ & $(\tilde{I}_1 - 3)^2(\tilde{I}_2 - 3)$ & $\theta_{14}$ & $(\tilde{I}_2 - 3)^4$ & $-$ & $-$ \\
            \bottomrule
        \end{tabular}
        \caption{List of features in the library set \( \boldsymbol{Q} \) used to represent the Neo-Hookean (\ref{eq28}), Isihara (\ref{eq29}), Haines-Wilson (\ref{eq30}), Ogden (\ref{eq_ogden}) and Holzapfel-Gasser-Ogden (\ref{eq_hgo}) forms of constitutive models.}
        \label{table_feature-list}
    \end{table}

\subsection{Robustness to noise}
\label{noise-robustness}
    To evaluate the robustness of the Hetero-EUCLID framework, we introduce artificial noise into the synthetic surface displacement data obtained from finite element simulations. This was done to mimic experimental displacement measurement noise that would arise due to digital image correlation (DIC) and camera resolution limitations. For the specimen shown in \autoref{hetero_pattern}, the reference configuration measures 50 mm\(\times\) 50 mm \(\times\) 1 mm and is subjected to non-equi-biaxial mechanical loading, with final stretch ratios of \( \lambda_{x} = 1.6 \) and \( \lambda_{y} = 2.2 \). The planar area of the final deformed plate would be \(80 \times 110\) mm. A typical DIC setup with a 1-megapixel camera would result in an approximate normalized field of view of 0.1 mm/pixel. As reported in \citet{bornert2009assessment}, the noise level in DIC setups typically varies between 0.01 and 0.05 pixels, which corresponds to $1 \times 10^{-3}$ mm and $5 \times 10^{-3}$ mm of noise for the geometry used in the current study. For the current study, we consider two cases of moderate and high noise level by introducing Gaussian noise having zero mean and standard deviation of \( \sigma_{u,\text{mod}} = 2 \times 10^{-3} \) mm and \(\sigma_{u,\text{high}} = 5 \times 10^{-3}\) mm, respectively. Hence, the displacement field is modified as follows
    \begin{equation}
        u_{i}^{a} = (u_{i}^{a})_{\text{FEM}} + (u_{i}^{a})_{\text{noise}} \quad \text{with} \quad (u_{i}^{a})_{\text{noise}} \sim \mathcal{N}(0,\sigma_{u}^{2}) \quad \forall \quad (a,i) \in \mathcal{D}
        \label{eq26}
    \end{equation}
    where \((u_{i}^{a})_{\text{FEM}}\) denotes the measured input displacement field without noise and \( (u_{i}^{a})_{\text{noise}} \) represents the artificially introduced uncorrelated Gaussian noise at each node \( a \) and direction \( i \).
    Next, as would be done with experimental data, the noisy displacement field \( \mathcal{U} \) is denoised using Kernel Ridge Regression (KRR) \citep{saunders1998ridge}  having a Radial Basis Function (RBF) kernel with hyperparameters tuned by random search to minimize the standard deviation of the denoised signal. In \ref{app:withoutKRR} we repeat these studies without performing any KRR denoising on the noisy displacement field, while in \ref{app:corr_noise} we repeat the study for the Type-I geometry (\autoref{hetero_pattern:type1}) while adding spatially correlated displacement noise.\par
    
    Within the Bayesian-EUCLID framework for solving the constitutive model (represented by \( \boldsymbol{\theta} \) in \autoref{eq17}), as was done in \citet{joshi2022bayesian}, we implement a random sub-sampling of rows in $\boldsymbol{A}^{\text{free}}$, which is equivalent to sub-sampling nodes from the interior of the domain. The set of nodes $\boldsymbol{N}_{\text{free}}$ and $\boldsymbol{N}_{\text{flag}}$ were defined earlier in \autoref{eq:Nfree} and in \autoref{alg:segm}, respectively. We subsample rows of $\boldsymbol{A}^{\text{free}}$ corresponding to subsets of $\boldsymbol{N}_{\text{free}}$ and $\boldsymbol{N}_{\text{flag}}$ as follows
    
    \begin{itemize}
        \item \textbf{Free nodes}: For nodes in \( \boldsymbol{{N}_{\text{free}}} \) within the reference domain \( \Omega \), we randomly sample a subset of inner nodes such that \( n_{\text{free}} < |\mathcal{D}^{\text{free}}| \). Typically, \( n_{\text{free}} \) is chosen as 2\% of the total free degrees of freedom. However, in cases involving complex heterogeneity such as Cahn–Hilliard patterns with added displacement noise, the value of \( n_{\text{free}} \) may be increased up to 10\% to ensure sufficient coverage.
    
        \item \textbf{Inter-segment nodes}: For nodes in \( \boldsymbol{{N}_{\text{flag}}} \), we first solve for $\boldsymbol{b}_{\text{het}}^{\text{residue}}$ as follows
        \begin{equation*}
            \boldsymbol{b}_{\text{het}}^{\text{residue}} = \boldsymbol{A^{\text{free}}}\left((\boldsymbol{A}^T\boldsymbol{A})^{-1}\boldsymbol{A}^T\boldsymbol{b}\right)
        \end{equation*}
        Next, we extract $(f^a_{\text{res}})_{\text{het}}$ from $\boldsymbol{b}_{\text{het}}^{\text{residue}}$ as described in \autoref{eq19} and \autoref{fig:ResForce_xyz}(a). We subsample $(f^a_{\text{res}})_{\text{het}}$ for each inter-segment boundary and chose 20\% of the nodes with the least $(f^a_{\text{res}})_{\text{het}}$ for every inter-segment boundary. The use of this scheme for subsampling nodes from the inter-segment boundaries is expected to decrease the number of mislabeled (i.e., assigned to the wrong material segment) nodes that contribute to solving for $\boldsymbol{\theta}$, thereby improving its accuracy and noise robustness. A detailed study regarding the effect of inter-segment node sub-sampling is provided in \ref{flagged-subsample}.
    \end{itemize}
    
    The sub-sampling strategy described above significantly reduces the dimensionality of the matrices and vectors involved in \autoref{eq17}, thereby substantially improving the computational speed. While the expression in \autoref{eq16} remains unchanged, the system defined in \autoref{eq13} is modified to account for the reduced number of degrees of freedom as follows
    \begin{align}
        &\boldsymbol{A}^{\text{fix}} \boldsymbol{\theta} = \boldsymbol{b}^{\text{fix}} \quad \text{with} \quad \boldsymbol{A}^{\text{fix}} \in \mathbb{R}^{n_{b} \times \mathit{n_{f}n_c}}, \quad \boldsymbol{b}^{\text{fix}} \in \mathbb{R}^{n_{b}} \nonumber \\
        &\boldsymbol{A}^{\text{free}} \boldsymbol{\theta} = \boldsymbol{b}^{\text{free}} \quad \text{with} \quad \boldsymbol{A}^{\text{free}} \in \mathbb{R}^{(n_{\text{free}} + n_{\text{flag}}) \times \mathit{n_{f}n_c}}, \quad \boldsymbol{b}^{\text{free}} \in \mathbb{R}^{(n_{\text{free}} + n_{\text{flag}})}
        \label{eq27}
    \end{align}
    
    \autoref{eq27} defines the updated force balance equations, incorporating subsampled nodal contributions from the free, flagged, and fixed regions of the domain.

\subsection{Pattern case study}
\label{case_study}

    \subsubsection*{Pattern Type 1: Cross-shaped inclusion}
    \label{subsubsec:Crossinc}
    
        \autoref{hetero_pattern:type1} illustrates a heterogeneous specimen with a cross-shaped inclusion embedded in a surrounding matrix material. Through this configuration, we aim to examine the effectiveness of the Hetero-EUCLID framework; however, it presents a non-trivial test case due to the presence of sharp corners in inter-segment interfaces. For the cross-shaped inclusion pattern, we begin the analysis for a hyperelastic bi-material setup where both the inclusion and the matrix follow Neo-Hookean constitutive models (\(\text{NH2}_{\text{a}}\)–\(\text{NH2}_{\text{b}}\)). The domain, modeled as a single-element-thick plate, is discretized using 10,442 uniformly sized wedge-type elements. As depicted in \autoref{hetero_pattern:type1}, the specimen is subjected to quasi-static, displacement-controlled non-equi-biaxial tensile loading over \( n_t = 6 \) time steps, reaching final directional stretch ratios of \( \lambda_x = 1.6 \) and \( \lambda_y = 2.2 \). Only displacement data from the final load-step $n_t=6$ is considered for further analysis.\par
        
\begin{figure}[H]
    \centering
    \includegraphics[width=\textwidth]{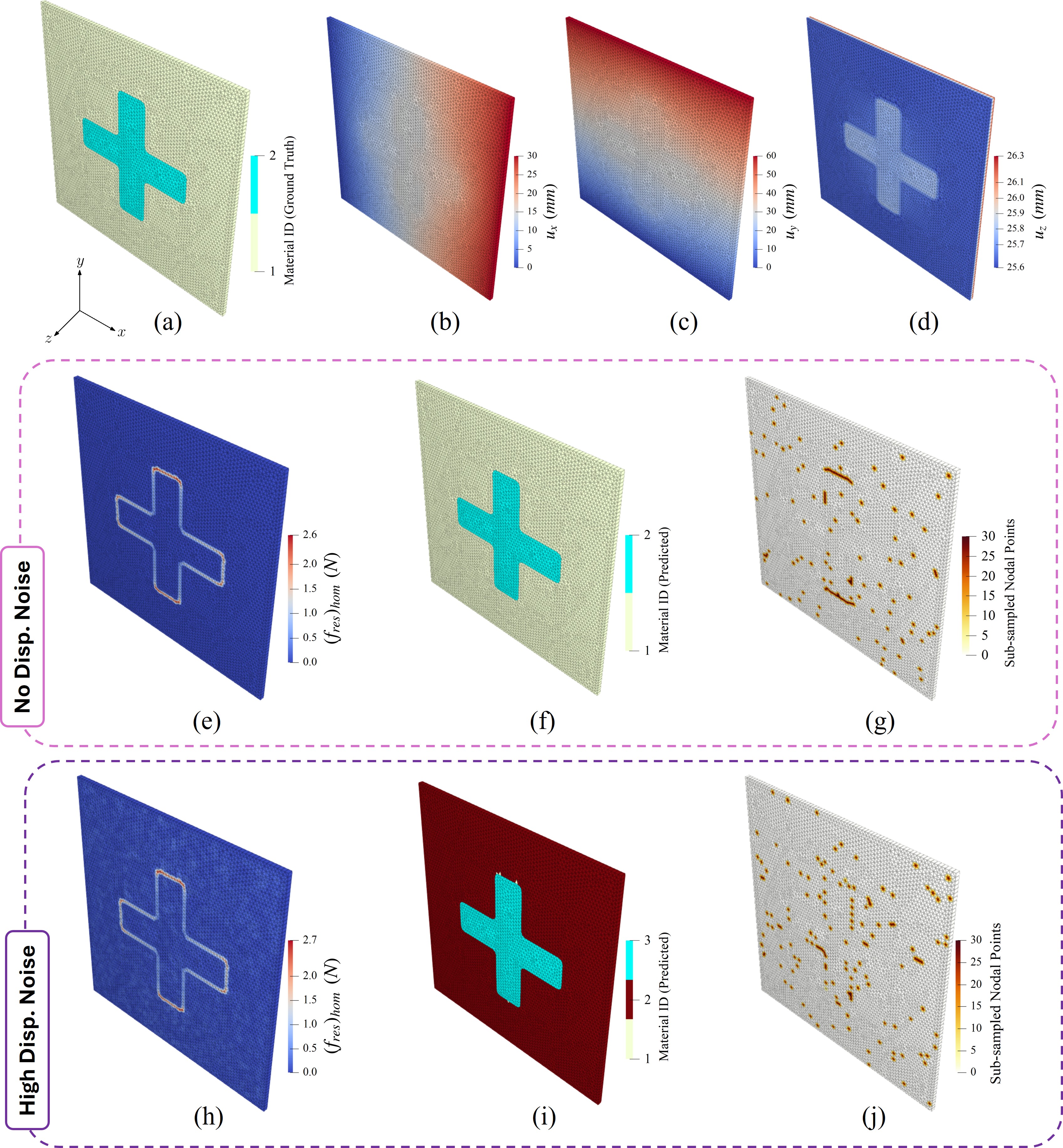}
    \caption{Model discovery for the cross-shaped inclusion pattern  (\autoref{hetero_pattern:type1}). The ground truth material segment IDs are shown in (a), and the displacement fields in the \(x\), \(y\), and \(z\) directions are provided in (b)--(d). For the \textit{zero-noise} case, (e) depicts the residual force norm distribution (\autoref{eq19}) from the homogenized model, (f) illustrates the material segments identified through the segmentation algorithm, and (g) highlights the sub-sampled nodes from the \(\mathcal{D}^{\text{free}}\) set used to construct the heterogenized model. The \textit{high-noise} case follows a similar structure in panels (h)--(j), showing how the residual force field, segmentation, and sub-sampling behave under high noise levels. In both scenarios, the final heterogenized model is solved using the Bayesian framework to recover the region-wise constitutive parameters as listed in \autoref{table_Type1_none_high_10k}.}
    \label{Type1_none_high_10k}
\end{figure}

        \autoref{Type1_none_high_10k} summarizes the performance of Hetero-EUCLID in identifying interpretable material models for the cross-shaped inclusion specimen. \autoref{Type1_none_high_10k}a shows the ground truth, indicating both the number of material segments and their true geometrical boundaries. The full-field displacement fields obtained from FEM simulations are visualized in \autoref{Type1_none_high_10k}b-d. Subsequently, solving the homogenized model using ordinary least squares regression yields the residual force distribution \( f^a_{\text{res}} \), as shown in \autoref{Type1_none_high_10k}e. The presence of high residual force resembles the imbalance at the material interface and aids in flagging the interface nodes based on the thresholding procedure described in \autoref{eq21}. Next, the domain is segmented using the random seed-based island growth algorithm described in \autoref{segmentation}. \autoref{Type1_none_high_10k}f illustrates clean and precise segmentation of the heterogeneous pattern within the specimen. The algorithm systematically identifies both the number of distinct material sub-domains and their corresponding element sets. Using the segmentation results, the heterogenized model is assembled using \autoref{eq17} (refer to \autoref{Atheta_b}), with nodal data selected via random sub-sampling as outlined in \autoref{noise-robustness}. The nodes included for solving \autoref{eq27} using the Bayesian-EUCLID framework are highlighted in \autoref{Type1_none_high_10k}g.\par

We then solve the system using a Bayesian framework with spike-slab priors, employing three parallel MCMC chains (Markov Chain Monte Carlo) of length 500 each (refer \ref{BayesianEUCLID} and \citet{joshi2022bayesian} for details on the Bayesian framework). The outcome of this procedure is a segment-wise interpretable discovery of material models. \autoref{table_Type1_none_high_10k} benchmarks the predicted model coefficients \( \theta_{1}\) and \( \theta_{19} \) against the ground truth values for the \(\text{NH2}_{\text{a}}\)–\(\text{NH2}_{\text{b}}\) material combination (\autoref{eq28}). It can be observed that, with only 2\% of the domain nodal data sampled, the identified physically interpretable coefficients, namely shear and bulk moduli, closely match their true values. For coefficients \( \theta_{2} \) to \( \theta_{5} \), which are not expected to contribute under the Neo-Hookean model formulation, the Bayesian framework predicts zero values due to the sparsity-promoting nature of the spike-slab prior. The small standard deviation values further indicate high confidence in the inferred parameters. \par

\begin{table}[H]
    \centering
    \begin{adjustbox}{width=\textwidth}
    \begin{tabular}{@{} l c c c c c c c c c @{}}
    \toprule[1.5pt]
    \multicolumn{10}{l}{\textbf{Pattern Type 1: Cross-shaped inclusion} $\vert$ Mesh elements (C3D6): \textbf{10,442} $\vert$ Activated feature library terms, $\boldsymbol{n_{\mathit{f}} = 6}$} \\
    \midrule[1.5pt]
    \addlinespace
    \begin{tabular}{@{}l@{}}Case\end{tabular}
    & \begin{tabular}{@{}c@{}}Material\\ID\end{tabular} 
    & \begin{tabular}{@{}c@{}}Constitutive\\model\end{tabular} 
    & \begin{tabular}{@{}c@{}}$\theta_{1}$\\(MPa)\end{tabular} 
    & \begin{tabular}{@{}c@{}}$\theta_{2}$\\(MPa)\end{tabular} 
    & \begin{tabular}{@{}c@{}}$\theta_{3}$\\(MPa)\end{tabular} 
    & \begin{tabular}{@{}c@{}}$\theta_{4}$\\(MPa)\end{tabular} 
    & \begin{tabular}{@{}c@{}}$\theta_{5}$\\(MPa)\end{tabular} 
    & \begin{tabular}{@{}c@{}}$\theta_{19}$\\(MPa)\end{tabular} 
    & \begin{tabular}{@{}c@{}}Free nodes\\sub-sampling\end{tabular} \\
    \midrule[1.0pt]
    \multirow{2}{*}{\begin{tabular}{@{}l@{}}Ground Truth\\ \textit{Refer} \ref{Type1_none_high_10k}(a)\end{tabular}} 
    & 1 & \(\text{NH2}_{\text{a}}\) & 1.80 & $-$ & $-$ & $-$ & $-$ & 6.00 & \multirow{2}{*}{$-$} \\
    
    & 2 & \(\text{NH2}_{\text{b}}\) & 5.40 & $-$ & $-$ & $-$ & $-$ & 15.00 & \\
    \midrule[1.0pt]
    \multirow{2}{*}{\begin{tabular}{@{}l@{}}No Noise\\ \textit{Refer} \ref{Type1_none_high_10k}(f)\end{tabular}} 
    & 1 & \(\text{NH2}_{\text{a}}\) & 1.79 $\pm$ 0.00 & 0.00 $\pm$ 0.00 & 0.00 $\pm$ 0.00 & 0.00 $\pm$ 0.00 & 0.00 $\pm$ 0.00 & 6.00 $\pm$ 0.00 & \multirow{2}{*}{2\%} \\
    
    & 2 & \(\text{NH2}_{\text{b}}\) & 5.46 $\pm$ 0.01 & 0.00 $\pm$ 0.00 & 0.00 $\pm$ 0.00 & 0.00 $\pm$ 0.00 & 0.00 $\pm$ 0.00 & 15.13 $\pm$ 0.04 & \\
    \midrule[1.0pt]
    \multirow{2}{*}{\begin{tabular}{@{}l@{}}Moderate Noise, S1(f) \\($\sigma_u = 2\times10^{-3}$) \end{tabular}} 
    & 1 & \(\text{NH2}_{\text{a}}\) & 1.81 $\pm$ 0.00 & 0.00 $\pm$ 0.00 & 0.00 $\pm$ 0.00 & 0.00 $\pm$ 0.00 & 0.00 $\pm$ 0.00 & 6.01 $\pm$ 0.00 & \multirow{2}{*}{2\%} \\
    
    & 2 & \(\text{NH2}_{\text{b}}\) & 5.42 $\pm$ 0.07 & 0.01 $\pm$ 0.03 & 0.00 $\pm$ 0.00 & 0.00 $\pm$ 0.00 & 0.00 $\pm$ 0.00 & 14.92 $\pm$ 0.13 & \\
    \midrule[1.0pt]
    \multirow{2}{*}{\begin{tabular}{@{}l@{}}High Noise, \ref{Type1_none_high_10k}(i) \\($\sigma_u = 5\times10^{-3}$) \end{tabular}} 
    & 2 & \(\text{NH2}_{\text{a}}\) & 1.81 $\pm$ 0.00 & 0.00 $\pm$ 0.00 & 0.00 $\pm$ 0.00 & 0.00 $\pm$ 0.00 & 0.00 $\pm$ 0.00 & 5.98 $\pm$ 0.00 & \multirow{2}{*}{2\%} \\
    
    & 3 & \(\text{NH2}_{\text{b}}\) & 5.25 $\pm$ 0.09 & 0.00 $\pm$ 0.01 & 0.00 $\pm$ 0.00 & 0.00 $\pm$ 0.00 & 0.00 $\pm$ 0.00 & 14.40 $\pm$ 0.22 & \\
    \bottomrule[1.5pt]
    \end{tabular}
    \end{adjustbox}
    \caption{True vs. predicted material parameters for cross-shaped inclusion pattern (\autoref{hetero_pattern:type1}) across different noise levels using Bayesian-EUCLID framework.}
    \label{table_Type1_none_high_10k}
\end{table}

The influence of noise on Hetero-EUCLID for cross-shaped inclusion is reported in \autoref{Type1_none_high_10k}. Under high noise case \( (\sigma_{u,\text{high}} = 5 \times 10^{-3}) \), the effect on the residual force norm \( (f^a_{\text{res}}) \) is shown in \autoref{Type1_none_high_10k}h, which may be visually compared with the noise-free case in \autoref{Type1_none_high_10k}e. Despite the added noise, the material region segmentation remains accurate as illustrated in \autoref{Type1_none_high_10k}i. Supplementary Figure S1 shows the corresponding visualization results for moderate noise condition \( (\sigma_{u,\text{mod}} = 2 \times 10^{-3}) \). \autoref{table_Type1_none_high_10k} reports the predicted material parameters for both moderate and high noise cases, where the maximum absolute error remains below 4\% compared to the ground truth. Supplementary section S2 presents an empirical validation study by comparing the strain energy density along six different deformation paths between the discovered material model and the corresponding ground truth. Supplementary Figure S7 shows the comparative strain energy density plots for the cross-shaped inclusion pattern under high noise conditions. The results indicate good agreement between the predicted and true strain energy densities, demonstrating the robustness of the proposed method under noise conditions. In \ref{study_low_elastic}, we repeat the study for a Type-1 geometry (cross-shaped inclusion), with the matrix and inclusion materials having similar material properties to evaluate the ability of the Hetero-EUCLID framework to accurately characterize and segment materials with similar properties.\par

    \subsubsection*{Pattern Type 2: Split-domain pattern}
    
\begin{figure}[H]
    \centering
    \includegraphics[width=\textwidth]{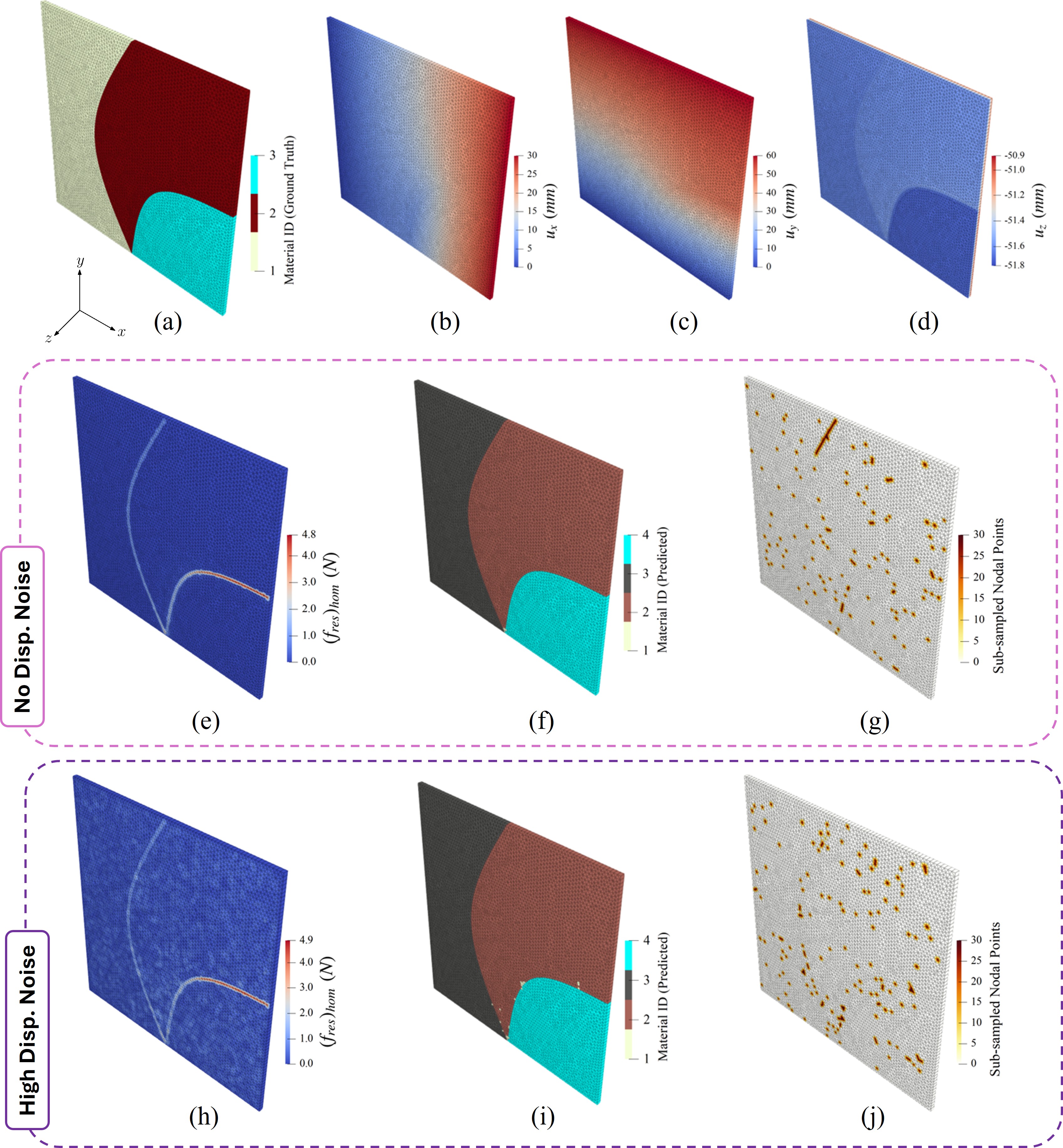}
    \caption{Model discovery for the split-domain pattern  (\autoref{hetero_pattern:type2}). The ground truth material segment IDs are shown in (a), and the displacement fields in the \(x\), \(y\), and \(z\) directions are provided in (b)--(d). For the \textit{zero-noise} case, (e) depicts the residual force norm distribution (\autoref{eq19}) from the homogenized model, (f) illustrates the material segments identified through the segmentation algorithm, and (g) highlights the sub-sampled nodes from the \(\mathcal{D}^{\text{free}}\) set used to construct the heterogenized model. The \textit{high-noise} case follows a similar structure in panels (h)--(j), showing how the residual force field, segmentation, and sub-sampling behave under high noise levels. In both scenarios, the final heterogenized model is solved using the Bayesian framework to recover the region-wise constitutive parameters as listed in \autoref{table_Type2_none_high_10k}.}
    \label{Type2_none_high_10k}
\end{figure}

        We further investigate a more complex heterogeneous configuration, featuring a split-domain pattern with three distinct material regions whose interfaces converge at a common junction located on one of the external boundaries, as illustrated in \autoref{hetero_pattern:type2}. The domain is discretized using 10,972 wedge elements, and the underlying ground truth material model comprises a combination of Neo-Hookean, Isihara, and Haines–Wilson constitutive laws. Using the same non-equi-biaxial loading conditions as in the previous case, we perform inverse parameter identification with the Hetero-EUCLID framework under noise-free, high-noise, and moderate-noise conditions. The corresponding results are presented in \autoref{Type2_none_high_10k}e-g, \ref{Type2_none_high_10k}h-j and Supplementary Figure S2e-g, respectively. In each of the above scenarios, the residual force distribution \( (f^a_{\text{res}}) \) reveals a clear mechanical imbalance at the material interfaces, as shown in \autoref{Type2_none_high_10k}e, \ref{Type2_none_high_10k}h and S2e. Next, the segmentation step is executed using the seed-based growth algorithm with a tunable threshold parameter \( \lambda \) (refer to \autoref{eq21}), and the resulting predicted material segments are shown in \autoref{Type2_none_high_10k}f, \ref{Type2_none_high_10k}i, and S2f. In each case, a small degree of misclustering is observed at the junction of the material interfaces, affecting only a few elements.\par

        \autoref{table_Type2_none_high_10k} compares the predicted material parameters \( \boldsymbol{\theta} \) under noise-free conditions against the ground truth values. The parameters corresponding to the Neo-Hookean (NH2: \( \theta_{1,19} \)), Isihara (ISH: \( \theta_{1-3,19} \)), and Haines–Wilson (HW: \( \theta_{1,2,4,5,19} \)) models are all recovered with high accuracy and low standard deviation. Under noisy conditions, as reported in \autoref{table_Type2_none_high_10k}, the region governed by the NH2 model is predicted accurately. However, the regions corresponding to ISH and HW models exhibit some deviations in the predicted parameters, likely due to the compounded effect of noise and minor misclustering. The noise-induced deviation can also be observed from the strain energy density plots for HW under high noise conditions, as shown in Supplementary Figure S8. The prediction accuracies can be expected to improve with the use of data from multiple experiments/load-steps instead of a single non-equi-biaxial stretch test. \par

\begin{table}[H]
    \centering
    \begin{adjustbox}{width=\textwidth}
    \begin{tabular}{@{} l c c c c c c c c c @{}}
    \toprule[1.5pt]
    \multicolumn{10}{l}{\textbf{Pattern Type 2: Split-domain} $\vert$ Mesh elements (C3D6): \textbf{10,972} $\vert$ Activated feature library terms, $\boldsymbol{n_{\mathit{f}} = 6}$} \\
    \midrule[1.5pt]
    \addlinespace
    \begin{tabular}{@{}l@{}}Case\end{tabular}
    & \begin{tabular}{@{}c@{}}Material\\ID\end{tabular} 
    & \begin{tabular}{@{}c@{}}Constitutive\\model\end{tabular} 
    & \begin{tabular}{@{}c@{}}$\theta_{1}$\\(MPa)\end{tabular} 
    & \begin{tabular}{@{}c@{}}$\theta_{2}$\\(MPa)\end{tabular} 
    & \begin{tabular}{@{}c@{}}$\theta_{3}$\\(MPa)\end{tabular} 
    & \begin{tabular}{@{}c@{}}$\theta_{4}$\\(MPa)\end{tabular} 
    & \begin{tabular}{@{}c@{}}$\theta_{5}$\\(MPa)\end{tabular} 
    & \begin{tabular}{@{}c@{}}$\theta_{19}$\\(MPa)\end{tabular} 
    & \begin{tabular}{@{}c@{}}Free nodes\\sub-sampling\end{tabular} \\
    \midrule[1.0pt]
    \multirow{3}{*}{\begin{tabular}{@{}l@{}}Ground\\Truth\\ \textit{Refer} \ref{Type2_none_high_10k}(a)\end{tabular}} 
    & 1 & NH2 & 6.00 & $-$ & $-$ & $-$ & $-$ & 32.00 &  \\
    & 2 & ISH & 4.00 & 0.50 & 0.30 & $-$ & $-$ & 21.00 & $-$ \\
    & 3 & HW & 1.00 & 0.15 & $-$ & 0.02 & 0.00 & 10.00 & \\

    \midrule[1.0pt]
    \multirow{3}{*}{\begin{tabular}{@{}l@{}}No Noise \\($\sigma_u = 0$) \\ \textit{Refer} \ref{Type2_none_high_10k}(f)\end{tabular}} 
    & 3 & NH2 & 6.00 $\pm$ 0.00 & 0.00 $\pm$ 0.00 & 0.00 $\pm$ 0.00 & 0.00 $\pm$ 0.00 & 0.00 $\pm$ 0.00 & 32.00 $\pm$ 0.00 &  \\
    & 2 & ISH & 3.99 $\pm$ 0.03 & 0.51 $\pm$ 0.01 & 0.29 $\pm$ 0.00 & 0.00 $\pm$ 0.00 & 0.00 $\pm$ 0.00 & 20.99 $\pm$ 0.01 & 2\% \\
    & 4 & HW & 0.92 $\pm$ 0.12 & 0.22 $\pm$ 0.03 & 0.00 $\pm$ 0.02 & 0.02 $\pm$ 0.00 & 0.00 $\pm$ 0.00 & 10.05 $\pm$ 0.01 & \\
    
    \midrule[1.0pt]
    \multirow{3}{*}{\begin{tabular}{@{}l@{}}Moderate Noise\\($\sigma_u = 2\times 10^{-3}$) \\ \textit{Refer} S2(f) \end{tabular}} 
    & 3 & NH2 & 5.99 $\pm$ 0.01 & 0.00 $\pm$ 0.00 & 0.00 $\pm$ 0.00 & 0.00 $\pm$ 0.00 & 0.00 $\pm$ 0.00 & 31.81 $\pm$ 0.02 &  \\
    & 2 & ISH & 5.25 $\pm$ 0.10 & 0.21 $\pm$ 0.03 & 0.21 $\pm$ 0.02 & 0.00 $\pm$ 0.00 & 0.00 $\pm$ 0.00 & 21.86 $\pm$ 0.03 & 2\% \\
    & 4 & HW & 1.59 $\pm$ 0.09 & 0.28 $\pm$ 0.01 & 0.00 $\pm$ 0.01 & 0.00 $\pm$ 0.00 & 0.00 $\pm$ 0.00 & 9.71 $\pm$ 0.02 & \\
    
    \midrule[1.0pt]
    \multirow{3}{*}{\begin{tabular}{@{}l@{}}High Noise\\($\sigma_u = 5\times 10^{-3}$) \\ \textit{Refer} \ref{Type2_none_high_10k}(i) \end{tabular}} 
    & 3 & NH2 & 5.99 $\pm$ 0.03 & 0.00 $\pm$ 0.00 & 0.00 $\pm$ 0.00 & 0.00 $\pm$ 0.00 & 0.00 $\pm$ 0.00 & 31.63 $\pm$ 0.04 &  \\
    & 2 & ISH & 6.12 $\pm$ 0.22 & 0.01 $\pm$ 0.02 & 0.07 $\pm$ 0.08 & 0.00 $\pm$ 0.00 & 0.01 $\pm$ 0.01 & 21.79 $\pm$ 0.07 & 2\% \\
    & 4 & HW & 1.69 $\pm$ 0.04 & 0.25 $\pm$ 0.01 & 0.00 $\pm$ 0.00 & 0.00 $\pm$ 0.00 & 0.00 $\pm$ 0.00 & 9.56 $\pm$ 0.04 & \\
    \bottomrule[1.5pt]
    \end{tabular}
    \end{adjustbox}
    \caption{True vs. predicted material parameters for split-domain pattern (\autoref{hetero_pattern:type2}) across different noise levels using Bayesian-EUCLID framework.}
    \label{table_Type2_none_high_10k}
\end{table}

    \subsubsection*{Pattern Type 3: Multiple inner inclusions}
    
\begin{figure}[H]
    \centering
    \includegraphics[width=\textwidth]{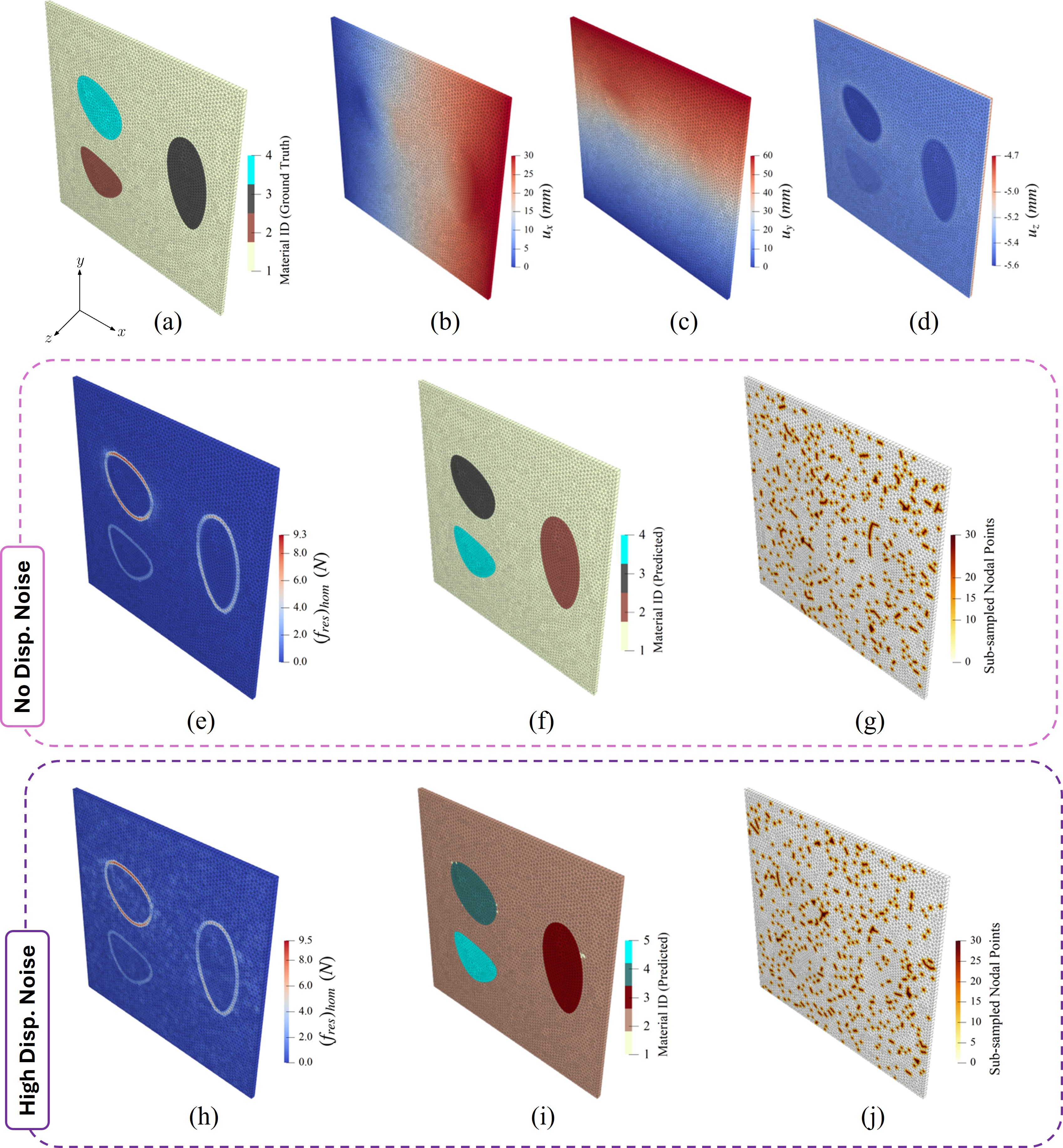}
    \caption{Model discovery for the multiple inner inclusions pattern  (\autoref{hetero_pattern:type3}). The ground truth material segment IDs are shown in (a), and the displacement fields in the \(x\), \(y\), and \(z\) directions are provided in (b)--(d). For the \textit{zero-noise} case, (e) depicts the residual force norm distribution (\autoref{eq19}) from the homogenized model, (f) illustrates the material segments identified through the segmentation algorithm, and (g) highlights the sub-sampled nodes from the \(\mathcal{D}^{\text{free}}\) set used to construct the heterogenized model. The \textit{high-noise} case follows a similar structure in panels (h)--(j), showing how the residual force field, segmentation, and sub-sampling behave under high noise levels. In both scenarios, the final heterogenized model is solved using the Bayesian framework to recover the region-wise constitutive parameters as listed in \autoref{table_Type3_none_high_10k}.}
    \label{Type3_none_high_10k}
\end{figure}

        Next, we extend our investigation to a heterogeneous configuration featuring multiple inclusions with non-circular geometries. \autoref{hetero_pattern:type3} illustrates the case where the matrix follows an Isihara (ISH) material model, and three embedded inclusions are governed by Neo-Hookean (\(\text{NH2}_{\text{a}}, \text{NH2}_{\text{b}}\)) and Haines–Wilson (HW) constitutive laws. Using this type of configuration, we aim to evaluate the proposed framework's ability to simultaneously segment and characterize multiple non-circular inclusions with varying classes of material models. The domain is discretized using 10,816 uniformly spaced wedge-type elements and subjected to the same non-equi-biaxial displacement-controlled loading conditions used in previous cases. We investigated the Type 3 specimen across different noise levels. \autoref{Type3_none_high_10k} presents the results of material region segmentation and parameter identification under noise-free conditions. Based on the residual force distribution shown in \autoref{Type3_none_high_10k}e, and applying the random seed-based island growth algorithm, the predicted material segments align well with the true inclusions, as illustrated in \autoref{Type3_none_high_10k}f. \autoref{table_Type3_none_high_10k} reports the accurate recovery of material parameters for all regions in the ISH–\(\text{NH2}_{\text{a}}\)–HW–\(\text{NH2}_{\text{b}}\) configuration under noise-free condition, with low standard deviations indicating high confidence in the prediction. \par

\begin{table}[H]
    \centering
    \begin{adjustbox}{width=\textwidth}
    \begin{tabular}{@{} l c c c c c c c c c @{}}
    \toprule[1.5pt]
    \multicolumn{10}{l}{\textbf{Pattern Type 3: Multiple inner inclusions} $\vert$ Mesh elements (C3D6): \textbf{10,816} $\vert$ Activated feature library terms, $\boldsymbol{n_{\mathit{f}} = 6}$} \\
    \midrule[1.5pt]
    \addlinespace
    \begin{tabular}{@{}l@{}}Case\end{tabular}
    & \begin{tabular}{@{}c@{}}Material\\ID\end{tabular} 
    & \begin{tabular}{@{}c@{}}Constitutive\\model\end{tabular} 
    & \begin{tabular}{@{}c@{}}$\theta_{1}$\\(MPa)\end{tabular} 
    & \begin{tabular}{@{}c@{}}$\theta_{2}$\\(MPa)\end{tabular} 
    & \begin{tabular}{@{}c@{}}$\theta_{3}$\\(MPa)\end{tabular} 
    & \begin{tabular}{@{}c@{}}$\theta_{4}$\\(MPa)\end{tabular} 
    & \begin{tabular}{@{}c@{}}$\theta_{5}$\\(MPa)\end{tabular} 
    & \begin{tabular}{@{}c@{}}$\theta_{19}$\\(MPa)\end{tabular} 
    & \begin{tabular}{@{}c@{}}Free nodes\\sub-sampling\end{tabular} \\
    \midrule[1.0pt]
    \multirow{3}{*}{\begin{tabular}{@{}l@{}}Ground\\Truth\\ \textit{Refer} \ref{Type3_none_high_10k}(a)\end{tabular}} 
    & 1 & ISH & 4.00 & 0.50 & 0.30 & $-$ & $-$ & 21.00 & \multirow{4}{*}{$-$} \\
    & 2 & \(\text{NH2}_{\text{a}}\) & 5.40 & $-$ & $-$ & $-$ & $-$ & 15.00 & \\
    & 3 & HW & 1.00 & 0.15 & $-$ & 0.02 & 0.00 & 10.00 & \\
    & 4 & \(\text{NH2}_{\text{b}}\) & 1.80 & $-$ & $-$ & $-$ & $-$ & 6.00 &  \\

    \midrule[1.0pt]
    \multirow{3}{*}{\begin{tabular}{@{}l@{}}No Noise\\ ($\sigma_u = 0$)\\ \textit{Refer} \ref{Type3_none_high_10k}(f)\end{tabular}} 
    & 1 & ISH & 4.00 $\pm$ 0.00 & 0.50 $\pm$ 0.00 & 0.30 $\pm$ 0.00 & 0.00 $\pm$ 0.00 & 0.00 $\pm$ 0.00 & 21.00 $\pm$ 0.00 & \multirow{4}{*}{10\%} \\
    & 4 & \(\text{NH2}_{\text{a}}\) & 5.40 $\pm$ 0.01 & 0.00 $\pm$ 0.00 & 0.00 $\pm$ 0.00 & 0.00 $\pm$ 0.00 & 0.00 $\pm$ 0.00 & 14.99 $\pm$ 0.02 &  \\
    & 2 & HW & 0.72 $\pm$ 0.01 & 0.22 $\pm$ 0.00 & 0.00 $\pm$ 0.00 & 0.03 $\pm$ 0.00 & 0.00 $\pm$ 0.00 & 9.81 $\pm$ 0.02 &  \\
    & 3 & \(\text{NH2}_{\text{b}}\) & 1.75 $\pm$ 0.00 & 0.00 $\pm$ 0.00 & 0.00 $\pm$ 0.00 & 0.00 $\pm$ 0.00 & 0.00 $\pm$ 0.00 & 5.83 $\pm$ 0.01 & \\
                
    \midrule[1.0pt]
    \multirow{3}{*}{\begin{tabular}{@{}l@{}}Moderate Noise\\($\sigma_u = 2\times 10^{-3}$)\\ \textit{Refer} S3(f) \end{tabular}} 
    & 2 & ISH & 4.10 $\pm$ 0.11 & 0.51 $\pm$ 0.00 & 0.29 $\pm$ 0.02 & 0.00 $\pm$ 0.00 & 0.00 $\pm$ 0.00 & 21.12 $\pm$ 0.05 & \multirow{4}{*}{10\%} \\
    & 5 & \(\text{NH2}_{\text{a}}\) & 5.34 $\pm$ 0.12 & 0.00 $\pm$ 0.01 & 0.00 $\pm$ 0.01 & 0.00 $\pm$ 0.00 & 0.00 $\pm$ 0.00 & 14.83 $\pm$ 0.22 &  \\
    & 3 & HW & 0.41 $\pm$ 6.61 & 0.52 $\pm$ 0.07 & 0.01 $\pm$ 0.01 & 0.00 $\pm$ 0.00 & 0.00 $\pm$ 0.00 & 9.24 $\pm$ 9.14 &  \\
    & 4 & \(\text{NH2}_{\text{b}}\) & 0.06 $\pm$ 0.24 & 0.08 $\pm$ 0.02 & 0.02 $\pm$ 0.01 & 0.00 $\pm$ 0.00 & 0.00 $\pm$ 0.00 & 2.81 $\pm$ 0.42 &  \\
    \midrule[1.0pt]
    
    \multirow{3}{*}{\begin{tabular}{@{}l@{}}High Noise\\ ($\sigma_u = 5\times10^{-3}$)\\ \textit{Refer} \ref{Type3_none_high_10k}(i) \end{tabular}} 
    & 2 & ISH & 5.16 $\pm$ 2.65 & 0.48 $\pm$ 0.10 & 0.17 $\pm$ 0.07 & 0.00 $\pm$ 0.01 & 0.00 $\pm$ 0.00 & 21.02 $\pm$ 0.27 & \multirow{4}{*}{10\%} \\
    & 5 & \(\text{NH2}_{\text{a}}\) & 4.91 $\pm$ 0.34 & 0.00 $\pm$ 0.03 & 0.00 $\pm$ 0.03 & 0.26 $\pm$ 6.29 & 0.00 $\pm$ 0.00 & 13.84 $\pm$ 0.92 &  \\
    & 3 & HW & 1.92 $\pm$ 7.30 & 0.62 $\pm$ 0.09 & 0.00 $\pm$ 0.04 & 0.03 $\pm$ 0.07 & 0.00 $\pm$ 0.00 & 6.15 $\pm$ 1.58 & \\
    & 4 & \(\text{NH2}_{\text{b}}\) & 0.67 $\pm$ 0.51 & 0.05 $\pm$ 0.05 & 0.00 $\pm$ 0.00 & 0.00 $\pm$ 0.00 & 0.00 $\pm$ 0.00 & 3.13 $\pm$ 0.79 &  \\
  
    \bottomrule[1.5pt]
    \end{tabular}
    \end{adjustbox}
    \caption{True vs. predicted material parameters for multiple inner inclusions pattern (\autoref{hetero_pattern:type3}) across different noise levels using Bayesian-EUCLID framework.}
    \label{table_Type3_none_high_10k}
\end{table}
        
        Under high-noise conditions \( (\sigma_{u,\text{high}} = 5 \times 10^{-3}) \), it can be observed that material region prediction of the embedded inclusions remains accurate with a few omissions of elements from the base matrix as illustrated in \autoref{Type3_none_high_10k}i. Supplementary Figure S3 displays the corresponding results with moderate noise condition  \( (\sigma_{u,\text{mod}} = 2 \times 10^{-3}) \). \autoref{table_Type3_none_high_10k} summarizes the identified material parameters for each region under the moderate and high-noise settings. For segments corresponding to ISH and \(\text{NH2}_{\text{a}}\) materials (material IDs 2 and 5), the predicted parameters \( \theta_{1-5,19} \) remain close to their ground truth values. In contrast, the material parameter prediction for the HW and \(\text{NH2}_{\text{b}}\) regions (material segment IDs 3 and 4, respectively) shows large deviation from the ground truth with higher standard deviations, indicating a possible breakdown of the proposed method under strong noise interference. Supplementary Figure S9 compares the predicted strain energy density along six distinct deformation paths with the ground truth for the ISH–\(\text{NH2}_{\text{a}}\)–HW–\(\text{NH2}_{\text{b}}\) configuration for high-noise. The plots show that the energy density predictions for the ISH and \(\text{NH2}_{\text{a}}\) segments exhibit higher accuracy, whereas the predictions for HW and \(\text{NH2}_{\text{b}}\) show lower \(R^2\) scores and wider percentile uncertainty bands. The results suggest that additional experimental data, particularly under varied loading conditions, may be necessary to better capture the shear-dominated material response, as the volumetric stiffnesses corresponding to $\theta_{19}$ are accurately predicted for each material.\par

    \subsubsection*{Pattern Type 4: Cahn-Hilliard pattern}
    
\begin{figure}[H]
    \centering
    \includegraphics[width=\textwidth]{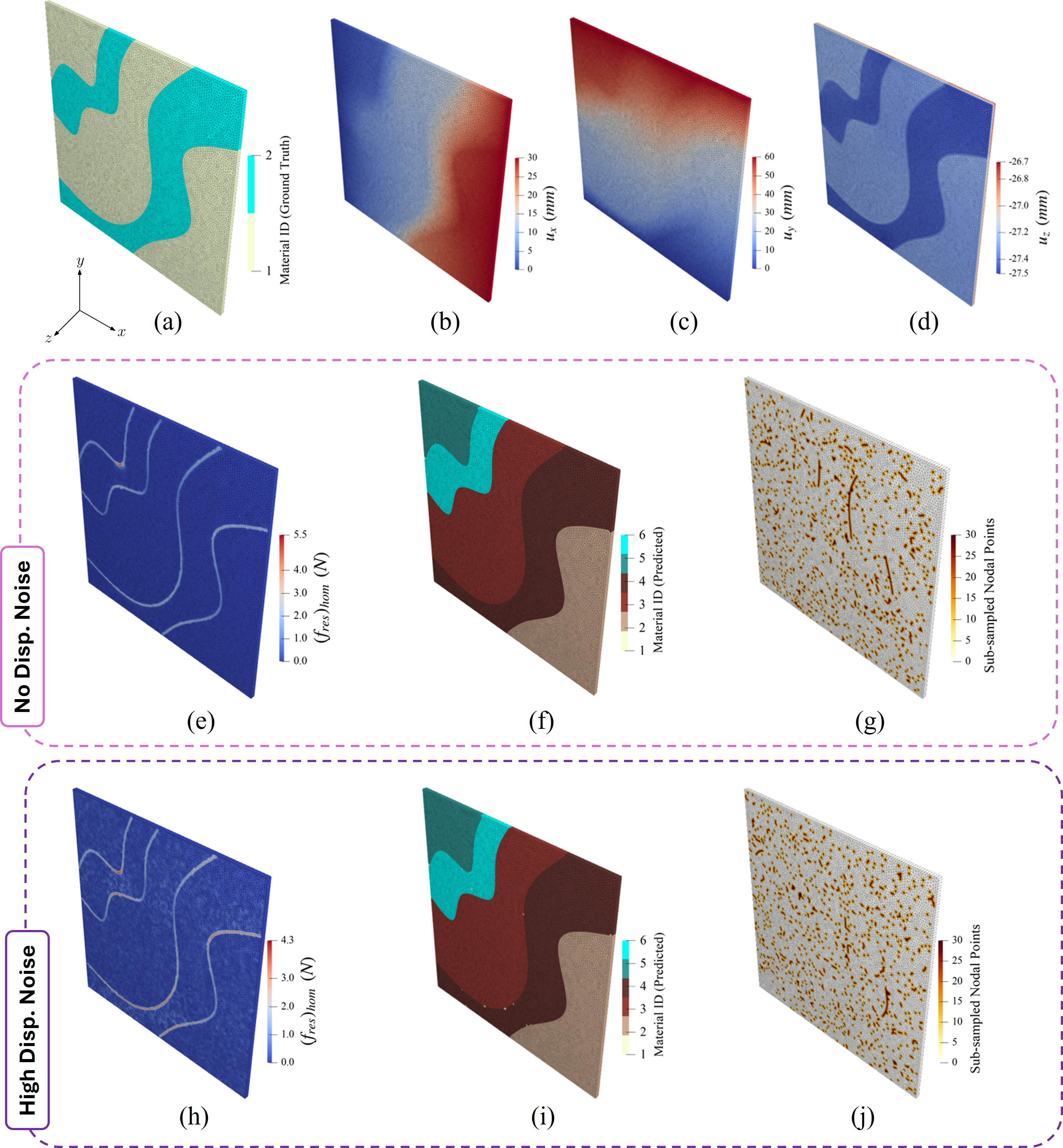}
    \caption{Model discovery for the  Cahn-Hilliard pattern  (\autoref{hetero_pattern:type4}). The ground truth material segment IDs are shown in (a), and the displacement fields in the \(x\), \(y\), and \(z\) directions are provided in (b)--(d). For the \textit{zero-noise} case, (e) depicts the residual force norm distribution (\autoref{eq19}) from the homogenized model, (f) illustrates the material segments identified through the segmentation algorithm, and (g) highlights the sub-sampled nodes from the \(\mathcal{D}^{\text{free}}\) set used to construct the heterogenized model. The \textit{high-noise} case follows a similar structure in panels (h)--(j), showing how the residual force field, segmentation, and sub-sampling behave under high noise levels. In both scenarios, the final heterogenized model is solved using the Bayesian framework to recover the region-wise constitutive parameters as listed in \autoref{table_Type4_none_high_20k}.}
    \label{Type4_none_high_20k}
\end{figure}

        In this type of domain heterogeneity, we investigate the Cahn-Hilliard pattern, which is often used to represent the biological pattern formation in soft tissue mechanics. \autoref{hetero_pattern:type4} illustrates this configuration, featuring five adjacent material regions alternately composed of two Neo-Hookean (\(\text{NH2}_{\text{a}},\text{NH2}_{\text{b}}\)) constitutive models. This specimen serves to evaluate the ability of Hetero-EUCLID to distinguish material sub-regions when multiple zones share identical mechanical properties. The domain is meshed with 21,040 uniformly distributed wedge-type elements. Using ABAQUS/CAE, we generate the full-field computational displacement dataset for quasi-static non-equi-biaxial loading as shown in \autoref{Type4_none_high_20k}b-d. The Cahn–Hilliard heterogeneous configuration is analyzed under noise-free, high, and moderate-noise conditions, with the corresponding results summarized in \autoref{Type4_none_high_20k}e-g, \ref{Type4_none_high_20k}h-j, and Supplementary Figure S4e-g, respectively. \par
        
        Using the homogenized model formulation, a distinct imbalance in the residual force norm is observed along the complex Cahn–Hilliard pattern boundaries, as shown in \autoref{Type4_none_high_20k}e, \ref{Type4_none_high_20k}h, and S4e. Material region segmentation is performed using the threshold parameter \( \lambda = 1.5 \), resulting in the prediction of six material segments. The segmentation output is visualized in \autoref{Type4_none_high_20k}f, \ref{Type4_none_high_20k}i, and S4f, where segment ID 1 contains a small number of misclustered elements under both noise-free and noisy conditions. The remaining segments (IDs 2-6) exhibit well-defined and accurate material interfaces. Subsequently, the material parameter identification is carried out using 5\% sub-sampling of the free nodes, and the noise-free results are reported in \autoref{table_Type4_none_high_20k}. Relative to the ground truth, material segments with IDs 2, 3, and 5 recover \(\text{NH2}_{\text{a}}\) model coefficients \( \theta_{1} \) and \( \theta_{19} \) close to 5.40 MPa and 15.00 MPa, respectively.\par
\begin{table}[H]
    \centering
    \begin{adjustbox}{width=\textwidth}
    \begin{tabular}{@{} l c c c c c c c c c @{}}
    \toprule[1.5pt]
    \multicolumn{10}{l}{\textbf{Pattern Type 4: Cahn-Hilliard} $\vert$ Mesh elements (C3D6): \textbf{21,040} $\vert$ Activated feature library terms, $\boldsymbol{n_{\mathit{f}} = 6}$} \\
    \midrule[1.5pt]
    \addlinespace
    \begin{tabular}{@{}l@{}}Case\end{tabular}
    & \begin{tabular}{@{}c@{}}Material\\ID\end{tabular} 
    & \begin{tabular}{@{}c@{}}Constitutive\\model\end{tabular} 
    & \begin{tabular}{@{}c@{}}$\theta_{1}$\\(MPa)\end{tabular} 
    & \begin{tabular}{@{}c@{}}$\theta_{2}$\\(MPa)\end{tabular} 
    & \begin{tabular}{@{}c@{}}$\theta_{3}$\\(MPa)\end{tabular} 
    & \begin{tabular}{@{}c@{}}$\theta_{4}$\\(MPa)\end{tabular} 
    & \begin{tabular}{@{}c@{}}$\theta_{5}$\\(MPa)\end{tabular} 
    & \begin{tabular}{@{}c@{}}$\theta_{19}$\\(MPa)\end{tabular} 
    & \begin{tabular}{@{}c@{}}Free nodes\\sub-sampling\end{tabular} \\
    \midrule[1.0pt]
    \multirow{2}{*}{\begin{tabular}{@{}l@{}}Ground Truth\\ \textit{Refer} \ref{Type4_none_high_20k}(a)\end{tabular}} 
    & 1 & \(\text{NH2}_{\text{a}}\) & 5.40 & $-$ & $-$ & $-$ & $-$ & 15.00 & \multirow{2}{*}{$-$} \\
    & 2 & \(\text{NH2}_{\text{b}}\) & 1.80 & $-$ & $-$ & $-$ & $-$ & 6.00 & \\

    \midrule[1.0pt]
    \multirow{3}{*}{\begin{tabular}{@{}l@{}}No Noise\\($\sigma_u = 0$)\\ \textit{Refer} \ref{Type4_none_high_20k}(f)\end{tabular}} 
    & 5 & \(\text{NH2}_{\text{a}}\) & 5.44 $\pm$ 0.01 & 0.00 $\pm$ 0.00 & 0.00 $\pm$ 0.00 & 0.00 $\pm$ 0.00 & 0.00 $\pm$ 0.00 & 15.08 $\pm$ 0.06 &  \\
    & 6 & \(\text{NH2}_{\text{b}}\) & 1.79 $\pm$ 0.01 & 0.00 $\pm$ 0.00 & 0.00 $\pm$ 0.00 & 0.00 $\pm$ 0.00 & 0.00 $\pm$ 0.00 & 5.97 $\pm$ 0.02 & \\
    & 3 & \(\text{NH2}_{\text{a}}\) & 5.41 $\pm$ 0.03 & 0.00 $\pm$ 0.00 & 0.00 $\pm$ 0.00 & 0.00 $\pm$ 0.00 & 0.00 $\pm$ 0.00 & 15.04 $\pm$ 0.05 & 5\%  \\
    & 4 & \(\text{NH2}_{\text{b}}\) & 1.80 $\pm$ 0.00 & 0.00 $\pm$ 0.00 & 0.00 $\pm$ 0.00 & 0.00 $\pm$ 0.00 & 0.00 $\pm$ 0.00 & 5.99 $\pm$ 0.01 & \\
    & 2 & \(\text{NH2}_{\text{a}}\) & 5.43 $\pm$ 0.01 & 0.00 $\pm$ 0.00 & 0.00 $\pm$ 0.00 & 0.00 $\pm$ 0.00 & 0.00 $\pm$ 0.00 & 15.03 $\pm$ 0.01 & \\
    
    \midrule[1.0pt]
    \multirow{4}{*}{\begin{tabular}{@{}l@{}}Moderate \\ Noise\\($\sigma_u = 2 \times 10^{-3}$)\\ \textit{Refer} S4(f) \end{tabular}} 
    & 5 & \(\text{NH2}_{\text{a}}\) & 6.36 $\pm$ 0.03 & 0.00 $\pm$ 0.00 & 0.00 $\pm$ 0.00 & 0.00 $\pm$ 0.00 & 0.00 $\pm$ 0.00 & 16.88 $\pm$ 0.11 &  \\
    & 6 &  \(\text{NH2}_{\text{b}}\) & 1.84 $\pm$ 0.01 & 0.00 $\pm$ 0.00 & 0.00 $\pm$ 0.00 & 0.00 $\pm$ 0.00 & 0.00 $\pm$ 0.00 & 6.07 $\pm$ 0.04 & \\
    & 3 & \(\text{NH2}_{\text{a}}\) & 5.31 $\pm$ 0.03 & 0.00 $\pm$ 0.00 & 0.00 $\pm$ 0.00 & 0.00 $\pm$ 0.00 & 0.00 $\pm$ 0.00 & 14.67 $\pm$ 0.08 & 5\%  \\
    & 4 &  \(\text{NH2}_{\text{b}}\) & 1.78 $\pm$ 0.01 & 0.00 $\pm$ 0.00 & 0.00 $\pm$ 0.00 & 0.00 $\pm$ 0.00 & 0.00 $\pm$ 0.00 & 5.93 $\pm$ 0.03 & \\
    & 2 &  \(\text{NH2}_{\text{a}}\) & 4.76 $\pm$ 0.05 & 0.37 $\pm$ 0.01 & 0.07 $\pm$ 0.01 & 0.00 $\pm$ 0.00 & 0.00 $\pm$ 0.00 & 15.48 $\pm$ 0.05 & \\
    
    \midrule[1.0pt]
    \multirow{3}{*}{\begin{tabular}{@{}l@{}}High Noise\\($\sigma_u = 5 \times 10^{-3}$)\\ \textit{Refer} \ref{Type4_none_high_20k}(i) \end{tabular}} 
    & 5 & \(\text{NH2}_{\text{a}}\) & 6.74 $\pm$ 0.22 & 0.00 $\pm$ 0.00 & 0.10 $\pm$ 0.10 & 0.00 $\pm$ 0.00 & 0.00 $\pm$ 0.01 & 18.32 $\pm$ 0.23 & \\
    & 6 & \(\text{NH2}_{\text{b}}\) & 1.78 $\pm$ 0.03 & 0.00 $\pm$ 0.00 & 0.00 $\pm$ 0.00 & 0.00 $\pm$ 0.00 & 0.00 $\pm$ 0.00 & 5.86 $\pm$ 0.08 & \\
    & 3 & \(\text{NH2}_{\text{a}}\) & 5.03 $\pm$ 0.06 & 0.00 $\pm$ 0.00 & 0.00 $\pm$ 0.00 & 0.00 $\pm$ 0.00 & 0.00 $\pm$ 0.00 & 13.81 $\pm$ 0.16 & 10\% \\
    & 4 & \(\text{NH2}_{\text{b}}\) & 1.67 $\pm$ 0.02 & 0.00 $\pm$ 0.00 & 0.00 $\pm$ 0.00 & 0.00 $\pm$ 0.00 & 0.00 $\pm$ 0.00 & 5.51 $\pm$ 0.07 & \\
    & 2 & \(\text{NH2}_{\text{a}}\) & 5.45 $\pm$ 0.11 & 0.17 $\pm$ 0.04 & 0.02 $\pm$ 0.02 & 0.00 $\pm$ 0.00 & 0.00 $\pm$ 0.00 & 16.28 $\pm$ 0.12 & \\
    \bottomrule[1.5pt]
    \end{tabular}
    \end{adjustbox}
    \caption{True vs. predicted material parameters for the Cahn-Hilliard pattern (\autoref{hetero_pattern:type4}) across different noise levels using the Bayesian-EUCLID framework.}
    \label{table_Type4_none_high_20k}
\end{table}
      
        Similarly, the corresponding predictions of \( \theta_{1} \) and \( \theta_{19} \) for material segments 4 and 6 closely match their \(\text{NH2}_{\text{b}}\) ground truth values of 1.80 MPa and 6.00 MPa, respectively. All remaining material parameters \( \theta_{2-5} \) are correctly predicted as zero with low standard deviation. A key advantage of the Hetero-EUCLID framework lies in its interpretability. While the segmentation algorithm identifies five major material regions, the predicted model parameters provide meaningful physical insight into material similarity. It can be inferred that segments 2, 3, and 5 exhibit nearly identical Neo-Hookean model coefficients, indicating that these segments represent the same material. Similarly, material regions with IDs 4 and 6 are identified as belonging to a second distinct material type. Thus, Hetero-EUCLID segments the domain and enables grouping of regions based on interpretable material parameters, effectively identifying five spatial zones that represent two unique material classes.\par

        \autoref{table_Type4_none_high_20k} summarizes a comparative study of material parameter prediction performance for the Cahn-Hilliard pattern specimen under moderate and high noise settings. The resulting predictions of \( \theta_{1} \) and \( \theta_{19} \) across all identified Neo-Hookean regions lie within an acceptable tolerance of the ground truth and are associated with low standard deviations, indicating high confidence. Additionally, the remaining coefficients \( \theta_{2-5} \) are correctly predicted as zero, promoting sparsity. Supplementary Figure S10 demonstrates strong agreement between the strain energy density plots computed from the predicted model parameters and those from the corresponding ground truth. These results demonstrate the robustness and interpretability of the Hetero-EUCLID framework for complex heterogeneous configurations under high noise conditions.\par


\subsection{Non-native mesh}
\label{non_native_study}
    \begin{figure}[H]
        \centering
        \includegraphics[width=\textwidth]{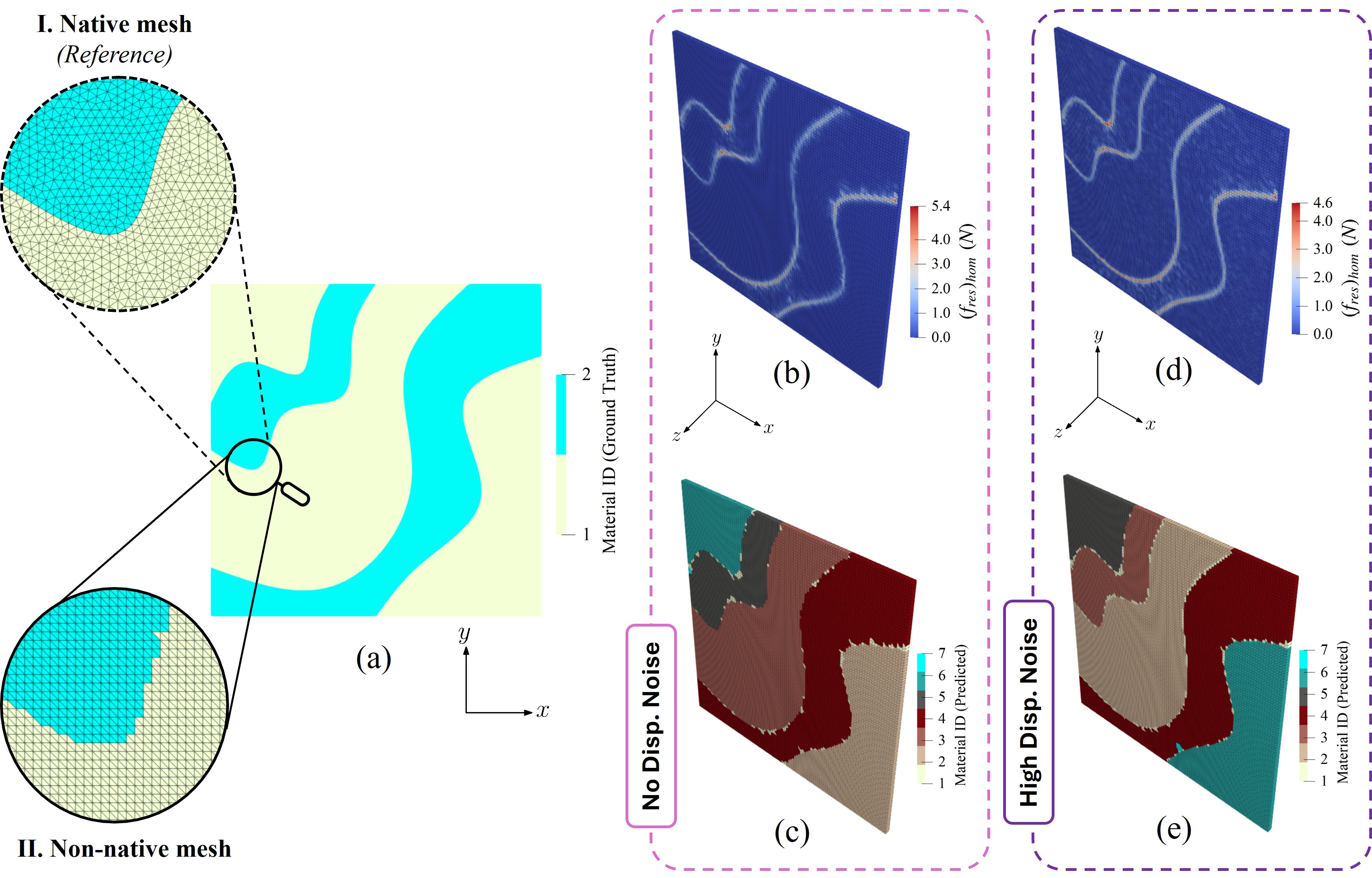} 
        \caption{Model discovery for the  Cahn-Hilliard pattern  (\autoref{hetero_pattern:type4})  using \textit{non-native} meshing. (a) shows the ground truth material segment IDs with visual comparison of (I) native and (II) non-native mesh views. For the \textit{zero-noise} case, (b) depicts the residual force norm distribution (\autoref{eq19}) from the homogenized model, and (c) illustrates the material segments identified through the segmentation algorithm. The \textit{high-noise} case follows a similar structure in panels (d)--(e), showing how the residual force field and segmentation behave under high noise levels. In both scenarios, the final heterogenized model is solved using the Bayesian framework to recover the region-wise constitutive parameters as listed in \autoref{table_Type4_none_20k_non-native}.}
        \label{Type4_none_20k_non-native}
    \end{figure}

    In \autoref{case_study}, we evaluated the performance of the Hetero-EUCLID framework across various case studies. For all configurations, the computational domain was discretized using wedge elements based on a known element connectivity matrix. Notably, the same mesh (referred to as the \textit{native mesh}) used in the forward FEM simulations for generating displacement data was also used for the inverse formulation. In this section, we investigate further to emulate DIC experimental conditions more pragmatically, by evaluating the Hetero-EUCLID framework on a \textit{non-native mesh}. A non-native mesh refers to a discretization where the element edges do not necessarily align with the underlying material interface boundaries. To investigate this scenario, we use the complex Type 4 Cahn–Hilliard heterogeneous pattern and discretize the domain using a uniform grid and generate wedge elements using Delaunay triangulation \citep{lee1980two} on the surface, resulting in 19,602 uniformly distributed wedge elements. \autoref{Type4_none_20k_non-native}a provides a comparative visualization between the native (I) and non-native (II) meshes, showing a zoomed-in view near a material interface. It can be observed that the native mesh aligns precisely with the material boundary, while the non-native mesh does not. In DIC experiments, the material interface geometry is inherently unknown during meshing; therefore, this non-native mesh study is crucial for evaluating the robustness and practical applicability of the Hetero-EUCLID framework under realistic experimental conditions. \par
    
    Using the displacement field generated from ABAQUS/CAE for the Cahn–Hilliard pattern under non-equi-biaxial loading, we perform FE-based interpolation to compute full-field displacements at the nodes of the non-native mesh. Such an FE-based interpolation is expected to add noise to the displacement field. In this context, we evaluate the performance of our Hetero-EUCLID framework for two cases-- one without any additional displacement noise (apart from the inherent interpolation noise) and another with additional, high, uncorrelated displacement noise \( (\sigma_{u,\text{high}} = 5 \times 10^{-3}) \). Subsequently, the segmentation is performed using the random seed-based island growth algorithm, and the resulting material segments are shown in \autoref{Type4_none_20k_non-native}c and \ref{Type4_none_20k_non-native}e. While the algorithm successfully identifies five major material zones, the interfaces appear less sharp, which is expected and justified given the non-aligned structure of the non-native mesh grid compared to those observed with the native mesh configuration.\par

    \begin{table}[H]
        \centering
        \begin{adjustbox}{width=\textwidth}
        \begin{tabular}{@{} l c c c c c c c c c @{}}
        \toprule[1.5pt]
        \multicolumn{10}{l}{\textbf{Pattern Type 4: Cahn-Hilliard} $\vert$ \textit{Non-native} Mesh elements (C3D6): \textbf{19,602} $\vert$ Activated feature library terms, $\boldsymbol{n_{\mathit{f}} = 6}$} \\
        \midrule[1.5pt]
        \addlinespace
        \begin{tabular}{@{}l@{}}Case\end{tabular}
        & \begin{tabular}{@{}c@{}}Material\\ID\end{tabular} 
        & \begin{tabular}{@{}c@{}}Constitutive\\model\end{tabular} 
        & \begin{tabular}{@{}c@{}}$\theta_{1}$\\(MPa)\end{tabular} 
        & \begin{tabular}{@{}c@{}}$\theta_{2}$\\(MPa)\end{tabular} 
        & \begin{tabular}{@{}c@{}}$\theta_{3}$\\(MPa)\end{tabular} 
        & \begin{tabular}{@{}c@{}}$\theta_{4}$\\(MPa)\end{tabular} 
        & \begin{tabular}{@{}c@{}}$\theta_{5}$\\(MPa)\end{tabular} 
        & \begin{tabular}{@{}c@{}}$\theta_{19}$\\(MPa)\end{tabular} 
        & \begin{tabular}{@{}c@{}}Free nodes\\sub-sampling\end{tabular} \\
        \midrule[1.0pt]
        \multirow{2}{*}{\begin{tabular}{@{}l@{}}Ground\\Truth, \ref{Type4_none_20k_non-native}(a)\end{tabular}} 
        & 1 & \(\text{NH2}_{\text{a}}\) & 5.40 & $-$ & $-$ & $-$ & $-$ & 15.00 & \multirow{2}{*}{$-$} \\
        & 2 & \(\text{NH2}_{\text{b}}\) & 1.80 & $-$ & $-$ & $-$ & $-$ & 6.00 & \\
        \midrule[1.0pt]

        \multirow{5}{*}{\begin{tabular}{@{}l@{}}No Noise \\($\sigma_u = 0$)\\ Bayesian\\ framework\\ \textit{Refer} \ref{Type4_none_20k_non-native}(c)  \end{tabular}} 
        & 6 & \(\text{NH2}_{\text{a}}\) & 6.05 $\pm$ 0.96 & 0.12 $\pm$ 0.43 & 0.01 $\pm$ 0.01 & 0.00 $\pm$ 0.01 & 0.00 $\pm$ 0.00 & 17.00 $\pm$ 1.62 &  \\
        & 5 & \(\text{NH2}_{\text{b}}\) & 1.29 $\pm$ 0.09 & 0.00 $\pm$ 0.00 & 0.00 $\pm$ 0.00 & 0.00 $\pm$ 0.00 & 0.00 $\pm$ 0.00 & 4.12 $\pm$ 0.27 &  \\
        & 3 & \(\text{NH2}_{\text{a}}\) & 4.97 $\pm$ 0.99 & 0.12 $\pm$ 0.08 & 0.01 $\pm$ 0.03 & 0.00 $\pm$ 0.00 & 0.00 $\pm$ 0.00 & 14.19 $\pm$ 2.50 & 10\% \\
        & 4 & \(\text{NH2}_{\text{b}}\) & 1.77 $\pm$ 0.01 & 0.00 $\pm$ 0.00 & 0.00 $\pm$ 0.00 & 0.00 $\pm$ 0.00 & 0.00 $\pm$ 0.00 & 5.61 $\pm$ 0.03 & \\
        & 2 & \(\text{NH2}_{\text{a}}\) & 5.89 $\pm$ 0.09 & 0.00 $\pm$ 0.00 & 0.00 $\pm$ 0.00 & 0.00 $\pm$ 0.02 & 0.00 $\pm$ 0.00 & 15.25 $\pm$ 0.04 & \\
        \midrule[1.0pt]

        \multirow{5}{*}{\begin{tabular}{@{}l@{}}High Noise\\ ($\sigma_u = 5 \times 10^{-3}$)\\ (\textit{Ordinary} \\ \textit{Least Squares}\\ \textit{Regression})\end{tabular}} 
        & 5 & \(\text{NH2}_{\text{a}}\) & 12.84 & -2.99 & -3.09 & 1.35 & 0.41 & 18.76 & \\
        & 3 & \(\text{NH2}_{\text{b}}\) & 1.31 & 0.24 & -0.06 & 0.06 & 0.00 & 3.68 & \\
        & 2 & \(\text{NH2}_{\text{a}}\) & 1.86 & 1.05 & 0.71 & -0.13 & -0.07 & 9.59 & $-$\\
        & 4 & \(\text{NH2}_{\text{b}}\) & 1.71 & -0.02 & -0.06 & 0.08 & 0.00 & 3.50 & \\
        & 6 & \(\text{NH2}_{\text{a}}\) & 8.38 & -2.31 & -1.92 & 1.94 & -0.02 & 13.65 & \\
        \midrule[1.0pt]

        \multirow{5}{*}{\begin{tabular}{@{}l@{}} High Noise\\ ($\sigma_u = 5 \times 10^{-3}$)\\Bayesian\\ framework \\ \textit{Refer} \ref{Type4_none_20k_non-native}(e) \end{tabular}} 
        & 5 & \(\text{NH2}_{\text{a}}\) & 7.69 $\pm$ 0.07 & 0.00 $\pm$ 0.00 & 0.04 $\pm$ 0.04 & 0.00 $\pm$ 0.00 & 0.07 $\pm$ 0.01 & 18.82 $\pm$ 0.16 &  \\
        & 3 & \(\text{NH2}_{\text{b}}\) & 1.55 $\pm$ 0.04 & 0.12 $\pm$ 0.01 & 0.00 $\pm$ 0.00 & 0.03 $\pm$ 0.00 & 0.00 $\pm$ 0.00 & 3.49 $\pm$ 0.04 &  \\
        & 2 & \(\text{NH2}_{\text{a}}\) & 3.36 $\pm$ 0.11 & 0.42 $\pm$ 0.06 & 0.26 $\pm$ 0.03 & 0.01 $\pm$ 0.02 & 0.00 $\pm$ 0.00 & 10.27 $\pm$ 0.10 & 10\% \\
        & 4 & \(\text{NH2}_{\text{b}}\) & 1.74 $\pm$ 0.02 & 0.00 $\pm$ 0.01 & 0.00 $\pm$ 0.00 & 0.04 $\pm$ 0.00 & 0.00 $\pm$ 0.00 & 3.46 $\pm$ 0.04 & \\
        & 6 & \(\text{NH2}_{\text{a}}\) & 5.66 $\pm$ 0.07 & 0.00 $\pm$ 0.00 & 0.13 $\pm$ 0.04 & 0.00 $\pm$ 0.00 & 0.00 $\pm$ 0.00 & 13.21 $\pm$ 0.09 & \\

        \bottomrule[1.5pt]
        \end{tabular}
        \end{adjustbox}
        \caption{True vs. predicted material parameters for the \textit{non-natively} meshed Cahn-Hilliard pattern. The results are presented for three cases, (i) no additional displacement noise, solved using Bayesian regression framework, (ii) High uncorrelated displacement noise  ($\sigma_{u,\mathrm{high}} = 5 \times 10^{-3}$) solved using the OLS regression approach, and (iii) High uncorrelated displacement noise ($\sigma_{u,\mathrm{high}} = 5 \times 10^{-3}$) solved using the Bayesian regression approach.}
        \label{table_Type4_none_20k_non-native}
    \end{table}
    
     \autoref{table_Type4_none_20k_non-native} presents the model discovery results for the non-native mesh under noise-free and high noise conditions, and compares the material parameter predictions obtained using ordinary least squares (OLS) regression and the Bayesian framework for the high noise case. When the heterogeneous model (refer to \autoref{eq17}) is solved using OLS, the predicted NH2 model coefficients \( (\theta_{1,19})_{\text{OLS}} \) exhibit notable deviations from the ground truth. Moreover, the intermediate model parameters \( (\theta_{2-5})_{\text{OLS}} \) yield negative values, which may render the strain energy density non-polyconvex and reflect overfitting under noise. In contrast, the Bayesian framework enforces positivity on the material parameters \( \boldsymbol{\theta} \) via truncated normal distribution priors, ensuring physical admissibility. The material parameters predictions via the Bayesian approach are closer to the ground truth and are associated with lower standard deviations, indicating a higher level of statistical confidence. Additionally, the remaining coefficients \( \theta_{2-5} \) are correctly predicted as close to zero, promoting sparsity. The strain-energy plots corresponding to the predicted material parameters possess high \( R^2 \) scores as shown in Supplementary Figure S11. While the segmentation is predicted accurately for the high displacement noise study, the predicted material parameters for segment 2 and segment 5 are considerably different from the ground truth, presumably due to the additional displacement noise added by the FE interpolation step (compared to results from \autoref{table_Type4_none_high_20k}). In the noise-free case, regions labeled 2, 3, and 6 can be seen to possess similar material properties, while segments 4 and 5 possess material properties similar to each other. These results demonstrate that the Hetero-EUCLID framework remains accurate and robust under reasonably high noise conditions, for a non-native displacement field mesh. \par

\subsection{Benchmarking study with feature exclusion}
\label{benchmark_study}

In this section, we further evaluate the performance of the proposed Hetero-EUCLID framework for complex hyperelastic material models. We consider two specific classes of materials-- one governed by the Ogden (OG) model \citep{ogden1972large} and the other by the Holzapfel--Gasser--Ogden (HGO) model \citep{holzapfel2000new}. In the two case studies the OG material and the HGO material are assumed to form a circular inclusion in a matrix made of Neo-Hookean (NH2) material. As before, we simulate non-equibiaxial stretch of these geometries in ABAQUS to generate the surface displacement data, and the boundary-averaged normal reaction force data. While performing model discovery using Hetero-EUCLID, the specific OG and HGO-type terms were omitted from the feature library \(Q\) (refer to \autoref{eq6} and \autoref{table_feature-list}) to evaluate the performance of the Hetero-EUCLID framework under feature exclusion. \par

\subsubsection*{Pattern Type 5: Circular inclusion (Ogden material) in a matrix (NH2 material)}
    
    We consider a hyperelastic heterogeneous specimen having a circular inclusion embedded in a surrounding matrix material as illustrated in \autoref{hetero_pattern:type5}. The matrix follows the Neo-Hookean model (NH2) (\autoref{eq28}) and the circular inclusion is described by the Ogden (OG) model (\autoref{eq_ogden}). The domain is discretized using 10,594 wedge elements, and the displacement data is generated under non-equi-biaxial loading conditions with final directional stretch ratios of \( \lambda_x = 1.6 \) and \( \lambda_y = 2.2 \).

\begin{figure}[H]
    \centering
    \includegraphics[width=\textwidth]{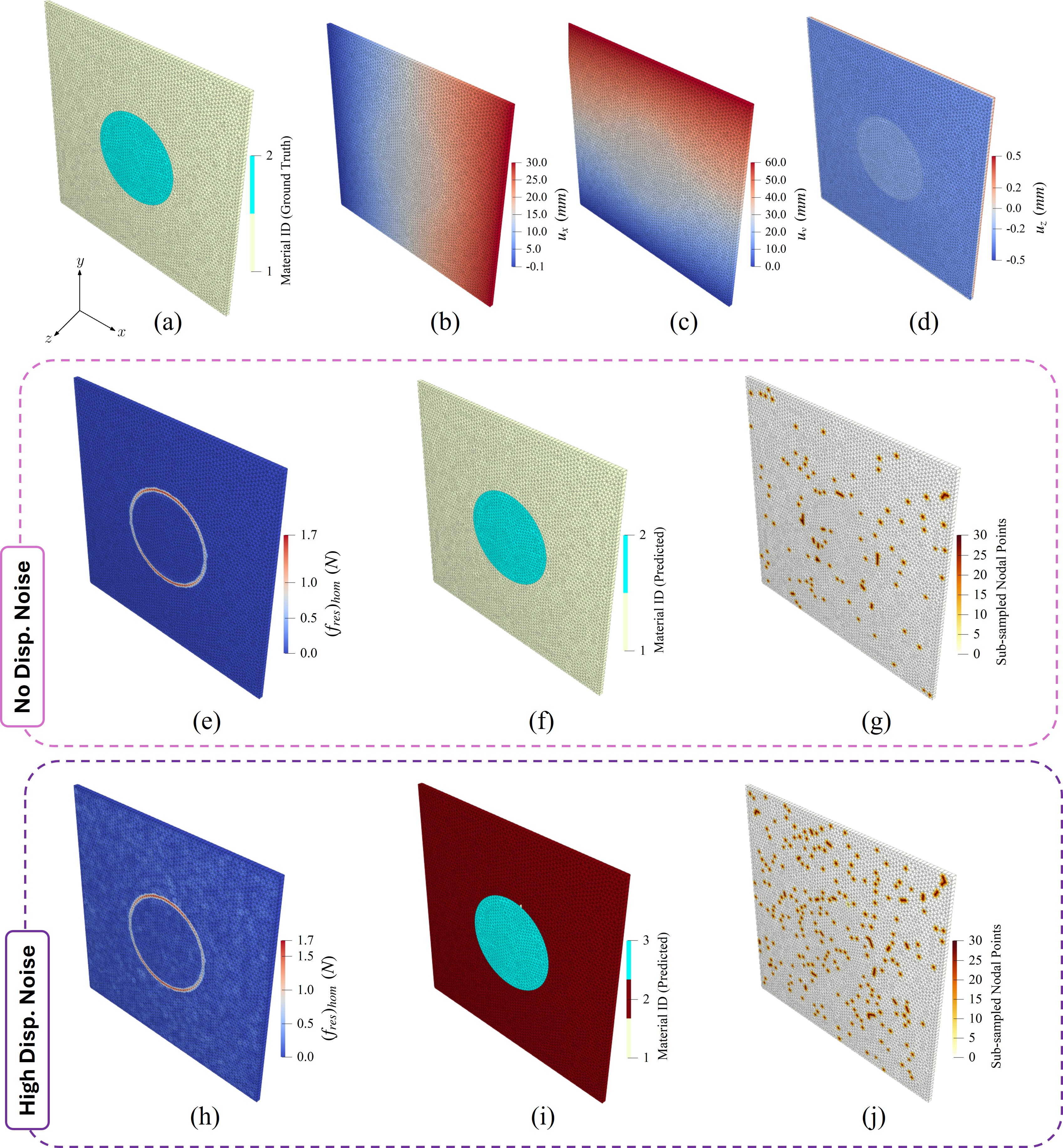}
    \caption{Model discovery for the  circular inclusion--Ogden pattern  (\autoref{hetero_pattern:type5}). The ground truth material segment IDs are shown in (a), and the displacement fields in the \(x\), \(y\), and \(z\) directions are provided in (b)--(d). For the \textit{zero-noise} case, (e) depicts the residual force norm distribution (\autoref{eq19}) from the homogenized model, (f) illustrates the material segments identified through the segmentation algorithm, and (g) highlights the sub-sampled nodes from the \(\mathcal{D}^{\text{free}}\) set used to construct the heterogenized model. The \textit{high-noise} case follows a similar structure in panels (h)--(j), showing how the residual force field, segmentation, and sub-sampling behave under high noise levels. In both scenarios, the final heterogenized model is solved using the Bayesian framework to recover the region-wise constitutive parameters as listed in \autoref{table_Type5_10k_Ogden}.}
    \label{Type5_none_high_10k}
\end{figure}

    The OG-type term \(\frac{2\mu}{\eta}(\tilde{\lambda}_{1}^{\eta} + \tilde{\lambda}_{2}^{\eta} + \tilde{\lambda}_{3}^{\eta} - 3)\) was intentionally omitted from the feature library \(Q\) defined in \autoref{eq6}. For the Type~5 (Ogden) case,  all 19 terms of the feature library (\(n_f = 19\)) listed in \autoref{table_feature-list} were selected for inverse parameter identification using the Hetero-EUCLID framework. The specimen is subjected to the same non-equi-biaxial loading conditions as in the previous case, and the analysis was performed under noise-free, moderate, and high noise levels to examine the framework’s robustness. The corresponding results are presented in \autoref{Type5_none_high_10k}e--g, \ref{Type5_none_high_10k}h--j, and Supplementary Figure~S5e-g, respectively. In each scenario, the residual force distribution \( (f^a_{\text{res}}) \) highlights a distinct mechanical imbalance at the material interfaces, as observed in \autoref{Type5_none_high_10k}e, \ref{Type5_none_high_10k}h, and S5e. Subsequently, the segmentation was performed using the seed-based growth algorithm with the tunable threshold parameter \( \lambda \) (refer to \autoref{eq21}), and the resulting material partitions are shown in \autoref{Type5_none_high_10k}f, \ref{Type5_none_high_10k}i, and S5f. The predicted material segments align well with the true segmentation (\autoref{Type5_none_high_10k}a), with a few elements being mis-labelled (Segment 1) under high-noise conditions (\(\sigma_{u,\text{high}} = 5\times 10^{-3}\)).\par 

\begin{figure}[H]
    \centering
    \begin{subfigure}[b]{0.24\textwidth}
        \includegraphics[width=0.95\textwidth]{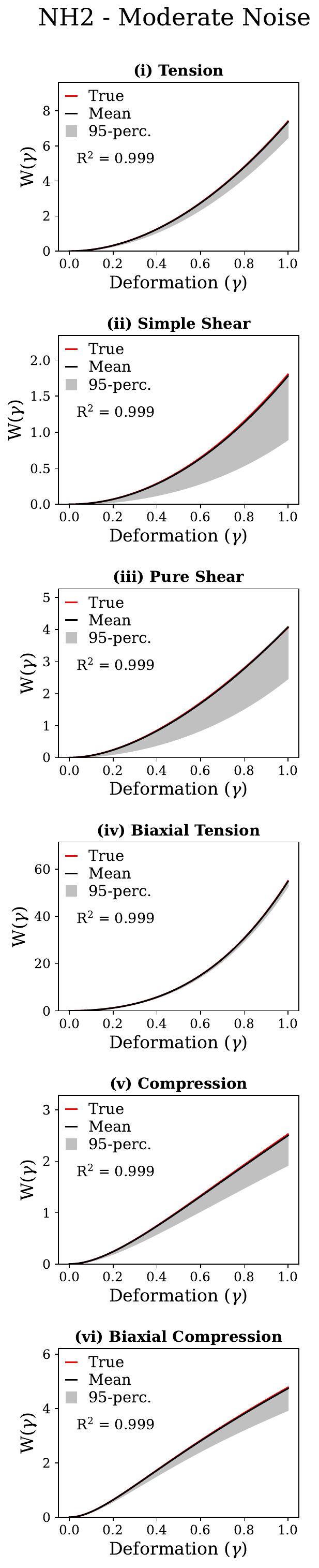}
        \caption{Material ID 1}
        \label{W_gamma_Type5_mod:ID1}
    \end{subfigure}
    \hfill
    \begin{subfigure}[b]{0.240\textwidth}
        \includegraphics[width=0.95\textwidth]{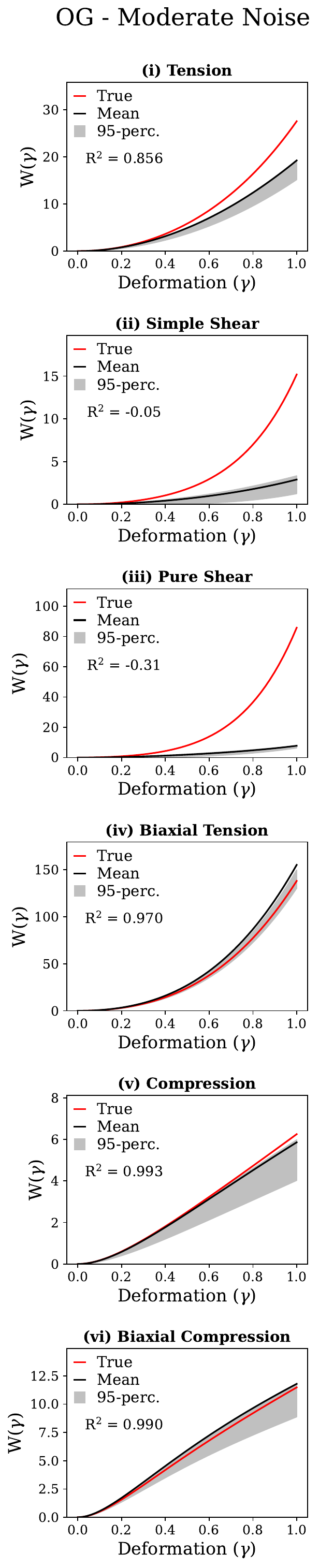}
        \caption{Material ID 2}
        \label{W_gamma_Type5_mod:ID2}
    \end{subfigure}
    \hfill
    \begin{subfigure}[b]{0.240\textwidth}
        \includegraphics[width=0.92\textwidth]{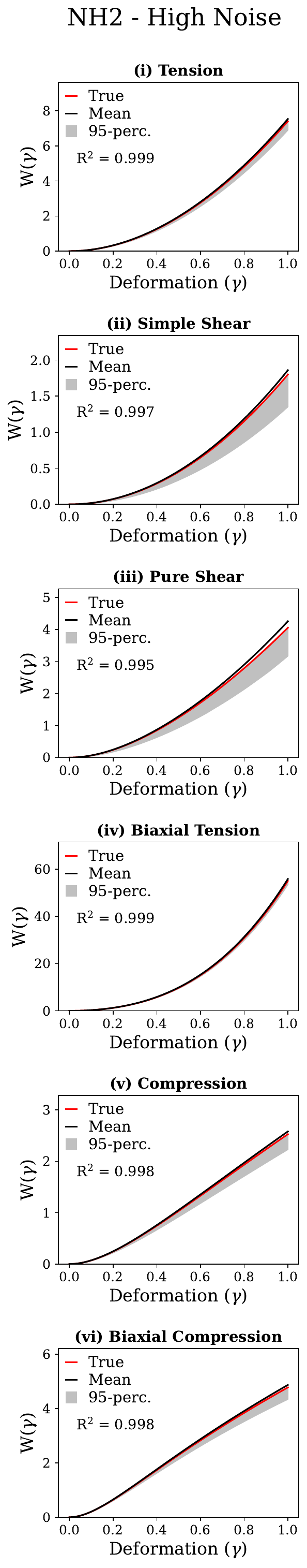}
        \caption{Material ID 2}
        \label{W_gamma_Type5_high:ID2}
    \end{subfigure}
    \hfill
    \begin{subfigure}[b]{0.239\textwidth}
        \includegraphics[width=0.95\textwidth]{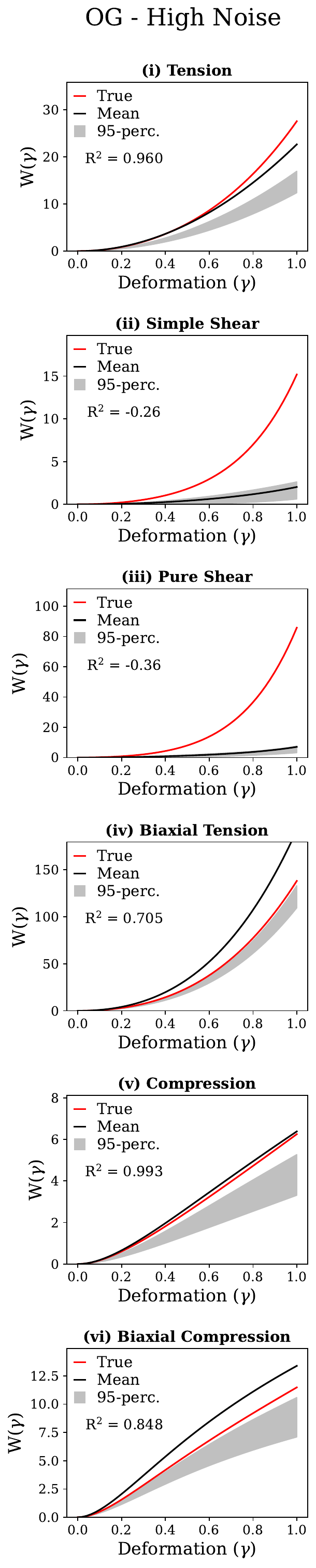}
        \caption{Material ID 3}
        \label{W_gamma_Type5_high:ID3}
    \end{subfigure}
    \caption{Strain energy density plots for the circular inclusion pattern (\autoref{hetero_pattern:type5}). Panels (a) and (b) correspond to the identified material sub-domains under moderate noise, while (c) and (d) show the results for the high noise case as reported in \autoref{table_Type5_10k_Ogden}. For each identified material segment ID (a)--(d), subplots (i)--(vi) show strain energy density plots \( W(\gamma) \) along various deformation paths (\autoref{eq31}), providing a comparison between the discovered and true models.}
    \label{W_gamma_Type5_mod_high}
\end{figure}

    \autoref{table_Type5_10k_Ogden} summarizes the discovery of material parameters for the NH2--OG regions for different noise levels. The parameters \((\theta_{1}, \theta_{19})\) corresponding to the base material NH2 are accurately predicted with low standard deviations under noise-free, moderate noise \((\sigma_{u,\text{mod}} = 2\times10^{-3})\) and high noise conditions \((\sigma_{u,\text{mod}} = 5\times10^{-3})\), indicating high confidence in the identified parameters. Noticeably, as the materials being considered are all isotropic, the parameters corresponding to anisotropic directional stretches-- $\theta_{15}$, $\theta_{16}$ and $\theta_{17}$ are all discovered to be 0 for all models at all noise levels. Since the OG-type term was intentionally omitted from the feature library, we evaluate the accuracy of the predicted parameters by comparing the true and predicted strain energy densities along six different deformation paths:

\begin{align}
    \label{eq31}
    \boldsymbol{F}^{\text{UT}}(\gamma) &= 
    \begin{bmatrix}
        1+\gamma & 0\\ 
        0 & 1
    \end{bmatrix}, &
    \boldsymbol{F}^{\text{UC}}(\gamma) &= 
    \begin{bmatrix}
        1/(1+\gamma) & 0\\ 
        0 & 1
    \end{bmatrix}, &
    \boldsymbol{F}^{\text{SS}}(\gamma) &= 
    \begin{bmatrix}
        1 & \gamma\\ 
        0 & 1
    \end{bmatrix},   \\ \nonumber
    \boldsymbol{F}^{\text{BT}}(\gamma) &= 
    \begin{bmatrix}
        1+\gamma & 0\\ 
        0 & 1 + \gamma
    \end{bmatrix}, &
    \boldsymbol{F}^{\text{BC}}(\gamma) &= 
    \begin{bmatrix}
        1/(1+\gamma) & 0\\ 
        0 & 1/(1+\gamma)
    \end{bmatrix}, &
    \boldsymbol{F}^{\text{PS}}(\gamma) &= 
    \begin{bmatrix}
        1+\gamma & 0\\ 
        0 & 1/(1+\gamma)
    \end{bmatrix}
\end{align}
where \( \gamma \in [0,1] \) denotes deformation parameter. The abbreviations correspond to different loading modes: uniaxial tensile (UT), uniaxial compressive (UC), biaxial tensile (BT), biaxial compressive (BC), simple shear (SS), and pure shear (PS). For each loading path, we report the predicted strain energy density along with the 95 percentile uncertainty band, and quantify the accuracy using the coefficient of determination \( R^2 \) between the predicted and true energy values.
    We compare the predicted and true strain energy densities for both NH2 and OG materials, for moderate and high noise displacement data. \autoref{W_gamma_Type5_mod_high} presents these results, where we have a close agreement between the predicted and true strain energy densities for the NH2 model (\autoref{W_gamma_Type5_mod:ID1} and \ref{W_gamma_Type5_high:ID2}) across all deformation paths, for both-- moderate and high noise cases. For the OG model, \autoref{W_gamma_Type5_mod:ID2} and \ref{W_gamma_Type5_high:ID3} exhibit good agreement between the predicted and true strain energy densities, except in shear-dominated deformation paths, where there are significant deviations between the predicted and true values. The shear-response accuracy of the OG model can be further enhanced by incorporating additional data from shear-specific loading tests alongside the biaxial stretch experiments. The cause for the unrealistically high standard deviations reported in \autoref{table_Type5_10k_Ogden} for the OG material can be understood by observing the strain energy plots in \autoref{W_gamma_Type5_mod:ID2} and \ref{W_gamma_Type5_high:ID3}. It can be observed that the predicted mean strain energy curve lies outside the 95 percentile region (shaded region). This occurs only if the distribution is extremely skewed, in which case the standard deviation is a less effective measure to predict spread in the distribution. In summary, the predicted mean results demonstrate that Hetero-EUCLID performs reliably across different noise levels and remains effective even when key Ogden-type features are excluded from the library.\par

\begin{table}[H]
    \centering
    \begin{adjustbox}{width=\textwidth}
    \begin{tabular}{@{} l c c *{6}{>{\centering\arraybackslash}p{2.3cm}} @{}}
    \toprule[1.5pt]
    \multicolumn{9}{l}{\textbf{Pattern Type 5: Circular inclusion – Ogden} $\vert$ Mesh elements (C3D6): \textbf{10,594}} \\
    \midrule[1.5pt]
    \multirow{2}{*}{\textbf{Case}} 
        & \multicolumn{2}{c}{Ground} 
        & \multicolumn{2}{c}{No Noise} 
        & \multicolumn{2}{c}{Moderate Noise} 
        & \multicolumn{2}{c}{High Noise} \\
        & \multicolumn{2}{c}{Truth} 
        & \multicolumn{2}{c}{($\sigma_u = 0$)} 
        & \multicolumn{2}{c}{($\sigma_u = 2 \times 10^{-3}$)} 
        & \multicolumn{2}{c}{($\sigma_u = 5 \times 10^{-3}$)} \\
    \midrule[1.0pt]
    
    \begin{tabular}{@{}l@{}}Constitutive\\model\end{tabular} 
        & NH2 & OG & NH2 & OG & NH2 & OG & NH2 & OG \\
    \midrule[1.0pt]

    Material ID & 1 & 2 & 1 & 2 & 1 & 2 & 2 & 3 \\
    \midrule[1.0pt]

    $\theta_{1}$ (MPa) & 1.80 & $-$ &
        1.80 $\pm$ 0.00 & 2.60 $\pm$ 0.38 &
        1.76 $\pm$ 0.47 & 2.59 $\pm$ 1.63 &
        1.84 $\pm$ 1.83 & 1.50 $\pm$ 4.03 \\
    $\theta_{2}$ (MPa) & $-$ & $-$ &
        0.00 $\pm$ 0.00 & 0.03 $\pm$ 0.30 &
        0.00 $\pm$ 0.00 & 0.04 $\pm$ 1.18 &
        0.00 $\pm$ 0.00 & 0.10 $\pm$ 3.14 \\
    $\theta_{3}$ (MPa) & $-$ & $-$ &
        0.00 $\pm$ 0.00 & 0.26 $\pm$ 0.11 &
        0.01 $\pm$ 0.04 & 0.21 $\pm$ 0.45 &
        0.00 $\pm$ 0.06 & 0.27 $\pm$ 0.45 \\
    $\theta_{4}$ (MPa) & $-$ & $-$ &
        0.00 $\pm$ 0.00 & 0.00 $\pm$ 0.00 &
        0.00 $\pm$ 0.00 & 0.01 $\pm$ 0.04 &
        0.01 $\pm$ 0.11 & 0.07 $\pm$ 0.19 \\
    $\theta_{5}$ (MPa) & $-$ & $-$ &
        0.00 $\pm$ 0.00 & 0.00 $\pm$ 0.00 &
        0.00 $\pm$ 0.00 & 0.00 $\pm$ 0.00 &
        0.00 $\pm$ 0.02 & 0.00 $\pm$ 0.01 \\
    $\theta_{6}$ (MPa) & $-$ & $-$ &
        0.00 $\pm$ 0.00 & 0.01 $\pm$ 0.03 &
        0.00 $\pm$ 0.00 & 0.02 $\pm$ 0.21 &
        0.00 $\pm$ 0.00 & 0.02 $\pm$ 0.07 \\
    $\theta_{7}$ (MPa) & $-$ & $-$ &
        0.00 $\pm$ 0.00 & 0.00 $\pm$ 0.00 &
        0.00 $\pm$ 0.00 & 0.01 $\pm$ 0.16 &
        0.00 $\pm$ 0.00 & 0.01 $\pm$ 0.06 \\
    $\theta_{8}$ (MPa) & $-$ & $-$ &
        0.00 $\pm$ 0.00 & 0.00 $\pm$ 0.00 &
        0.00 $\pm$ 0.00 & 0.00 $\pm$ 0.01 &
        0.00 $\pm$ 0.09 & 0.00 $\pm$ 0.01 \\
    $\theta_{9}$ (MPa) & $-$ & $-$ &
        0.00 $\pm$ 0.00 & 0.00 $\pm$ 0.00 &
        0.00 $\pm$ 0.00 & 0.00 $\pm$ 0.00 &
        0.00 $\pm$ 0.00 & 0.00 $\pm$ 0.00 \\
    $\theta_{10}$ (MPa) & $-$ & $-$ &
        0.00 $\pm$ 0.00 & 0.00 $\pm$ 0.00 &
        0.00 $\pm$ 0.00 & 0.01 $\pm$ 0.25 &
        0.00 $\pm$ 0.00 & 0.01 $\pm$ 0.04 \\
    $\theta_{11}$ (MPa) & $-$ & $-$ &
        0.00 $\pm$ 0.00 & 0.00 $\pm$ 0.05 &
        0.00 $\pm$ 0.00 & 0.00 $\pm$ 0.00 &
        0.00 $\pm$ 0.00 & 0.01 $\pm$ 0.02 \\
    $\theta_{12}$ (MPa) & $-$ & $-$ &
        0.00 $\pm$ 0.00 & 0.00 $\pm$ 0.00 &
        0.00 $\pm$ 0.00 & 0.00 $\pm$ 0.00 &
        0.00 $\pm$ 0.00 & 0.00 $\pm$ 0.01 \\
    $\theta_{13}$ (MPa) & $-$ & $-$ &
        0.00 $\pm$ 0.00 & 0.00 $\pm$ 0.00 &
        0.00 $\pm$ 0.00 & 0.00 $\pm$ 0.00 &
        0.00 $\pm$ 0.00 & 0.00 $\pm$ 0.00 \\
    $\theta_{14}$ (MPa) & $-$ & $-$ &
        0.00 $\pm$ 0.00 & 0.00 $\pm$ 0.00 &
        0.00 $\pm$ 0.00 & 0.00 $\pm$ 0.00 &
        0.00 $\pm$ 0.00 & 0.03 $\pm$ 0.97 \\
    $\theta_{15}$ (MPa) & $-$ & $-$ &
        0.00 $\pm$ 0.00 & 0.00 $\pm$ 0.00 &
        0.00 $\pm$ 0.01 & 0.00 $\pm$ 0.00 &
        0.00 $\pm$ 0.00 & 0.00 $\pm$ 0.00 \\
    $\theta_{16}$ (MPa) & $-$ & $-$ &
        0.00 $\pm$ 0.00 & 0.00 $\pm$ 0.00 &
        0.00 $\pm$ 0.13 & 0.00 $\pm$ 0.03 &
        $-$ & $-$ \\
    $\theta_{17}$ (MPa) & $-$ & $-$ &
        0.00 $\pm$ 0.00 & 0.00 $\pm$ 0.00 &
        0.00 $\pm$ 0.01 & 0.00 $\pm$ 0.01 &
        $-$ & $-$ \\
    $\theta_{18}$ (MPa) & $-$ & $-$ &
        1.35 $\pm$ 0.04 & 2.58 $\pm$ 5.87 &
        $-$ & $-$ &
        $-$ & $-$ \\
    $\theta_{19}$ (MPa) & 6.00 & 15.00 &
        4.85 $\pm$ 0.03 & 13.88 $\pm$ 6.41 &
        5.98 $\pm$ 0.08 & 17.06 $\pm$ 19.15 &
        6.09 $\pm$ 1.95 & 21.22 $\pm$ 219.96 \\

    $\mu$ (MPa) & $-$ & 5.40 & $-$ & $-$ & $-$ & $-$ & $-$ & $-$ \\
    $\eta$ & $-$ & 4.00 & $-$ & $-$ & $-$ & $-$ & $-$ & $-$ \\

    \midrule[1.0pt]
    \begin{tabular}{@{}l@{}}Free nodes\\sub-sampling\end{tabular}
        & \multicolumn{2}{c}{$-$}
        & \multicolumn{2}{c}{5\%}
        & \multicolumn{2}{c}{5\%}
        & \multicolumn{2}{c}{5\%} \\
    
    \bottomrule[1.5pt]
    \end{tabular}
    \end{adjustbox}
    \caption{True vs. predicted material parameters for the circular inclusion--Ogden pattern (\autoref{hetero_pattern:type5}) across different noise levels using the Hetero-EUCLID framework. Feature 18 (corresponding to $\theta_{18}$) contained divergent values for moderate noise displacement data, and features 16, 17 and 18 (refer \autoref{table_feature-list}) contained similar divergent values for high displacement noise data. Thus, for moderate and high noise cases, these features were suppressed in the Bayesian discovery process.}
    \label{table_Type5_10k_Ogden}
\end{table}

\subsubsection*{Pattern Type 6: Circular inclusion (Holzapfel-Gasser-Ogden material) in a NH2 matrix}
Next, we again consider a hyperelastic heterogeneous specimen having a circular inclusion embedded in a surrounding matrix material as illustrated in \autoref{hetero_pattern:type5}. The matrix follows the Neo-Hookean model (NH2) (\autoref{eq28}) and the circular inclusion is described by the Holzapfel-Gasser-Ogden (HGO) form \cite{holzapfel2000new} with one fiber family in $x-y$ plane at \(\alpha = 30^{\circ}\) to the $x$-axis. The domain is discretized using 10,617 wedge elements and the displacement data is generated under non-equi-biaxial loading conditions with final directional stretch ratios of \( \lambda_x = 1.6 \) and \( \lambda_y = 2.2 \).
 \begin{equation}
     W_{\text{HGO}}(\boldsymbol{C};\boldsymbol{\theta}) = \theta_{1}(\tilde{I}_{1}-3)  + \theta_{18}\left[\frac{(J^2 -1)}{2} -\ln{(J)}\right] + \underbrace{\frac{k_1}{2k_2}\left[e^{k_2(\tilde{I}_a - 1)^2} -1 \right]}_{\text{omitted from feature library}} 
     \label{eq_hgo1} 
 \end{equation}

\begin{figure}[H]
    \centering
    \includegraphics[width=\textwidth]{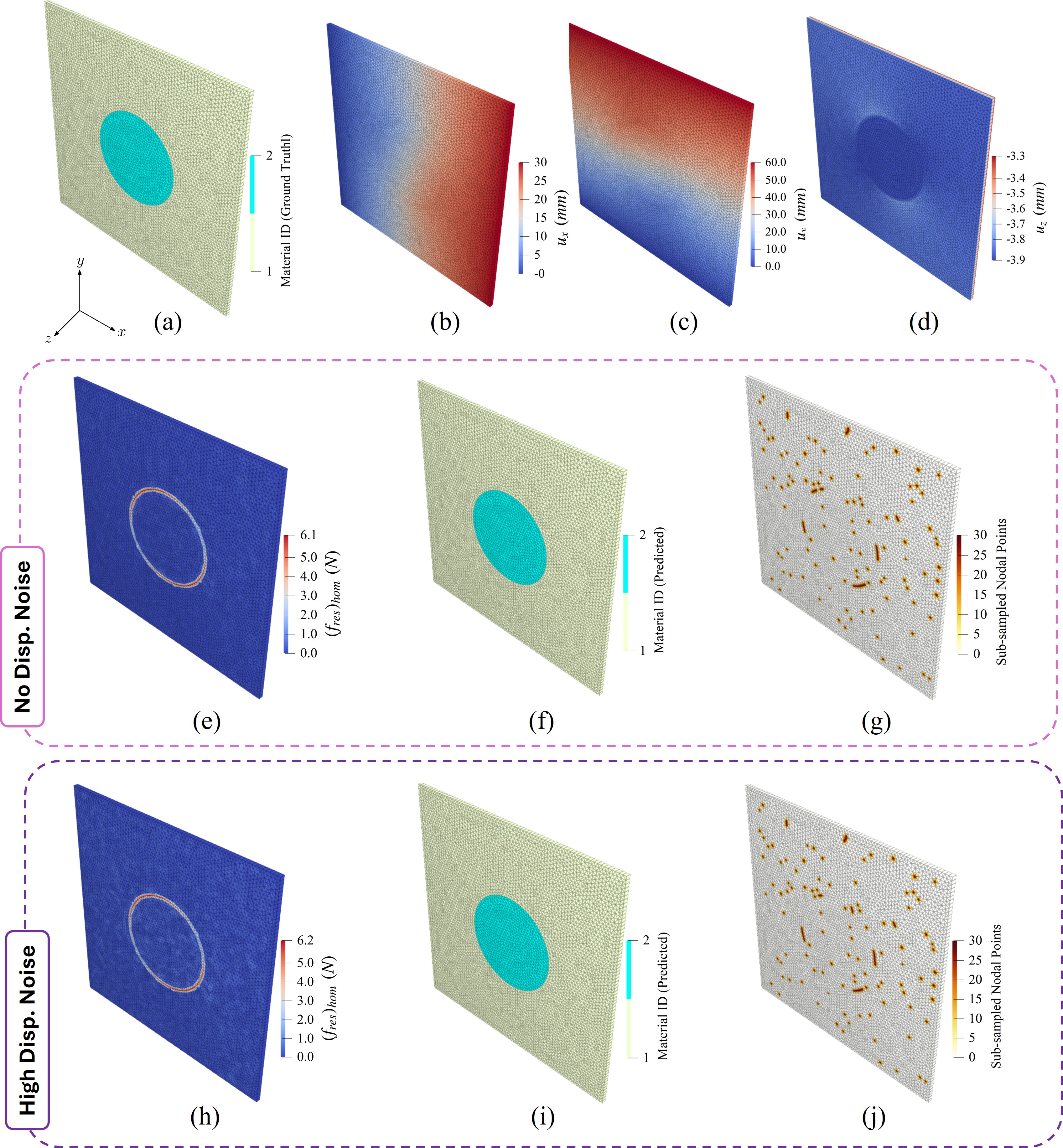}
    \caption{Model discovery for the circular inclusion with anisotropic fiber reinforcement pattern  (\autoref{hetero_pattern:type6}). The ground truth material segment IDs are shown in (a), and the displacement fields in the \(x\), \(y\), and \(z\) directions are provided in (b)--(d). For the \textit{zero-noise} case, (e) depicts the residual force norm distribution (\autoref{eq19}) from the homogenized model, (f) illustrates the material segments identified through the segmentation algorithm, and (g) highlights the sub-sampled nodes from the \(\mathcal{D}^{\text{free}}\) set used to construct the heterogenized model. The \textit{high-noise} case follows a similar structure in panels (h)--(j), showing how the residual force field, segmentation, and sub-sampling behave under high noise levels. In both scenarios, the final heterogenized model is solved using the Bayesian framework to recover the region-wise constitutive parameters as listed in \autoref{table_Type6_10k_HGO}.}
    \label{Type6_none_high_10k}
\end{figure}

 \autoref{eq_hgo1} refers to the HGO model where the term, \(\dfrac{k_1}{2k_2}[e^{(k_2(\tilde{I}_a - 1)^2)} -1 ]\) is intentionally omitted from the feature library \(Q\). All 19 terms of the feature library as listed in \autoref{table_feature-list} were employed for model discovery for this study. We investigated the Type 6 (HGO) specimen across different noise levels. \autoref{Type6_none_high_10k} presents the results of material segmentation for noise-free and high noise conditions, while Figure~S6 corresponds to the moderate noise case. The spatial distribution of the residual forces are shown in \autoref{Type6_none_high_10k}e and \ref{Type6_none_high_10k}h for noise-free and high noise conditions, respectively. The random seed-based island growth algorithm is applied to these residual force distributions. The predicted material segments
align well with the true segmentation (\autoref{Type6_none_high_10k}a), as illustrated in \autoref{Type6_none_high_10k}f and \ref{Type6_none_high_10k}i, respectively.\par

\begin{table}[H]
    \centering
    \begin{adjustbox}{width=\textwidth}
    \begin{tabular}{@{} l c c *{6}{>{\centering\arraybackslash}p{2.2cm}} @{}}
    \toprule[1.5pt]
    \multicolumn{9}{l}{\textbf{Pattern Type 6: Circular inclusion – HGO} $\vert$ Mesh elements (C3D6): \textbf{10,617} $\vert$ Activated feature library terms, $\boldsymbol{n_{\mathit{f}} = 19}$} \\
    \midrule[1.5pt]
    \multirow{2}{*}{\textbf{Case}} 
        & \multicolumn{2}{c}{Ground} 
        & \multicolumn{2}{c}{No Noise} 
        & \multicolumn{2}{c}{Moderate Noise} 
        & \multicolumn{2}{c}{High Noise} \\
        & \multicolumn{2}{c}{Truth} 
        & \multicolumn{2}{c}{($\sigma_u = 0$)} 
        & \multicolumn{2}{c}{($\sigma_u = 2 \times 10^{-3}$)} 
        & \multicolumn{2}{c}{($\sigma_u = 5 \times 10^{-3}$)} \\
    \midrule[1.0pt]
    
    \begin{tabular}{@{}l@{}}Constitutive\\model\end{tabular} 
        &\begin{tabular}{@{}l@{}}NH2\end{tabular}
        & \begin{tabular}{@{}l@{}}HGO\end{tabular}
        & \begin{tabular}{@{}l@{}}NH2\end{tabular}
        & \begin{tabular}{@{}l@{}}HGO\end{tabular}
        & \begin{tabular}{@{}l@{}}NH2\end{tabular}
        & \begin{tabular}{@{}l@{}}HGO\end{tabular}
        & \begin{tabular}{@{}l@{}}NH2\end{tabular}
        & \begin{tabular}{@{}l@{}}HGO\end{tabular} \\
    \midrule[1.0pt]
     Material ID & 1 & 2 & 1 & 2 & 1 & 2 & 1 & 2 \\
    \midrule[1.0pt]

    $\theta_{1}$ (MPa) & 5.40 & 1.80 & 5.40 $\pm$ 0.00 & 1.79 $\pm$ 0.01 & 5.33 $\pm$ 0.55 & 1.88 $\pm$ 0.28 & 5.38 $\pm$ 0.01 & 1.78 $\pm$ 0.10 \\
    $\theta_{2}$ (MPa) & $-$ & $-$ & 0.00 $\pm$ 0.00 & 0.00 $\pm$ 0.00 & 0.00 $\pm$ 0.00 & 0.00 $\pm$ 0.00 & 0.00 $\pm$ 0.00 & 0.00 $\pm$ 0.00 \\
    $\theta_{3}$ (MPa) & $-$ & $-$ & 0.00 $\pm$ 0.00 & 0.00 $\pm$ 0.00 & 0.02 $\pm$ 0.22 & 0.00 $\pm$ 0.00 & 0.00 $\pm$ 0.00 & 0.00 $\pm$ 0.00 \\
    $\theta_{4}$ (MPa) & $-$ & $-$ & 0.00 $\pm$ 0.00 & 0.00 $\pm$ 0.00 & 0.00 $\pm$ 0.00 & 0.00 $\pm$ 0.00 & 0.00 $\pm$ 0.00 & 0.00 $\pm$ 0.00 \\
    $\theta_{5}$ (MPa) & $-$ & $-$ & 0.00 $\pm$ 0.00 & 0.00 $\pm$ 0.00 & 0.00 $\pm$ 0.00 & 0.00 $\pm$ 0.00 & 0.00 $\pm$ 0.00 & 0.00 $\pm$ 0.00 \\
    $\theta_{6}$ (MPa) & $-$ & $-$ & 0.00 $\pm$ 0.00 & 0.00 $\pm$ 0.00 & 0.00 $\pm$ 0.00 & 0.00 $\pm$ 0.00 & 0.00 $\pm$ 0.00 & 0.00 $\pm$ 0.00 \\
    $\theta_{7}$ (MPa) & $-$ & $-$ & 0.00 $\pm$ 0.00 & 0.00 $\pm$ 0.00 & 0.00 $\pm$ 0.00 & 0.00 $\pm$ 0.00 & 0.00 $\pm$ 0.00 & 0.00 $\pm$ 0.00 \\
    $\theta_{8}$ (MPa) & $-$ & $-$ & 0.00 $\pm$ 0.00 & 0.00 $\pm$ 0.00 & 0.00 $\pm$ 0.00 & 0.00 $\pm$ 0.00 & 0.00 $\pm$ 0.00 & 0.00 $\pm$ 0.00 \\
    $\theta_{9}$ (MPa) & $-$ & $-$ & 0.00 $\pm$ 0.00 & 0.00 $\pm$ 0.00 & 0.00 $\pm$ 0.00 & 0.00 $\pm$ 0.00 & 0.00 $\pm$ 0.00 & 0.00 $\pm$ 0.00 \\
    $\theta_{10}$ (MPa) & $-$ & $-$ & 0.00 $\pm$ 0.00 & 0.00 $\pm$ 0.00 & 0.00 $\pm$ 0.00 & 0.00 $\pm$ 0.00 & 0.00 $\pm$ 0.00 & 0.00 $\pm$ 0.00 \\
    $\theta_{11}$ (MPa) & $-$ & $-$ & 0.00 $\pm$ 0.00 & 0.00 $\pm$ 0.00 & 0.00 $\pm$ 0.00 & 0.00 $\pm$ 0.00 & 0.00 $\pm$ 0.00 & 0.00 $\pm$ 0.00 \\
    $\theta_{12}$ (MPa) & $-$ & $-$ & 0.00 $\pm$ 0.00 & 0.00 $\pm$ 0.00 & 0.00 $\pm$ 0.00 & 0.00 $\pm$ 0.00 & 0.00 $\pm$ 0.00 & 0.00 $\pm$ 0.00 \\
    $\theta_{13}$ (MPa) & $-$ & $-$ & 0.00 $\pm$ 0.00 & 0.00 $\pm$ 0.00 & 0.00 $\pm$ 0.00 & 0.00 $\pm$ 0.00 & 0.00 $\pm$ 0.00 & 0.00 $\pm$ 0.00 \\
    $\theta_{14}$ (MPa) & $-$ & $-$ & 0.00 $\pm$ 0.00 & 0.00 $\pm$ 0.00 & 0.00 $\pm$ 0.00 & 0.00 $\pm$ 0.00 & 0.00 $\pm$ 0.00 & 0.00 $\pm$ 0.00 \\
    $\theta_{15}$ (MPa) & $-$ & $-$ & 0.00 $\pm$ 0.00 & 0.03 $\pm$ 0.08 & 0.00 $\pm$ 0.00 & 1.23 $\pm$ 0.38 & 0.00 $\pm$ 0.00 & 0.98 $\pm$ 0.46 \\
    $\theta_{16}$ (MPa) & $-$ & $-$ & 0.00 $\pm$ 0.00 & 0.02 $\pm$ 0.09 & 0.00 $\pm$ 0.01 & 0.05 $\pm$ 0.17 & 0.00 $\pm$ 0.01 & 0.13 $\pm$ 0.22 \\
    $\theta_{17}$ (MPa) & $-$ & $-$ & 0.00 $\pm$ 0.00 & 0.47 $\pm$ 0.07 & 0.00 $\pm$ 0.00 & 0.01 $\pm$ 0.08 & 0.00 $\pm$ 0.00 & 0.02 $\pm$ 0.06 \\
    $\theta_{18}$ (MPa) & $-$ & 6.00 & 0.00 $\pm$ 0.00 & 5.95 $\pm$ 0.13 & 6.17 $\pm$ 1.68 & 5.47 $\pm$ 2.09 & 11.31 $\pm$ 1.46 & 5.07 $\pm$ 1.40 \\
    $\theta_{19}$ (MPa) & 15.00 & $-$ & 15.00 $\pm$ 0.00 & 0.01 $\pm$ 0.10 & 9.83 $\pm$ 1.51 & 0.53 $\pm$ 1.29 & 5.47 $\pm$ 1.24 & 0.57 $\pm$ 1.08 \\
    $k_1$ & $-$ & 0.90 & $-$ & $-$ & $-$ & $-$ & $-$ & $-$ \\
    $k_2$ & $-$ & 0.80 & $-$ & $-$ & $-$ & $-$ & $-$ & $-$ \\
    $\alpha$ & $-$ & 30\textdegree & $-$ & $-$ & $-$ & $-$ & $-$ & $-$ \\

    \midrule[1.0pt]
    \begin{tabular}{@{}l@{}}Free nodes\\sub-sampling\end{tabular} & \multicolumn{2}{c}{$-$} & \multicolumn{2}{c}{2\%} & \multicolumn{2}{c}{2\%} & \multicolumn{2}{c}{2\%} \\
    
    \bottomrule[1.5pt]
    \end{tabular}
    \end{adjustbox}
    \caption{True vs. predicted material parameters for the circular inclusion with anisotropic fiber reinforcement (Holzapfel-Gasser-Ogden) pattern (\autoref{hetero_pattern:type6}) across different noise levels using the Hetero-EUCLID framework.}
    \label{table_Type6_10k_HGO}
\end{table}

\begin{figure}[H]
    \centering
    \begin{subfigure}[b]{0.242\textwidth}
        \includegraphics[width=0.95\textwidth]{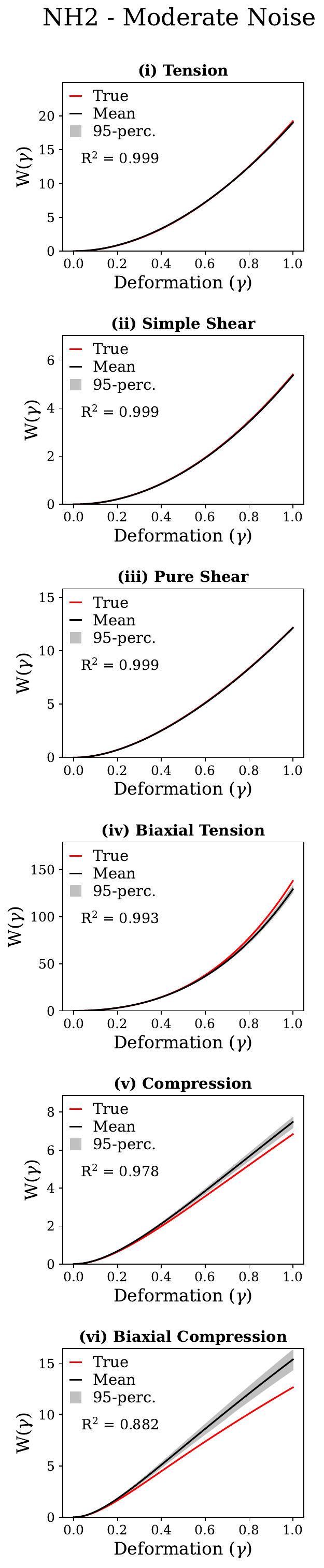}
        \caption{Material ID 1}
        \label{W_gamma_Type6_mod:ID1}
    \end{subfigure}
    \hfill
    \begin{subfigure}[b]{0.239\textwidth}
        \includegraphics[width=0.94\textwidth]{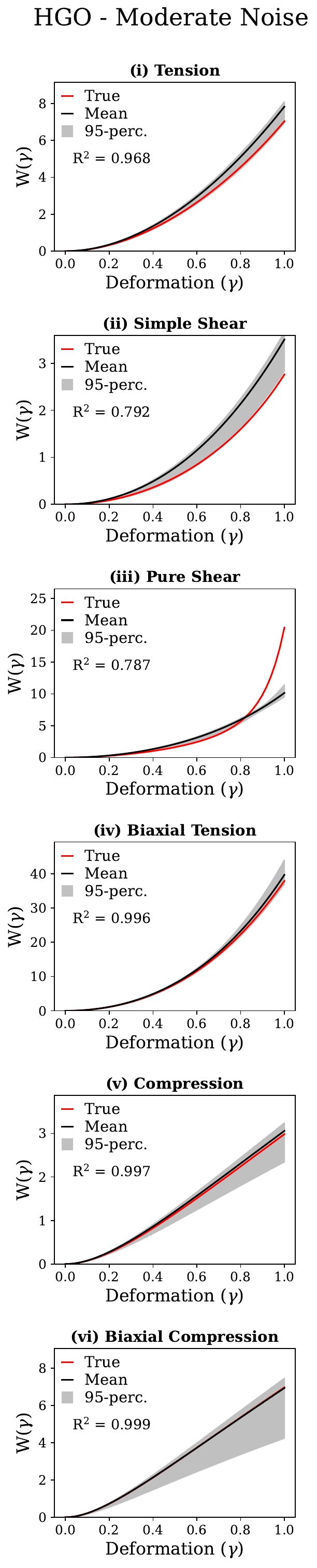}
        \caption{Material ID 2}
        \label{W_gamma_Type6_mod:ID2}
    \end{subfigure}
    \hfill
    \begin{subfigure}[b]{0.239\textwidth}
        \includegraphics[width=0.92\textwidth]{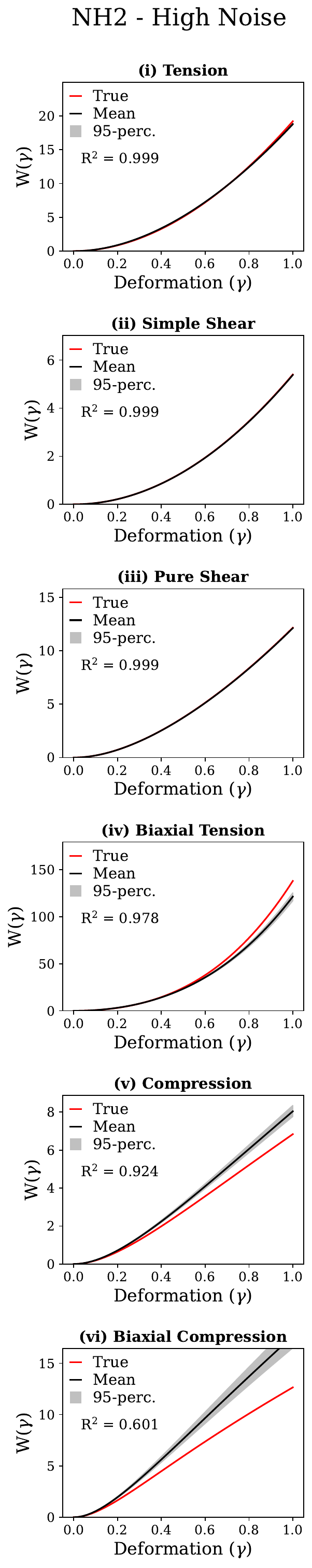}
        \caption{Material ID 1}
        \label{W_gamma_Type6_high:ID1}
    \end{subfigure}
    \hfill
    \begin{subfigure}[b]{0.237\textwidth}
        \includegraphics[width=0.91\textwidth]{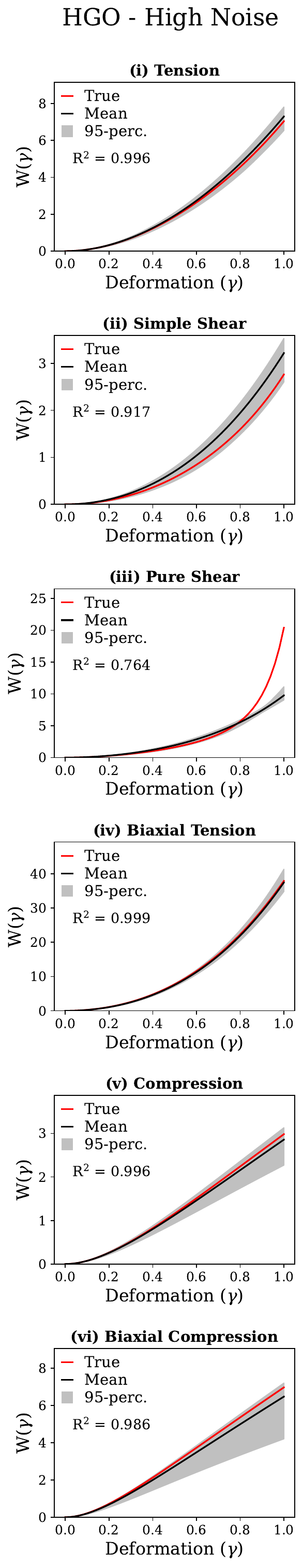}
        \caption{Material ID 2}
        \label{W_gamma_Type6_high:ID2}
    \end{subfigure}
    \caption{Strain energy density plots for the circular inclusion with anisotropic fibers pattern (\autoref{hetero_pattern:type6}). Panels (a) and (b) correspond to the identified material sub-domains under moderate noise, while (c) and (d) show the results for the high noise case as reported in \autoref{table_Type6_10k_HGO}. For each identified material segment ID (a)--(d), subplots (i)--(vi) show strain energy density plots \( W(\gamma) \) along various deformation paths (\autoref{eq31}), providing a comparison between the discovered and true models.}
    \label{W_gamma_Type6_mod_high}
\end{figure}

\autoref{table_Type6_10k_HGO} presents the recovered material parameters for the NH2--HGO configuration under noise-free, moderate \((\sigma_{u,\text{mod}} = 2\times10^{-3})\), and high noise \((\sigma_{u,\text{high}} = 5\times10^{-3})\) conditions. The material parameters \((\theta_{1}, \theta_{18})\) corresponding to the HGO model are accurately recovered with low standard deviations across all noise levels, indicating high confidence in the predictions. In addition, most parameters associated with the extended feature library are correctly identified as zero, reflecting the effectiveness of the sparsity-promotion of the Bayesian formulation. \autoref{W_gamma_Type6_mod_high} illustrates the comparative strain energy density plots for the Type~6 circular inclusion pattern under moderate and high noise conditions. \autoref{W_gamma_Type6_mod_high} shows a strong agreement between the strain energy densities computed using the predicted model parameters and those from the ground truth, with low standard deviations indicating a high level of statistical confidence across all deformation paths. This consistency is further supported by the high \(R^2\) scores observed across the strain energy density subplots. This implies that the parameters associated with the anisotropy features-- $\theta_{15}$, $\theta_{16}$ and $\theta_{17}$ adequately capture the anisotropy behavior of the excluded HGO term (\autoref{eq_hgo1}). These results demonstrate that the proposed Hetero-EUCLID framework remains robust across different noise conditions under feature exclusion, even for materials described by the anisotropic Holzapfel-Gasser-Ogden model. \par

\subsection{A comment on detecting noise}
\label{subsec:noise}
\begin{table}[H]
        \centering
        \begin{adjustbox}{max width=\textwidth}
        \small
        \begin{tabular}{@{}l c l c c c c c c@{}}
            \toprule[1.25pt]
            \multirow{3}{*}{\begin{tabular}{@{}l@{}}\textbf{Specimen}\\ \textbf{Type}\end{tabular}} & $R_{\mathrm{max}}$ [N] & \multirow{3}{*}{\begin{tabular}{@{}l@{}}\textbf{Noise}\\ \textbf{Level}\end{tabular}} & $\lambda$ & $\mu_{f_{\mathrm{res}}}$ [N] & $\sigma_{f_{\mathrm{res}}} [N] $ & \multirow{3}{*}{$\dfrac{\mu_{f_{\mathrm{res}}}}{\mathit{\sigma_{f_{\text{res}}}}}$} & \multirow{3}{*}{$\dfrac{\mathit{\sigma_{f_{\text{res}}}}}{R_{\mathrm{max}}}$}  & \multirow{3}{*}{\begin{tabular}{@{}c@{}}\textbf{CPU}\\\textbf{time}\\ (min)\end{tabular}} \\
             & \textbf{(Max. boundary} &  & \textbf{(Tunable } & \textbf{(Mean)} & \textbf{(Standard} &  &  & \\ 
             & \textbf{force)} &  & \textbf{scalar)} &  & \textbf{deviation)} &  &  & \\
            
            \midrule[1.0pt]
            \multirow{3}{*}{\begin{tabular}{@{}l@{}}Type 0\\(Homogeneous)\end{tabular}} 
            & \multirow{3}{*}{904.56} & Zero     & --   & $2.38 \times 10^{-5}$ & $1.48\times 10^{-5}$ & 1.608 & $1.64 \times 10^{-8}$ & $-$\\
            & & Moderate & --   & 0.122 & 0.074 & 1.664 & $8.13 \times 10^{-5}$ & $-$\\
            & & High     & --   & 0.247 & 0.135 & 1.826 & $1.49 \times 10^{-4}$ & $-$\\
            
            \midrule[1.0pt]
            \multirow{3}{*}{\begin{tabular}{@{}l@{}}Type 1\\(\autoref{hetero_pattern:type1})\end{tabular}} 
            & \multirow{3}{*}{344.44} & Zero     & 2.00 & 0.042 & 0.252 & 0.168 & $7.31 \times 10^{-4}$ & 1.4\\
            & & Moderate & 2.00 & 0.085 & 0.247 & 0.343 & $7.17 \times 10^{-4}$ & 2.0\\
            & & High     & 1.50 & 0.134 & 0.243 & 0.552 & $7.04 \times 10^{-4}$ & 2.5\\
            
            \midrule[1.0pt]
            \multirow{3}{*}{\begin{tabular}{@{}l@{}}Type 2\\(\autoref{hetero_pattern:type2})\end{tabular}} 
            & \multirow{3}{*}{879.28} & Zero     & 1.50 & 0.065 & 0.380 & 0.172 & $4.32 \times 10^{-4}$ & 11.3\\
            & & Moderate & 1.50 & 0.193 & 0.371 & 0.520 & $4.22 \times 10^{-4}$ & 4.6\\
            & & High     & 2.12 & 0.328 & 0.377 & 0.871 & $4.28 \times 10^{-4}$ & 4.1\\
            
            \midrule[1.0pt]
            \multirow{3}{*}{\begin{tabular}{@{}l@{}}Type 3\\(\autoref{hetero_pattern:type3})\end{tabular}} 
            & \multirow{3}{*}{969.58} & Zero     & 1.70 & 0.185 & 0.906 & 0.204 & $9.34 \times 10^{-4}$ & 61.7\\
            & & Moderate & 1.70 & 0.338 & 0.902 & 0.375 & $9.30 \times 10^{-4}$ & 77.6\\
            & & High     & 1.80 & 0.507 & 0.888 & 0.571 & $9.16 \times 10^{-4}$ & 113.2\\
            
            \midrule[1.0pt]
            \multirow{3}{*}{\begin{tabular}{@{}l@{}}Type 4\\(\autoref{hetero_pattern:type4})\end{tabular}} 
            & \multirow{3}{*}{562.26} & Zero     & 1.50 & 0.095 & 0.472 & 0.202 & $8.40 \times 10^{-4}$ & 108.5\\
            & & Moderate & 1.50 & 0.186 & 0.469 & 0.396 & $8.34 \times 10^{-4}$ & 113.3\\
            & & High     & 1.50 & 0.272 & 0.464 & 0.587 & $8.25 \times 10^{-4}$ & 241.1\\
            
            \midrule[1.0pt]
            \multirow{3}{*}{\begin{tabular}{@{}l@{}}Type 5\\(\autoref{hetero_pattern:type5})\end{tabular}} 
            & \multirow{3}{*}{330.54} & Zero     & 1.50 & 0.024 & 0.173 & 0.138 & $5.24 \times 10^{-4}$ & 20.7\\
            & & Moderate & 1.50 & 0.067 & 0.171 & 0.390 & $5.16 \times 10^{-4}$ & 32.0\\
            & & High     & 1.80 & 0.114 & 0.171 & 0.666 & $5.16 \times 10^{-4}$ & 39.1\\
            
            \midrule[1.0pt]
            \multirow{3}{*}{\begin{tabular}{@{}l@{}}Type 6\\(\autoref{hetero_pattern:type6})\end{tabular}} 
            & \multirow{3}{*}{795.96} & Zero     & 2.00 & 0.095 & 0.642 & 0.149 & $8.07 \times 10^{-4}$ & 3.6\\
            & & Moderate & 2.00 & 0.195 & 0.634 & 0.307 & $7.97 \times 10^{-4}$ & 10.1\\
            & & High     & 2.00 & 0.296 & 0.626 & 0.473 & $7.87 \times 10^{-4}$ & 5.0\\
            \bottomrule[1.25pt]
        \end{tabular}
        \end{adjustbox}
        \caption{Comparison of statistical measures (mean, standard deviation, and ratios) and computational time for each heterogeneous specimen type under different noise levels.}
        \label{table_res_noise}
    \end{table}

    \autoref{table_res_noise} provides statistical details of the 2D nodal residual force $f^a_{\text{res}}$ corresponding to each type of heterogeneity and noise-level considered in the study. Additionally, we provide statistical details of residual forces in a homogeneous specimen with the same thin square plate domain geometry in the same table for comparison. It is evident from the results that the dimensionless ratio: \(
    \left(\mu_{f_{\mathrm{res}}} / \mathit{\sigma_{f_{\text{res}}}}\right)\) is a measure of noise in the displacement data, with a higher ratio indicating a higher displacement noise or a lack of heterogeneity in the domain. Another important observation is that alongside the previous ratio, the dimensionless ratio \(\left(\mathit{\sigma_{f_{\text{res}}}}/R_{\mathrm{max}}\right)\) provides an indication of the existence of heterogeneity in the domain. If \(\left(\mathit{\sigma_{f_{\text{res}}}}/R_{\mathrm{max}}\right) \leq 10^{-5}\) and \(\left(\mu_{f_{\mathrm{res}}}/ \mathit{\sigma_{f_{\text{res}}}}\right) \geq 1\), it is reasonable to consider the domain to be nominally homogeneous. However, a further increase in displacement noise can be expected to increase both ratios, making it challenging to identify heterogeneities (or the lack thereof) in the domain. The statistical trends observed from $f^a_{\text{res}}$ are thus useful to characterize displacement noise and the existence of heterogeneities from actual experimental data.

\section{Conclusion and outlook}
\label{conclusion}
    In this study, we presented and validated Hetero-EUCLID, a physics-constrained, interpretable framework for discovering the segmentation and region-wise constitutive models in heterogeneous hyperelastic materials under 3D plane stress conditions. We demonstrated that Hetero-EUCLID accurately identifies both the material interface boundaries and the underlying constitutive laws for each region for different heterogeneity patterns, including cross-inclusion, split-domains, non-circular inclusions, and Cahn–Hilliard configurations. The Bayesian framework provided sparsity and a parsimonious model parameter under quantified uncertainty, enhancing confidence in predictions. Its performance and efficacy have been validated through empirical checks across various noise levels and non-native mesh discretizations, confirming the robustness and interpretability of the proposed approach. The computations were performed on a machine with 12\textsuperscript{th} Gen Intel\textsuperscript{$\circledR$} Core\textsuperscript{TM} i9-12900K processor having clock frequencies of 3.2 GHz and 32 GB RAM. We employed 3 cores in parallel, each corresponding to a parallel Markov chain. The average compute time was seen to be around 48 mins, with lower noise data requiring significantly lower compute time. The computation time can be further improved by using multiple shorter parallel chains.\par
    
    As part of future work, we aim to extend Hetero-EUCLID to accommodate more complex history-dependent constitutive behaviors, such as viscoelasticity, to model a broader range of soft tissues and composites, while accounting for spatial correlations in residual forces (see \autoref{subsec:corr_noise}). As mentioned in \citet{shi2025deep}, the paradigm of first segmenting the material and then identifying the constitutive parameters of each segment is highly dependent on the ability to have clean inter-material segments. However, clean inter-material segments are not available in situations involving functionally graded materials, and most practical experiments where the material segments have smooth (not sudden) transitions between properties. Furthermore, in the current study, we demonstrate that a sufficiently high displacement noise can undermine the accuracy of material segmentation, thereby further reducing the accuracy of predicting each material's properties. We therefore intend to explore frameworks that provide a position-dependent field of the material properties, as opposed to segment-specific properties. An experimental validation using DIC/Digital Volume Correlation (DVC) data will also provide further insights into the merits, practical limitations, and necessary improvements in the proposed framework.\par

\section*{Declaration of competing interest}
    The authors declare that they have no known competing financial interests or personal relationships that could have appeared to influence the work reported in this paper.  

\section*{Code and data availability}
    The codes and data generated during the current study are available from the authors upon reasonable request.

\section*{Acknowledgments}
    AJ would like to acknowledge funding by Anusandhan National Research Foundation (ANRF) through Grant No. ANRF/ECRG/2024/005303/ENS.

\appendix

\section{Overview of Bayesian-EUCLID framework}
\label{BayesianEUCLID}

\subsection{Parsimony-driven Model Prior}
    We adopt a probabilistic framework that simultaneously performs model selection and quantifies uncertainty to introduce sparsity and maintain interpretability in the discovered model. Instead of relying on conventional deterministic sparse regression, we use a hierarchical Bayesian formulation with spike--slab priors, initially introduced by \citet{nayek2021spike} in a supervised setting, and later adapted to an unsupervised formulation by \citet{joshi2022bayesian}. Spike--slab priors are characterized by a sharp peak (``\textit{spike}'') at zero and broad, shallow tails (``\textit{slab}''). A random variable may be drawn from either the spike or slab distribution. Sampling from the spike yields values at zero, reflecting our bias toward parsimonious (sparse) models, whereas sampling from the slab allows non-zero values with a near-uniform spread.\par
    
    For the model parameter vector $\boldsymbol{\theta}$, a binary indicator vector $\mathbf{z} \in [0,1]^{\mathit{n_f}}$ is introduced, along with random variables $\nu_s \ge 0$ and $\sigma^2 \ge 0$. The conditional prior for each component is
    \begin{equation}
        p(\theta_i \;| \; z_i, \nu_s, \sigma^2) =
        \begin{cases}
        p_{\text{spike}}(\theta_i) = \delta(\theta_i) \quad &\text{if} \quad z_i = 0, \\
        p_{\text{slab}}(\theta_i \;| \; \nu_s, \sigma^2) = \mathcal{N}_+(0, \nu_s \sigma^2) \quad &\text{if} \quad z_i = 1,
        \end{cases}
        \qquad \forall \quad i = 1,\dots,\mathit{n_f}
        \label{BE-eq1}
    \end{equation}
    where $\mathcal{N}_{+}$ denotes a non-negative truncated normal distribution with zero mean and variance \( \nu_s\sigma^2 \) (slab). The truncated normal ($\mathcal{N}_{+}$) is obtained from the standard normal distribution by restricting samples to non-negative values \( (\theta_i \geq 0) \), a requirement that ensures physical admissibility. When \( z_i = 0 \), the \( i^{\text{th}} \) feature is considered inactive, meaning \( \theta_i = 0 \), and its probability density reduces to a Dirac delta function (spike) centered at zero. Conversely, if \( z_i = 1 \), the feature is active and \( \theta_i \) is drawn from a truncated normal distribution \( \mathcal{N}^+ \) . Assuming that the elements of \( \boldsymbol{\theta} \) are independent and identically distributed (i.i.d.), the joint prior for \( \boldsymbol{\theta} \) can be expressed as
    \begin{equation}
        p(\boldsymbol{\theta} \;|\: \mathbf{z}, \nu_s, \sigma^2) = p_{\text{slab}}(\boldsymbol{\theta}_r \;| \; \nu_s, \sigma^2) \prod_{i:z_i=0} p_{\text{spike}}(\theta_i) \quad \text{with} \quad p_{\text{slab}}(\boldsymbol{\theta}_r \;| \; \nu_s, \sigma^2) = 
        \mathcal{N}_+(\boldsymbol{0}, \sigma^2 \nu_s \boldsymbol{I_r}) 
        \label{BE-eq2}
    \end{equation}
    where, \( \boldsymbol{\theta}_r \in \mathbb{R}^r_{+} \) denotes the reduced vector having components of \( \boldsymbol{\theta} \) corresponds to only the slab, or active features and \( \boldsymbol{I}_r \) is the \( r \times r \) identity matrix. Sampling from the multivariate truncated normal distribution is performed using the algorithm proposed by \citet{botev2017normal}. The binary variables $z_i$ are drawn from a Bernoulli distribution with activation probability $p_0$ as follows
    \begin{equation}
        z_i \: | \: \sim \mathrm{Bern}(p_0), \quad \text{where} \quad p_0 \in [0,1]  
        \label{BE-eq3}
    \end{equation}
    
    The hyper-priors for the random variables (\(\nu_s, \sigma^2,  p_0\)) are modeled as  
    \begin{equation}
        \nu_s \sim \mathcal{IG}(a_\nu, b_\nu), \quad
        \sigma^2 \sim \mathcal{IG}(a_\sigma, b_\sigma), \quad
        p_0 \sim \mathrm{Beta}(a_p, b_p)
        \label{BE-eq4}
    \end{equation}
    where \(\mathcal{IG}\) and Beta represent the Inverse gamma and Beta distribution, respectively, and \((a, b) \) denotes the set of hyperparameters for the random variables \(\nu_s, \sigma^2,\) and \(p_0\).
    
    \subsection{Model Identification through Posterior Estimation}
    The Bayesian learning task now reduces to estimating the joint posterior probability distribution of the random variables $\{\boldsymbol{\theta}, \mathbf{z}, p_0, \nu_s, \sigma^2\}$ conditioned on the observed data  \( \boldsymbol{A} \) and \( \boldsymbol{b} \) and the weak form of momentum balance. Applying Bayes’ theorem, the posterior can be expressed as: 
    \begin{equation}
        p(\boldsymbol{\theta}, \boldsymbol{z}, p_0, \nu_s, \sigma^2 \: | \: \boldsymbol{A}, \boldsymbol{b}) \propto \underbrace{p(\boldsymbol{b} \: | \: \boldsymbol{\theta}, \boldsymbol{z}, p_0, \nu_s, \sigma^2,  \boldsymbol{A})}_{\text{physics-constrained likelihood}} \quad \underbrace{p(\boldsymbol{\theta}, \boldsymbol{z}, p_0, \nu_s, \sigma^2)}_{\text{spike-slab model prior}}
        \label{BE-eq5}
    \end{equation}
    
    It is important to note that the direct analytical sampling from \autoref{BE-eq5} is challenging due to the presence of spike-slab priors; therefore, Markov Chain Monte Carlo (MCMC) methods are deployed. In this work, we adopt a Gibbs sampling strategy~\cite{casella1992explaining} for posterior distribution estimation. The Gibbs sampler requires the conditional posterior distributions of all random variables. Closed-form expressions for these conditional posteriors were derived by \citet{nayek2021spike}, and a concise summary of these results is provided in \autoref{cond-post-dist}:
    
    \begin{table}[H]
        \centering
        \renewcommand{\arraystretch}{1.5}
        \setlength{\tabcolsep}{6pt}
        \begin{tabularx}{\textwidth}{>{\raggedright\arraybackslash}m{4cm} >{\raggedright\arraybackslash}m{10cm}}
            \hline
            \textbf{Parameter/Variable} & \textbf{Conditional posterior distribution} \\
            \hline
            \multirow{2}{*}{$\boldsymbol{\theta_r}$} &
            $\boldsymbol{\theta}_r \mid z, p_0, \nu_s, \sigma^2, \boldsymbol{A}, \boldsymbol{b} 
            \sim \mathcal{N}_+(\boldsymbol{\mu}, \sigma^2 \boldsymbol{\mathit{\Sigma}})$ \\ &
            $\boldsymbol{\mathit{\Sigma}} = \left(\boldsymbol{A}_r^{T} \boldsymbol{A}_r + \nu_s^{-1} \boldsymbol{I}_r \right)^{-1},\quad
            \boldsymbol{\mu} = \boldsymbol{\mathit{\Sigma}}\, \boldsymbol{A}_r^{T} \boldsymbol{b}$ \\
            \hline
            $\sigma^2$ &
            $\sigma^2 \mid \boldsymbol{\theta}, z, p_0, \nu_s, \boldsymbol{A}, \boldsymbol{b} 
            \sim \mathcal{IG} \left( a_\sigma + \frac{N}{2},\; b_\sigma + \frac{1}{2} \left(\boldsymbol{b}^T \boldsymbol{b} - \boldsymbol{\mu}^T \mathit{\Sigma}^{-1} \boldsymbol{\mu} \right) \right)$ \\
            \hline
            $\nu_s$ &
            $\nu_s \mid \boldsymbol{\theta}, z, p_0, \sigma^2, \boldsymbol{A}, \boldsymbol{b} 
            \sim \mathcal{IG} \left( a_{\nu} + \frac{s_z}{2},\; b_{\nu} + \frac{\boldsymbol{\theta}_r^T \boldsymbol{\theta}_r}{2\sigma^2} \right)$ \\
            \hline
            $p_0$ &
            $p_0 \mid \boldsymbol{\theta}, z, \nu_s, \sigma^2, \boldsymbol{A}, \boldsymbol{b} 
            \sim \mathrm{Beta} \left( a_p + s_z,\; b_p + n_f - s_z \right)$ \\
            \hline
            \multirow{2}{*}{$z_i$} &
            $z_i \mid \boldsymbol{\theta}, \boldsymbol{z}_{-i}, p_0, \nu_s, \sigma^2, \boldsymbol{A}, \boldsymbol{b} 
            \sim \mathrm{Bern}(\xi_i)$ \\ &
            $\xi_i = p_0 \left[ p_0 + \dfrac{p\left( \boldsymbol{b} \mid z_i = 0, \boldsymbol{z}_{-i}, \nu_s, \boldsymbol{A} \right)}
            {p\left( \boldsymbol{b} \mid z_i = 1, \boldsymbol{z}_{-i}, \nu_s, \boldsymbol{A} \right)} (1 - p_0) \right]^{-1}$ \\
            \hline
        \end{tabularx}
        \caption{Full conditional posterior distributions for model parameters.}
        \label{cond-post-dist}
    \end{table}
    
    The matrix \( \boldsymbol{A}_r \in \mathbb{R}^{N \times s_z} \) is constructed by concatenating only those columns \( i = 1, \dots, n_f \) of \( \boldsymbol{A} \) for which \( z_i = 1 \); hence, the distribution applies exclusively and jointly to the active features \( \boldsymbol{\theta}_r \). Here, \( \boldsymbol{I}_r \) denotes the \( s_z \times s_z \) identity matrix, and \( s_z = \sum_{i=1}^{n_f} z_i \) is the total number of active features. For any \( z_i = 0 \), the associated \( \theta_i \) is set to zero. In our framework following \citet{joshi2022bayesian}, however, physical admissibility requires \( \boldsymbol{\theta} \ge 0 \), which modifies the conditional posterior on \(\boldsymbol{\theta}\) from an unconstrained multivariate normal distribution to a constrained (non-negative) multivariate normal \( \mathcal{N}^+ \), while leaving the other conditional distributions unchanged as formulated by \citet{nayek2021spike}.\par
    
    At each iteration of the Gibbs sampler, the components of 
    \(\boldsymbol{z}\) are updated in a random sequence. 
    The notation \(\boldsymbol{z}_{-i}\) refers to the vector \(\boldsymbol{z}\) 
    with its \(i\)-th component \(z_i\) removed. 
    The marginal likelihood \( p(\boldsymbol{b} \,|\, \boldsymbol{z}, \nu_s, \boldsymbol{A}) \) is then obtained by integrating out \(\boldsymbol{\theta}\) and \(\sigma^2\) from the likelihood in \autoref{eq20}, and is expressed as
    
    \begin{equation}
        \begin{aligned}
            p(\boldsymbol{b} \mid \boldsymbol{z}, \nu_s, \boldsymbol{A}) &=
            \frac{\Gamma\left(a_\sigma + 0.5N\right) \, (b_\sigma)^{a_\sigma}}
                 {(2\pi)^{N/2} (\nu_s)^{s_z/2} \, \Gamma(a_\sigma)}
                 \left[ \det\left( \left( \boldsymbol{A}_r^T \boldsymbol{A}_r + \nu_s^{-1} \boldsymbol{I}_r \right)^{-1} \right) \right]^{1/2} \\
            &\quad \times \left[ b_\sigma + 0.5 \, \boldsymbol{b}^T
                 \left( \boldsymbol{I}_N - \boldsymbol{A}_r
                 \left( \boldsymbol{A}_r^T \boldsymbol{A}_r + \nu_s^{-1} \boldsymbol{I}_r \right)^{-1}
                 \boldsymbol{A}_r^T \right) \boldsymbol{b} \right]^{-(a_\sigma + 0.5N)}.
        \end{aligned}
    \end{equation}
    
    where \(\Gamma(\cdot)\) is the Gamma function, \(N\) is the number of 
    observations, and \(\boldsymbol{A}_r\) corresponds to the active feature 
    sub-matrix defined by the current state of \(\boldsymbol{z}\).
    
    In the Gibbs sampling process, each variable in the set  
    \(\{\boldsymbol{\theta}, \boldsymbol{z}, p_0, \nu_s, \sigma^2\}\)  
    is iteratively updated by drawing from its corresponding conditional posterior distribution, as defined in \autoref{cond-post-dist}. This sequential updating generates a Markov chain of length \(N_G\), which serves as an empirical approximation of the joint posterior \(
    p(\boldsymbol{\theta}, \boldsymbol{z}, p_0, \nu_s, \sigma^2 \mid \boldsymbol{A}, \boldsymbol{b}) \)
    \begin{equation}
        \boldsymbol{\theta}^{(0)} \rightarrow \sigma^{2(0)} \rightarrow \nu_s^{(0)} \rightarrow p_0^{(0)} \rightarrow \boldsymbol{z}^{(0)}
        \rightarrow \ldots \rightarrow
        \boldsymbol{\theta}^{(N_G)} \rightarrow \sigma^{2(N_G)} \rightarrow \nu_s^{(N_G)} \rightarrow p_0^{(N_G)} \rightarrow \boldsymbol{z}^{(N_G)}
    \end{equation}
    To ensure the chain reflects the target distribution, the initial \(N_{\text{burn}}\) samples are discarded as burn-in, thereby mitigating the influence of potentially poor starting states. Additionally, \(N_{\text{chains}}\) independent Markov chains are generated, each initialized with a different randomly chosen starting point. After burn-in removal, these independent chains are concatenated to produce the final sample set used for posterior inference.

\section{Moran's \textit{I} test for assessing spatial correlations in residual force}
\label{MoransI_study}
We conduct a series of spatial correlation studies separately on the \(x\), \(y\), and \(z\) degrees of freedom of the residual force vector \(\boldsymbol{f}_{\mathrm{res,}\boldsymbol{\theta}} = \boldsymbol{A\theta - b} \), with $\boldsymbol{A}$, $\boldsymbol{\theta}$ and $\boldsymbol{b}$ as defined in \autoref{eq17} of the main paper. Further, \autoref{fig:ResForce_xyz} of the main paper illustrates the process to obtain  $\left(\boldsymbol{f}_{\mathrm{res,}\boldsymbol{\theta}}\right)_x$, $\left(\boldsymbol{f}_{\mathrm{res,}\boldsymbol{\theta}}\right)_y$ and $\left(\boldsymbol{f}_{\mathrm{res,}\boldsymbol{\theta}}\right)_z$, which are the $x$, $y$ and $z$ components of $\boldsymbol{f}_{\mathrm{res,}\boldsymbol{\theta}}$ vector. To examine spatial correlations in different components of the residual force vector, we make use of Moran’s index (\(I\)). \par

Given any residual vector $\boldsymbol{\phi}$, the Moran’s index (\(I\)) \citep{moran1950notes} is defined as follows:
\begin{equation}
    I = \frac{N}{W} \left[
    \frac{ \sum_{i=1}^{N} \sum_{j=1}^{N} (\phi_i - \mu) W_{ij} (\phi_j - \mu)}{\sum_{i=1}^{N} (\phi_i - \mu)^2
    }\right]
    \label{eq:MoransI}
\end{equation}
where \(\phi_i\) denotes the \(i^{\text{th}}\) component of the residual vector \(\boldsymbol{\phi}\),  \(  \mu = (1/N) \sum_{i=1}^{N} \phi_i\) and \( \boldsymbol{W}\) is a weight matrix indicating connectivity between nodes, i.e., \(W_{ij}\) defines the connection between the \(i^{\text{th}}\) and \(j^{\text{th}}\) component of the residual vector and \(W_{ii}=0\) (no summation), i.e, the weight matrix has all diagonal entries set to zero.\par

In this study, $\boldsymbol{\phi} = \left(\boldsymbol{f}_{\mathrm{res,}\boldsymbol{\theta}}\right)_x$, $\left(\boldsymbol{f}_{\mathrm{res,}\boldsymbol{\theta}}\right)_y$ or $\left(\boldsymbol{f}_{\mathrm{res,}\boldsymbol{\theta}}\right)_z$ (various components of the residual force vector), and $N$ would be equal to the number of nodes in the finite element mesh. \(\boldsymbol{W}\) is constructed using the finite element connectivity matrix by setting $W_{ij} =1$, only if the $i^{\text{th}}$ and $j^{\text{th}}$ nodes are connected, and $W_{ij} = 0$ otherwise. For a residual vector ($\boldsymbol{\phi}$) that is sampled from an independent and identical distribution (i.i.d.), there would be no correlations between different components of the residual vector. For such cases, Moran's index \(I\) would be close to zero. Conversely, a Moran's index close to 0 indicates the lack of spatial correlations in the residual vector, and a higher magnitude of the Moran's index indicates stronger spatial correlations between different components of the residual vector. \par

\subsection{Test study 1 - Evaluating \textit{I} for residual force vector sampled from an i.i.d.}
\label{Morans_test_study_1}
\begin{figure}[H]
    \centering
    \includegraphics[width=\textwidth]{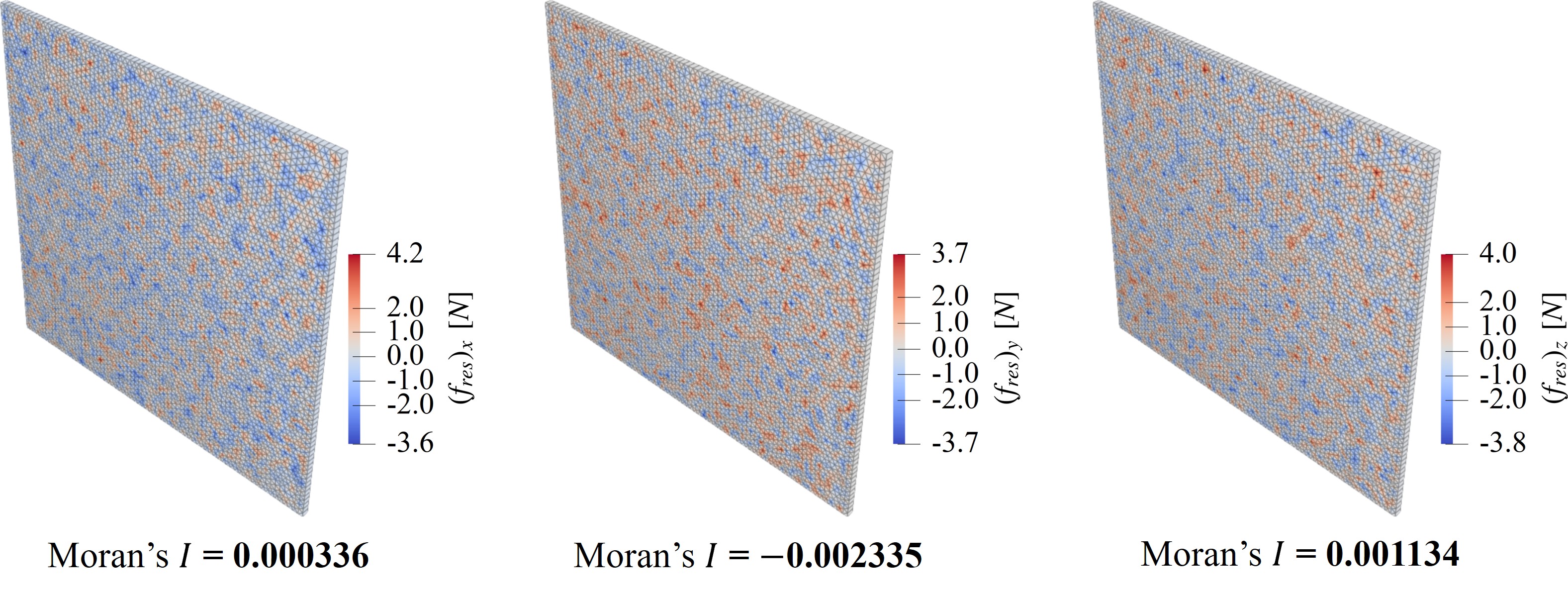}
    \caption{Residual force samples for the \(x\), \(y\), and \(z\) degrees of freedom sampled from an i.i.d. \( \mathcal{N}(\boldsymbol{0}, \boldsymbol{I}) \) distribution for Moran’s index (\textit{I}) evaluation, illustrating the expected near-zero Moran’s index for uncorrelated data.}
    \label{Type0_MoransI_uncorrelated}
\end{figure}

As the first benchmark study for Moran's index, a synthetic residual force vector is sampled from an independent and identically distributed (i.i.d.) normal distribution (\autoref{eq:Morans_resf_uncorrsampling}).
\begin{equation}
    \boldsymbol{f}_{\mathrm{res,}\boldsymbol{\theta}} = \boldsymbol{A \theta - b} \sim \mathcal{N}(\boldsymbol{0}, \boldsymbol{I}) 
    \label{eq:Morans_resf_uncorrsampling}
\end{equation}
Then, $I$ is evaluated for the residual vectors $\left(\boldsymbol{f}_{\mathrm{res,}\boldsymbol{\theta}}\right)_x$, $\left(\boldsymbol{f}_{\mathrm{res,}\boldsymbol{\theta}}\right)_y$, and $\left(\boldsymbol{f}_{\mathrm{res,}\boldsymbol{\theta}}\right)_z$ using \autoref{eq:MoransI}. As shown in  \autoref{Type0_MoransI_uncorrelated}, the resulting Moran’s index values remain close to zero for all components of the residual force vector. This is consistent with sampling the residual force vector from an i.i.d.

\subsection{Test study 2 - Evaluating \textit{I} for residual force vector sampled from a correlated normal distribution.}
\label{Morans_test_study_2}
As the next benchmark study for Moran's index, we generate a residual force vector from a spatially correlated zero-mean Gaussian distribution:
\begin{equation}
  \boldsymbol{f}_{\mathrm{res,}\boldsymbol{\theta}} = \boldsymbol{A \theta - b} \sim \mathcal{N}(\boldsymbol{0},\boldsymbol{\Sigma}) 
    \label{eq:Morans_resf_corrsampling}
\end{equation}

where \(\boldsymbol{\Sigma}\) denotes the covariance matrix. The entries of \(\boldsymbol{\Sigma}\) are defined through an exponential kernel so that nodes that are closer in space share more strongly correlated noise.
\begin{equation}
    \Sigma_{pq} = \exp\left(-\frac{d_{pq}}{\ell_c}\right),
    \qquad 
    d_{pq} = \|\boldsymbol{x}_p - \boldsymbol{x}_q\|_2 \quad \text{where}, \quad p,q \in \{1,\dots,n_n\}
    \label{eq:exp_kernel_Morans}
\end{equation}

 where \(d_{pq}\) represents the Euclidean distance between nodes $p$ and $q$, and \(\ell_c\)  denotes the correlation length.  For the \(50~\text{mm} \times 50~\text{mm} \times 1~\text{mm}\) hyperelastic specimen considered here, we choose \(\ell_c = 1.0~\text{mm}\)  equivalent to 2\% of the specimen size of in--plane dimension.  \par

\begin{figure}[H]
    \centering
    \includegraphics[width=\textwidth]{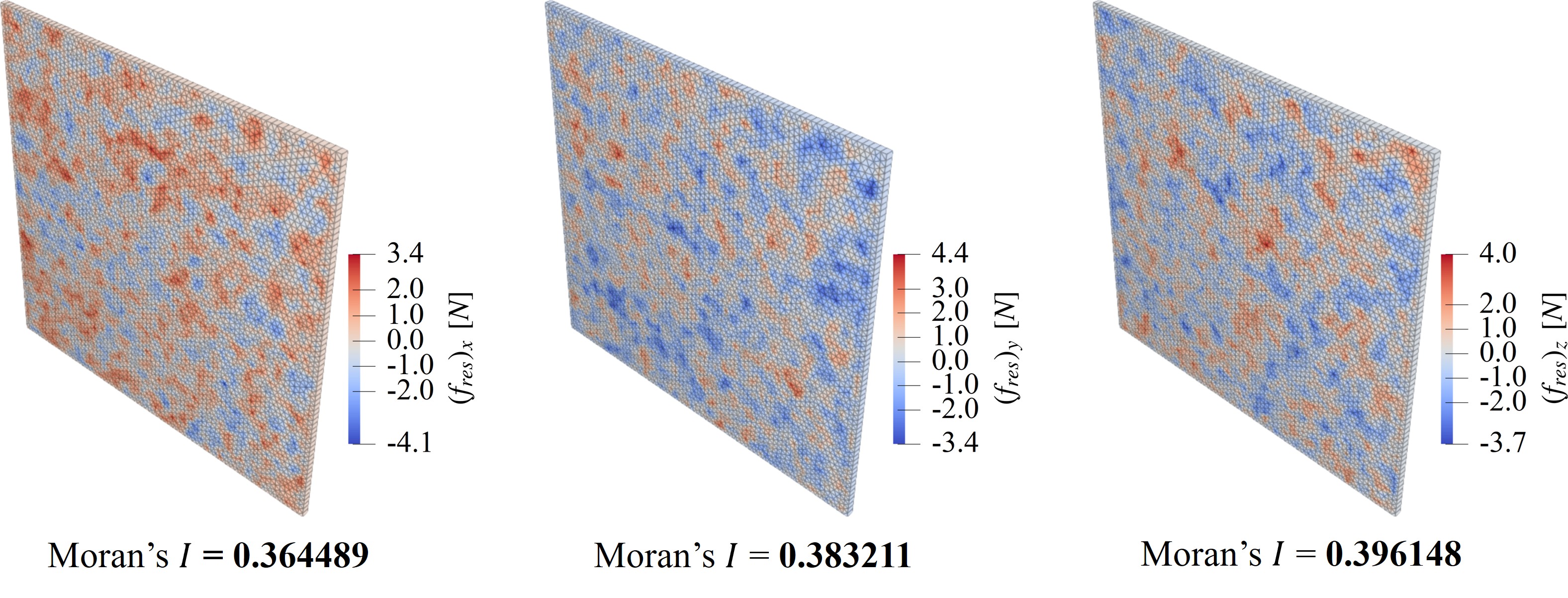}
    \caption{Residual force samples for the \(x\), \(y\), and \(z\) degrees of freedom drawn from a cross-correlated Gaussian distribution \( \mathcal{N}(\boldsymbol{0}, \boldsymbol{\Sigma}) \), illustrating the increase in Moran’s index (\textit{I}) due to spatial correlation embedded in the covariance structure.}
    \label{Type0_MoransI_correlated}
\end{figure}

The resulting $I$ values for $\left(\boldsymbol{f}_{\mathrm{res,}\boldsymbol{\theta}}\right)_x$, $\left(\boldsymbol{f}_{\mathrm{res,}\boldsymbol{\theta}}\right)_y$, and $\left(\boldsymbol{f}_{\mathrm{res,}\boldsymbol{\theta}}\right)_z$ (\autoref{Type0_MoransI_correlated}) are noticeably higher than those observed in test study 1 (\ref{Morans_test_study_1}), although the magnitudes of the residual forces are similar in both test studies. Thus, from test studies 1 and 2, we can confirm that the $I$ values for the residual forces increase with an increase in spatial correlation. Additionally, we obtain benchmarks to ascertain the extent of spatial correlation for a given residual force vector, i.e., $I$ values closer to those of test study 1 indicate the absence of spatial correlations. In contrast, $I$ values closer to those of test study 2 indicate the presence of significant spatial correlations. \par

\subsection{Test study 3 - Effect of uncorrelated displacement noise}
\label{Morans_test_study_3}
In this final test study, we calculate $I$ values for residual forces obtained from simulated biaxial tension tests on a thin, square-shaped, homogeneous hyperelastic specimen. The finite element simulation discretizes the geometry using 10,158 finite elements (C3D6 type). The hyperelastic material was prescribed a NH2 constitutive model with strain-energy given by $W_{\mathrm{NH2}} = \mu(\tilde{I}_{1}-3) + K(J-1)^2$. The residual force vector \( \boldsymbol{f}_{\mathrm{res,}\boldsymbol{\theta}} = \boldsymbol{A \theta - b}\) is obtained using the ground-truth material model $\boldsymbol{\theta}$ that was used in the simulations ($\mu=$ 5.4 and $K=$ 15.0).
\begin{figure}[H]
    \centering
    \includegraphics[width=\textwidth]{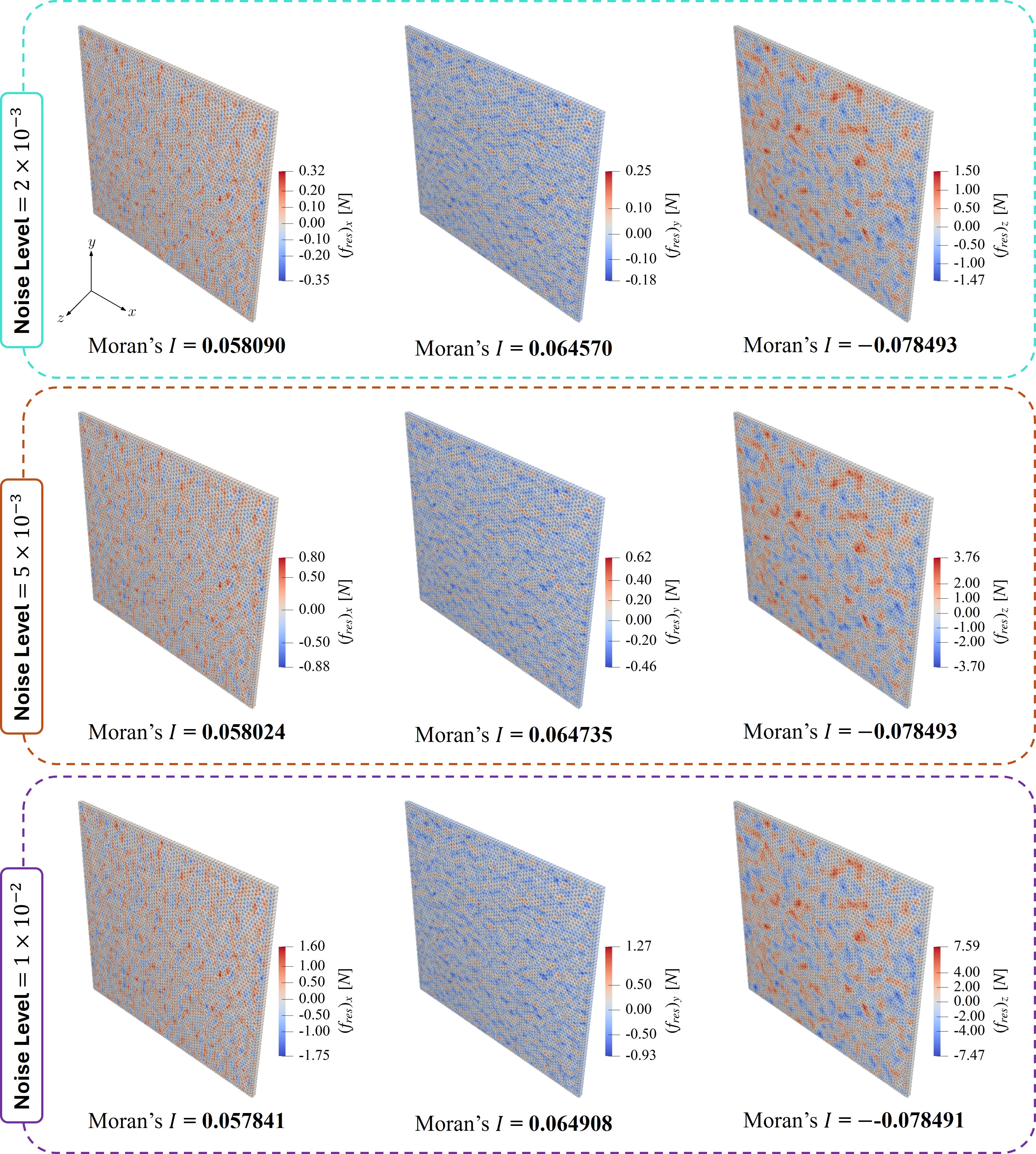}
    \caption{Moran's index (\(I\)) score for a homogeneous hyperelastic material subject to biaxial tension, using 10,158 elements under an uncorrelated noise (\(\sigma_{u}\)) with three noise levels as (\( 2\times10^{-3}\: , \: 5\times10^{-3} \: \text{and} \: 1\times10^{-2} \))  to the displacement field.}
    \label{MoransI_Type0_allNoise}
\end{figure}
An uncorrelated displacement noise $\sigma_u$ is added to the displacements obtained from these simulations. The addition of uncorrelated displacement noise (\(\sigma_{u} = 2\times10^{-3},~5\times10^{-3},~1\times10^{-2}\)) will only affect the $\boldsymbol{A}$ matrix, which is computed using the nodal displacement data and the connectivity table. The vector $\boldsymbol{b}$ is set to $\boldsymbol{0}$. The resulting $I$ values for $\left(\boldsymbol{f}_{\mathrm{res,}\boldsymbol{\theta}}\right)_x$, $\left(\boldsymbol{f}_{\mathrm{res,}\boldsymbol{\theta}}\right)_y$, and $\left(\boldsymbol{f}_{\mathrm{res,}\boldsymbol{\theta}}\right)_z$ (see \autoref{fig:ResForce_xyz}) are shown in \autoref{MoransI_Type0_allNoise}. The $I$ values, being an order of magnitude higher than those encountered in test study 1 (\ref{Morans_test_study_1}), indicate similar significant spatial correlations in residual forces for all noise cases.\par

Thus, the spatial correlations in the residual forces observed in test study 3 (\ref{Morans_test_study_3}) indicate that it would be inaccurate to sample the force vector \( \boldsymbol{f}_{\mathrm{res,}\boldsymbol{\theta}}\) from an independent and identical distribution (i.i.d.) as was done in the likelihood equation (Equation 33 in the Bayesian-EUCLID paper \citep{joshi2022bayesian} and \autoref{eq20} in this paper).

\section{Sampling residual forces from an i.i.d. Gaussian distribution is equivalent to solving OLS and minimizing MSE.}
\label{LikelihoodMSEOLS}
Given the non-negligible spatial correlations between components of the residual force vector ($\boldsymbol{f}_{\mathrm{res,}\boldsymbol{\theta}}$), in \ref{MoransI_study} we have established the inaccuracy of Equation 33 in the Bayesian-EUCLID paper \citep{joshi2022bayesian} and \autoref{eq20} in the current Hetero-EUCLID paper. However, in this section, we show that many conventional inverse material identification approaches implicitly make the same incorrect assumption. Specifically, all data-driven approaches that employ the Ordinary Least Squares (OLS) (such as the equilibrium-gap methods \citep{flaschel2021unsupervised, haustrate2024mechanical}) or that minimize the Mean-Squared Error (MSE) of residual forces (such as NN-based stress-unsupervised frameworks \citep{thakolkaran_nn-euclid_2022,lourenco_indirect_2024, meng_machine-learning-based_2025, shi2025deep}) implicitly assume a spatially uncorrelated residual force vector. To prove this, we first establish the equivalences between obtaining the material parameters $\boldsymbol{\theta}$ by maximizing the i.i.d. based likelihood (\autoref{eq20} of the Hetero-EUCLID framework), and obtaining $\boldsymbol{\theta}$ using OLS or by minimizing an MSE-based objective function.\par

In the Bayesian-EUCLID \citep{joshi2022bayesian} framework to model hyperelastic materials, we have the residual forces given by a linear function of the material parameters ($\boldsymbol{\theta}$), i.e, $\boldsymbol{f}_{\mathrm{res,}\boldsymbol{\theta}} = \boldsymbol{A \theta - b}$. However, in general, $\boldsymbol{f}_{\mathrm{res,}\boldsymbol{\theta}}$ can also be a non-linear function of the material parameters ($\boldsymbol{\theta}$), as would happen in the case of using neural networks to obtain the residual force \citep{thakolkaran_nn-euclid_2022,shi2025deep} or when dealing with non-hyperelastic materials \citep{flaschel2022discovering,lourenco_indirect_2024}. In the Bayesian-EUCLID formulation, if we were to impose a more generic likelihood (accounting for spatial correlations in the residual force) for satisfying the weak form of the linear momentum balance, we would have:
\begin{equation}\boldsymbol{f}_{\mathrm{res,}\boldsymbol{\theta}}  \sim \mathcal{N}(\boldsymbol{\mu}, \boldsymbol{\Sigma}),
    \label{eq:Atheta-b_sampling_corr}
\end{equation}
where \(\boldsymbol{\mu}\) is the mean vector and \(\boldsymbol{\Sigma}\) is the covariance matrix.

\autoref{eq:Atheta-b_sampling_corr} can be written as:
\begin{align}
P(\boldsymbol{f}_{\mathrm{res,}\boldsymbol{\theta}}) & = \mathcal{N}(\boldsymbol{\mu}, \boldsymbol{\Sigma}) \nonumber \\
   & = K \exp\left(- \frac{1}{2} [ \boldsymbol{f}_{\mathrm{res,}\boldsymbol{\theta}} - \boldsymbol{\mu}]^{T} \boldsymbol{\Sigma}^{-1} [ \boldsymbol{f}_{\mathrm{res,}\boldsymbol{\theta}} - \boldsymbol{\mu}] \right)
    \label{eq:P_f_actual_exp_corr}
\end{align}
where \(K\) is a proportionality constant that collects all factors independent of \(\boldsymbol{\theta}\).\par
To discover the material parameters ($\boldsymbol{\theta}^{*}$), we would have to maximize the likelihood of satisfying the weak form of linear momentum balance \autoref{eq:P_f_actual_exp_corr}. That is,
\begin{equation}
   \boldsymbol{\theta}^{*} = \operatorname*{argmax}_{\theta} \: P( \boldsymbol{f}_{\mathrm{res,}\boldsymbol{\theta}}) 
   \label{eq:argmax_P_f_theta}
\end{equation}

Maximizing the likelihood (\autoref{eq:argmax_P_f_theta}) is the same as maximizing its logarithm:
\begin{equation}
   \boldsymbol{\theta}^{*} = \operatorname*{argmax}_{\theta} \: P(\boldsymbol{f}_{\mathrm{res,}\boldsymbol{\theta}}) = \operatorname*{argmax}_{\theta} \: \log P(\boldsymbol{f}_{\mathrm{res,}\boldsymbol{\theta}})
   \label{eq:argmax_log_P_f_theta}
\end{equation}
The logarithm of the likelihood can be further simplified as:
\begin{equation}
    \log P(\boldsymbol{f}_{\mathrm{res,}\boldsymbol{\theta}}) = \log K - \frac{1}{2}[ \boldsymbol{f}_{\mathrm{res,}\boldsymbol{\theta}} - \boldsymbol{\mu}]^{T} \boldsymbol{\Sigma}^{-1} [ \boldsymbol{f}_{\mathrm{res,}\boldsymbol{\theta}} - \boldsymbol{\mu}]
    \label{eq:log-likelihood}
\end{equation}
Since $K$ is independent of $\boldsymbol{\theta}$, \autoref{eq:argmax_log_P_f_theta} can be written as
\begin{equation}
    \label{eq:argmin_corr}
    \boldsymbol{\theta}^{*} = \operatorname*{argmin}_{\theta} \: [ \boldsymbol{f}_{\mathrm{res,}\boldsymbol{\theta}} - \boldsymbol{\mu}]^{T} \boldsymbol{\Sigma}^{-1} [ \boldsymbol{f}_{\mathrm{res,}\boldsymbol{\theta}} - \boldsymbol{\mu}]
\end{equation}

In \autoref{eq:Atheta-b_sampling_corr}, if we were to set $\boldsymbol{\mu}=\boldsymbol{0}$ (zero-mean) and $\boldsymbol{\Sigma}=\sigma^2\boldsymbol{I}$ (i.i.d.), we would recover the likelihood for satisfying the weak form of linear momentum balance used in the Bayesian-EUCLID paper \citep{joshi2022bayesian}. For $\boldsymbol{\mu}=\boldsymbol{0}$ and $\boldsymbol{\Sigma}=\sigma^2\boldsymbol{I}$, \autoref{eq:argmin_corr} will simplify to:
\begin{equation}
    \label{eq:argminMSE}
    \boldsymbol{\theta}^{*} = \operatorname*{argmin}_{\theta} \: \left \lVert \boldsymbol{f}_{\mathrm{res,}\boldsymbol{\theta}} \right \rVert_2^{2}
\end{equation}

\autoref{eq:argminMSE} describes the process of minimizing the Mean Squared Error (MSE) of the residual forces ($\boldsymbol{f}_{\mathrm{res,}\boldsymbol{\theta}}$), and inherently assumes that the expected value of the residual forces is zero, i.e, $\boldsymbol{\mu}=\boldsymbol{0}$, and that there are no spatial correlations in the residual forces, i.e, $\boldsymbol{\Sigma}=\sigma^2\boldsymbol{I}$. Many works that use neural networks to surrogate the constitutive model of materials in a stress-unsupervised manner use the MSE loss of residual forces to train their neural networks \citep{thakolkaran_nn-euclid_2022,shi2025deep,meng_machine-learning-based_2025}. Thus, these works also implicitly assume that the residual forces ($\boldsymbol{f}_{\mathrm{res,}\boldsymbol{\theta}}$) are expected to be $\boldsymbol{0}$ and have no spatial correlations. In order to correctly account for spatial correlations in the residual force, these works (\citep{thakolkaran_nn-euclid_2022,shi2025deep,meng_machine-learning-based_2025}) should have used \autoref{eq:argmin_corr} as their loss function, instead of \autoref{eq:argminMSE}.

Further, in \autoref{eq:argminMSE}, if we substitute $\boldsymbol{f}_{\mathrm{res,}\boldsymbol{\theta}}=\boldsymbol{A\theta-b}$, we obtain:
\begin{equation}
    \label{eq:argminOLS}
    \boldsymbol{\theta}^{*} = \operatorname*{argmin}_{\theta} \: \left \lVert \boldsymbol{A\theta-b} \right \rVert_2^{2}
\end{equation}

\autoref{eq:argminOLS} defines the Ordinary Least Squares (OLS) solution to obtain the material parameters $\boldsymbol{\theta}^{*}$, and inherently assumes that the expected value of the residual forces is zero, i.e, $\boldsymbol{\mu=0}$ and that there are no spatial correlations in the residual forces, i.e, $\boldsymbol{\Sigma}=\sigma^2\boldsymbol{I}$. This approach is commonly used to obtain material parameters in equilibrium gap-based methods (\citep{haustrate2024mechanical,flaschel2021unsupervised}), which by extension, also implicitly assume lack of spatial correlations in the residual forces ($\boldsymbol{f}_{\mathrm{res,}\boldsymbol{\theta}}$).\par

\section{Optimal sub-sampling criterion at material interface boundaries}
\label{flagged-subsample}
    
    To assess the impact of interface node sub-sampling on the robustness of Hetero-EUCLID model predictions under high noise conditions (\( \sigma_{\text{u, high}} = 5 \times 10^{-3}\)), we conducted a parameter study by varying the subset of sampled flagged nodes and evaluating the percentage error \( \epsilon \), between the predicted and ground truth material parameters. Specifically, we define \( n_{\text{flag}} \) as a percentage of all flagged nodes \( \boldsymbol{{N}_{\text{flag}}} \), selected by sorting the entries of residual force norm vector \( (f^a_{\text{res}})_{\text{het}} \), in ascending order and choosing the lowest \( n_{\text{flag}} \%\) of the sorted column vector.
    \begin{equation}
        \epsilon_k = \left| \frac{\theta_k^{\text{predicted}} - \theta_k^{\text{true}}}{\theta_k^{\text{true}}} \right| \times 100 \;\% \,, \quad \text{where} \quad k = 1, \dots, n_f
        \label{eqA1}
    \end{equation}
    
\begin{figure}[H]
        \centering
        \includegraphics[width=0.95\textwidth]{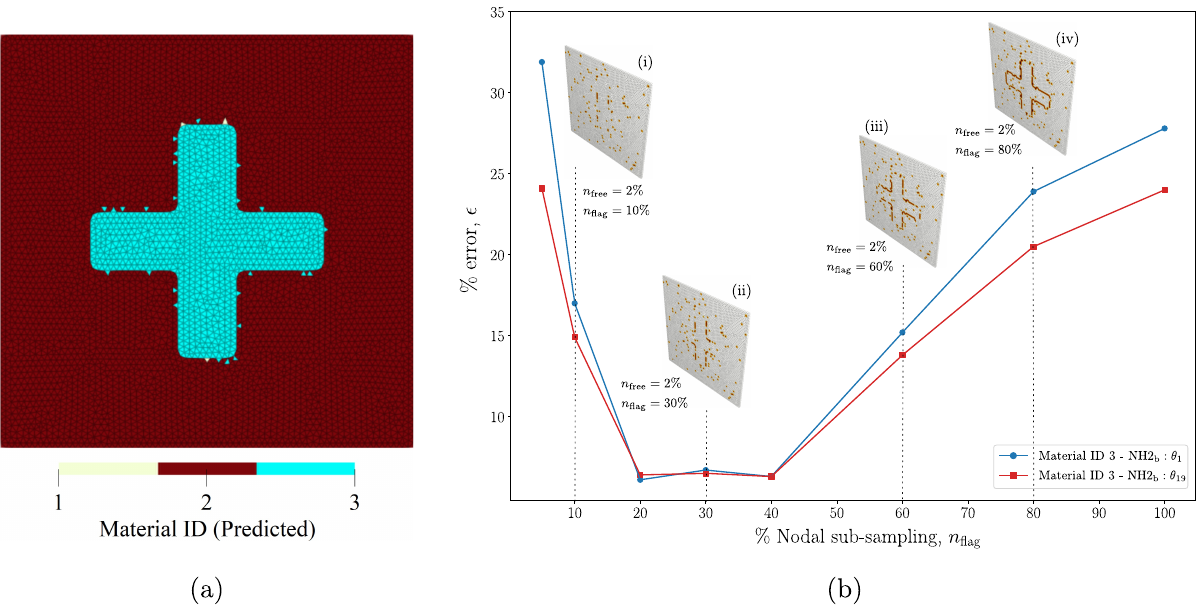}
        \caption{Study on the effect of node sub-sampling on misclustered inter-segment boundaries for Type 1 heterogeneous domain.}
        \label{mis_clustering}
    \end{figure}

    For the Type 1: Cross-shaped inclusion pattern (see \autoref{hetero_pattern:type1}) governed by the \(\text{NH2}_{\text{a}}\)–\(\text{NH2}_{\text{b}}\) model, we introduced a random misclustering of wedge elements near the material interface (see \autoref{mis_clustering}a) to simulate a realistic segmentation error. \autoref{mis_clustering}b shows how the percentage error in predicted material parameters (\(\theta_{1}\) and \(\theta_{19} \)) for cross inclusion changes as \( n_{\text{flag}} \) increases from 5\% to 100\%, while the free node sub-sampling rate \( n_{\text{free}} \) is kept constant at 2\%. A characteristic valley-shaped trend indicates that an intermediate sub-sampling level performs best under high noise conditions. The corresponding spatial distribution of the sampled interface nodes for 10\%, 30\%, 60\%, and 80\% is visualized in \autoref{mis_clustering}b(i)--(iv). It is evident that selecting 20\%--40\% of the nodes with the lowest residual forces leads to more accurate and stable parameter predictions, indicating the robustness of the Hetero-EUCLID framework. However, higher values of \( n_{\text{flag}} \) do not significantly improve accuracy and come with increased computational cost. Hence, we chose \( n_{\text{flag}} = 20\% \) as the optimal setting for interface node sub-sampling throughout the current study.

\section{Performance of Hetero-EUCLID on high-noise data without KRR denoising}
\label{app:withoutKRR}

\begin{figure}[H]
    \centering
    \includegraphics[width=0.95\textwidth]{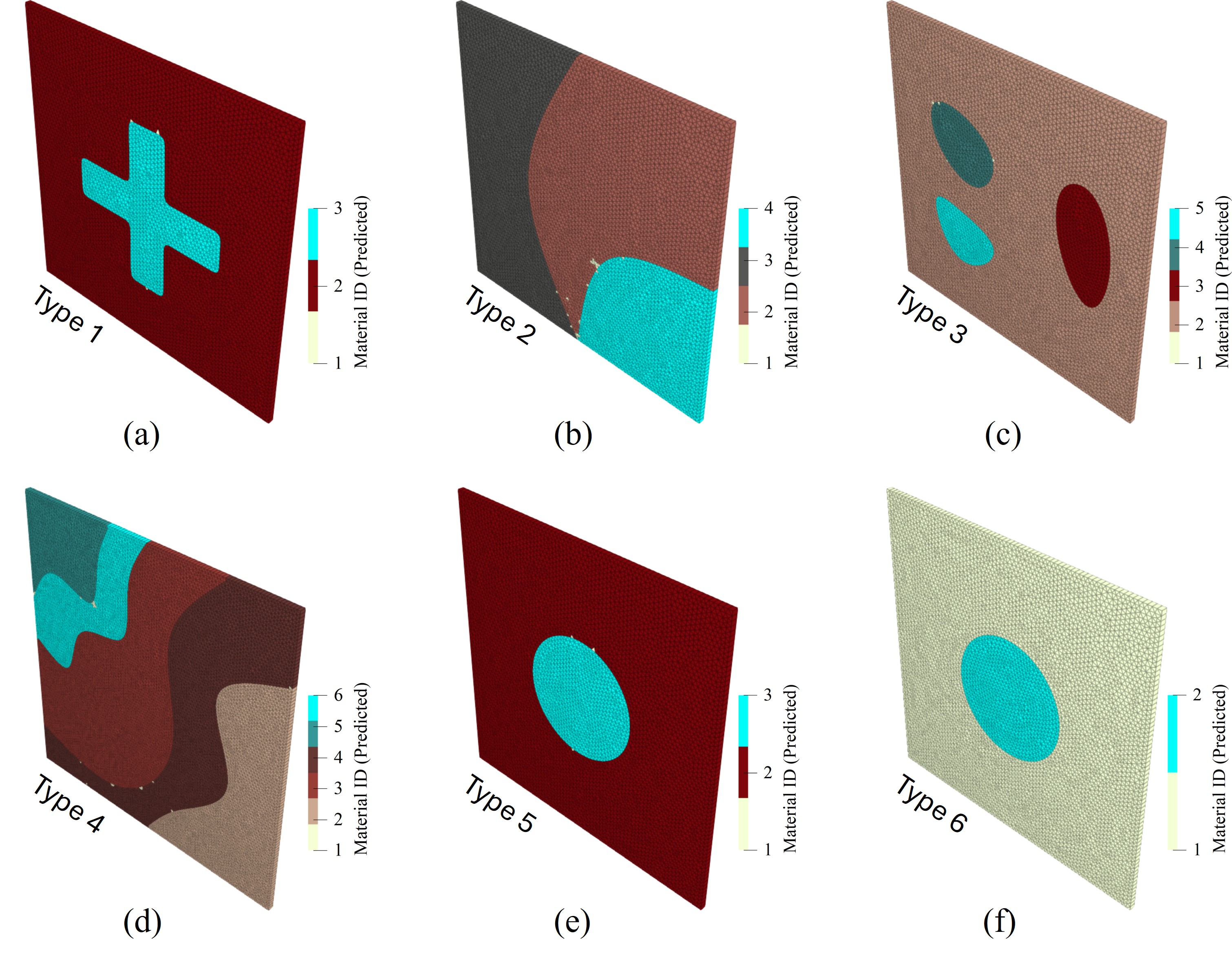}
    \caption{Segmentation outcomes for the heterogeneity patterns shown in \autoref{hetero_pattern}, evaluated under high displacement noise using Hetero-EUCLID without applying KRR-based denoising. The configurations include: (a) Type~1 - cross-shaped inclusion, (b) Type~2 - split-domain pattern, (c) Type~3 - multiple inner inclusions, (d) Type~4 - Cahn–Hilliard pattern, (e) Type~5 - circular inclusion, and (f) Type~6 - circular inclusion with anisotropic fiber reinforcement.}
    \label{Type_all_high_no_krr}
\end{figure}

To evaluate the performance of the Hetero-EUCLID framework and examine the role of Kernel Ridge Regression (KRR) based denoising, we conduct a study in which all heterogeneity patterns listed in \autoref{hetero_pattern} are subjected to high displacement noise 
\((\sigma_{u,\text{high}} = 5 \times 10^{-3})\). The inverse material parameter identification is then carried out using the Hetero-EUCLID framework without applying KRR-based denoising on the noisy displacement field. \autoref{Type_all_high_no_krr}a--f illustrates the segmentation outcomes for Type~1--6 configurations obtained using the seed-based growth algorithm (\autoref{alg:segm}). The segmentation results across all heterogeneity patterns are consistent with those obtained for the KRR-denoised data, as discussed in \autoref{case_study} and \ref{benchmark_study}. \par

\begin{table}[H]
    \centering
    \begin{adjustbox}{width=\textwidth}
    \begin{tabular}{@{} l c c c c c c c c c @{}}
    \toprule[1.5pt]
    \multicolumn{10}{l}{\textbf{Pattern Type 1: Cross-shaped inclusion} $\vert$ Mesh elements (C3D6): \textbf{10,442} $\vert$ Activated feature library terms, $\boldsymbol{n_{\mathit{f}} = 6}$} \\
    \midrule[1.5pt]
    \addlinespace
    \begin{tabular}{@{}l@{}}Case\end{tabular}
    & \begin{tabular}{@{}c@{}}Material\\ID\end{tabular} 
    & \begin{tabular}{@{}c@{}}Constitutive\\model\end{tabular} 
    & \begin{tabular}{@{}c@{}}$\theta_{1}$\\(MPa)\end{tabular} 
    & \begin{tabular}{@{}c@{}}$\theta_{2}$\\(MPa)\end{tabular} 
    & \begin{tabular}{@{}c@{}}$\theta_{3}$\\(MPa)\end{tabular} 
    & \begin{tabular}{@{}c@{}}$\theta_{4}$\\(MPa)\end{tabular} 
    & \begin{tabular}{@{}c@{}}$\theta_{5}$\\(MPa)\end{tabular} 
    & \begin{tabular}{@{}c@{}}$\theta_{19}$\\(MPa)\end{tabular} 
    & \begin{tabular}{@{}c@{}}Free nodes\\sub-sampling\end{tabular} \\
    \midrule[1.0pt]
    \multirow{2}{*}{\begin{tabular}{@{}l@{}}Ground Truth\\ \textit{Refer} \ref{Type1_none_high_10k}(a)\end{tabular}} 
    & 1 & \(\text{NH2}_{\text{a}}\) & 1.80 & $-$ & $-$ & $-$ & $-$ & 6.00 & \multirow{2}{*}{$-$} \\
    & 2 & \(\text{NH2}_{\text{b}}\) & 5.40 & $-$ & $-$ & $-$ & $-$ & 15.00 & \\
    \midrule[1.0pt]
    
   \multirow{2}{*}{\begin{tabular}{@{}l@{}}High Noise, \ref{Type_all_high_no_krr}(a) \\($\sigma_u = 5\times10^{-3}$) \end{tabular}} 
    & 2 & \(\text{NH2}_{\text{a}}\) & 1.80 $\pm$ 0.00 & 0.00 $\pm$ 0.00 & 0.00 $\pm$ 0.00 & 0.00 $\pm$ 0.02 & 0.00 $\pm$ 0.00 & 5.97 $\pm$ 0.00 & \multirow{2}{*}{2\%} \\
    & 3 & \(\text{NH2}_{\text{b}}\) & 5.20 $\pm$ 0.11 & 0.01 $\pm$ 0.04 & 0.00 $\pm$ 0.00 & 0.00 $\pm$ 0.02 & 0.00 $\pm$ 0.00 & 14.22 $\pm$ 0.21 & \\

    \bottomrule[1.5pt]
    \end{tabular}
    \end{adjustbox}
    \caption{True vs. predicted material parameters for cross-shaped inclusion pattern (\autoref{hetero_pattern:type1}) without KRR denoising under high noise condition 
    using the Bayesian-EUCLID framework.}
    \label{table:Type1_high_10k_noKRR}
\end{table}

\begin{table}[H]
    \centering
    \begin{adjustbox}{width=\textwidth}
    \begin{tabular}{@{} l c c c c c c c c c @{}}
    \toprule[1.5pt]
    \multicolumn{10}{l}{\textbf{Pattern Type 2: Split-domain} $\vert$ Mesh elements (C3D6): \textbf{10,972} $\vert$ Activated feature library terms, $\boldsymbol{n_{\mathit{f}} = 6}$} \\
    \midrule[1.5pt]
    \addlinespace
    \begin{tabular}{@{}l@{}}Case\end{tabular}
    & \begin{tabular}{@{}c@{}}Material\\ID\end{tabular} 
    & \begin{tabular}{@{}c@{}}Constitutive\\model\end{tabular} 
    & \begin{tabular}{@{}c@{}}$\theta_{1}$\\(MPa)\end{tabular} 
    & \begin{tabular}{@{}c@{}}$\theta_{2}$\\(MPa)\end{tabular} 
    & \begin{tabular}{@{}c@{}}$\theta_{3}$\\(MPa)\end{tabular} 
    & \begin{tabular}{@{}c@{}}$\theta_{4}$\\(MPa)\end{tabular} 
    & \begin{tabular}{@{}c@{}}$\theta_{5}$\\(MPa)\end{tabular} 
    & \begin{tabular}{@{}c@{}}$\theta_{19}$\\(MPa)\end{tabular} 
    & \begin{tabular}{@{}c@{}}Free nodes\\sub-sampling\end{tabular} \\
    \midrule[1.0pt]
    \multirow{3}{*}{\begin{tabular}{@{}l@{}}Ground\\Truth\\ \textit{Refer} \ref{Type2_none_high_10k}(a) \end{tabular}} 
    & 1 & NH2 & 6.00 & $-$ & $-$ & $-$ & $-$ & 32.00 &  \\
    
    & 2 & ISH & 4.00 & 0.50 & 0.30 & $-$ & $-$ & 21.00 & $-$ \\
    
    & 3 & HW & 1.00 & 0.15 & $-$ & 0.02 & 0.00 & 10.00 & \\
    
    \midrule[1.0pt]
    \multirow{3}{*}{\begin{tabular}{@{}l@{}}High Noise\\($\sigma_u = 5\times 10^{-3}$)\\ \textit{Refer} \ref{Type_all_high_no_krr}(b) \end{tabular}} 
    & 3 & NH2 & 1.27 $\pm$ 2.39 & 0.73 $\pm$ 1.06 & 0.45 $\pm$ 0.72 & 0.01 $\pm$ 0.21 & 0.02 $\pm$ 0.34 & 25.61 $\pm$ 14.87 &  \\
    & 2 & ISH & 6.33 $\pm$ 16.22 & 0.02 $\pm$ 0.17 & 0.28 $\pm$ 0.56 & 0.00 $\pm$ 0.00 & 0.00 $\pm$ 0.00 & 24.25 $\pm$ 46.33 & 5\% \\
    & 4 & HW & 3.28 $\pm$ 20.29 & 1.36 $\pm$ 10.34 & 0.16 $\pm$ 2.85 & 0.05 $\pm$ 0.35 & 0.01 $\pm$ 0.12 & 27.42 $\pm$ 113.87 & \\
    \bottomrule[1.5pt]
    \end{tabular}
    \end{adjustbox}
    \caption{True vs. predicted material parameters for split-domain pattern (\autoref{hetero_pattern:type2}) without KRR denoising under high noise condition using Bayesian-EUCLID framework.}
    \label{table:Type2_high_10k_noKRR}
\end{table}

\begin{table}[H]
    \centering
    \begin{adjustbox}{width=\textwidth}
    \begin{tabular}{@{} l c c c c c c c c c @{}}
    \toprule[1.5pt]
    \multicolumn{10}{l}{\textbf{Pattern Type 3: Multiple inner inclusions} $\vert$ Mesh elements (C3D6): \textbf{10,816} $\vert$ Activated feature library terms, $\boldsymbol{n_{\mathit{f}} = 6}$} \\
    \midrule[1.5pt]
    \addlinespace
    \begin{tabular}{@{}l@{}}Case\end{tabular}
    & \begin{tabular}{@{}c@{}}Material\\ID\end{tabular} 
    & \begin{tabular}{@{}c@{}}Constitutive\\model\end{tabular} 
    & \begin{tabular}{@{}c@{}}$\theta_{1}$\\(MPa)\end{tabular} 
    & \begin{tabular}{@{}c@{}}$\theta_{2}$\\(MPa)\end{tabular} 
    & \begin{tabular}{@{}c@{}}$\theta_{3}$\\(MPa)\end{tabular} 
    & \begin{tabular}{@{}c@{}}$\theta_{4}$\\(MPa)\end{tabular} 
    & \begin{tabular}{@{}c@{}}$\theta_{5}$\\(MPa)\end{tabular} 
    & \begin{tabular}{@{}c@{}}$\theta_{19}$\\(MPa)\end{tabular} 
    & \begin{tabular}{@{}c@{}}Free nodes\\sub-sampling\end{tabular} \\
    \midrule[1.0pt]
    \multirow{3}{*}{\begin{tabular}{@{}l@{}}Ground\\Truth\\\textit{Refer} \ref{Type3_none_high_10k}(a) \end{tabular}} 
    & 1 & ISH & 4.00 & 0.50 & 0.30 & $-$ & $-$ & 21.00 & \multirow{4}{*}{$-$} \\
    & 2 & \(\text{NH2}_{\text{a}}\) & 5.40 & $-$ & $-$ & $-$ & $-$ & 15.00 & \\
    & 3 & HW & 1.00 & 0.15 & $-$ & 0.02 & 0.00 & 10.00 & \\
    & 4 & \(\text{NH2}_{\text{b}}\) & 1.80 & $-$ & $-$ & $-$ & $-$ & 6.00 &  \\
    \midrule[1.0pt]
    
    \multirow{3}{*}{\begin{tabular}{@{}l@{}}High Noise\\ ($\sigma_u = 5\times10^{-3}$)\\\textit{Refer} \ref{Type_all_high_no_krr}(c) \end{tabular}} 
    & 2 & ISH & 4.82 $\pm$ 0.31 & 0.47 $\pm$ 0.02 & 0.13 $\pm$ 0.06 & 0.00 $\pm$ 0.00 & 0.00 $\pm$ 0.00 & 20.99 $\pm$ 0.03 & \multirow{4}{*}{10\%} \\
    & 5 & \(\text{NH2}_{\text{a}}\) & 4.42 $\pm$ 1.25 & 0.01 $\pm$ 0.04 & 0.04 $\pm$ 0.13 & 0.00 $\pm$ 0.00 & 0.02 $\pm$ 0.24 & 15.21 $\pm$ 20.01 &  \\
    & 3 & HW & 0.23 $\pm$ 0.30 & 0.66 $\pm$ 0.06 & 0.00 $\pm$ 0.00 & 0.00 $\pm$ 0.00 & 0.02 $\pm$ 0.01 & 5.35 $\pm$ 0.58 &  \\
    & 4 & \(\text{NH2}_{\text{b}}\) & 0.21 $\pm$ 0.40 & 0.07 $\pm$ 0.05 & 0.00 $\pm$ 0.00 & 0.00 $\pm$ 0.00 & 0.00 $\pm$ 0.00 & 2.12 $\pm$ 1.03 & \\
    \bottomrule[1.5pt]
    \end{tabular}
    \end{adjustbox}
    \caption{True vs. predicted material parameters for multiple inner inclusions pattern (\autoref{hetero_pattern:type3}) without KRR denoising under high noise condition using Bayesian-EUCLID framework.}
    \label{table:Type3_high_10k_noKRR}
\end{table}

\begin{table}[H]
    \centering
    \begin{adjustbox}{width=\textwidth}
    \begin{tabular}{@{} l c c c c c c c c c @{}}
    \toprule[1.5pt]
    \multicolumn{10}{l}{\textbf{Pattern Type 4: Cahn-Hilliard} $\vert$ Mesh elements (C3D6): \textbf{21,040} $\vert$ Activated feature library terms, $\boldsymbol{n_{\mathit{f}} = 6}$} \\
    \midrule[1.5pt]
    \addlinespace
    \begin{tabular}{@{}l@{}}Case\end{tabular}
    & \begin{tabular}{@{}c@{}}Material\\ID\end{tabular} 
    & \begin{tabular}{@{}c@{}}Constitutive\\model\end{tabular} 
    & \begin{tabular}{@{}c@{}}$\theta_{1}$\\(MPa)\end{tabular} 
    & \begin{tabular}{@{}c@{}}$\theta_{2}$\\(MPa)\end{tabular} 
    & \begin{tabular}{@{}c@{}}$\theta_{3}$\\(MPa)\end{tabular} 
    & \begin{tabular}{@{}c@{}}$\theta_{4}$\\(MPa)\end{tabular} 
    & \begin{tabular}{@{}c@{}}$\theta_{5}$\\(MPa)\end{tabular} 
    & \begin{tabular}{@{}c@{}}$\theta_{19}$\\(MPa)\end{tabular} 
    & \begin{tabular}{@{}c@{}}Free nodes\\sub-sampling\end{tabular} \\
    \midrule[1.0pt]
    \multirow{2}{*}{\begin{tabular}{@{}l@{}}Ground Truth\\\textit{Refer} \ref{Type4_none_high_20k}(a) \end{tabular}} 
    & 1 & \(\text{NH2}_{\text{a}}\) & 5.40 & $-$ & $-$ & $-$ & $-$ & 15.00 & \multirow{2}{*}{$-$} \\
    & 2 & \(\text{NH2}_{\text{b}}\) & 1.80 & $-$ & $-$ & $-$ & $-$ & 6.00 & \\
    \midrule[1.0pt]
    
    \multirow{3}{*}{\begin{tabular}{@{}l@{}}High Noise\\($\sigma_u = 5 \times 10^{-3}$)\\\textit{Refer} \ref{Type_all_high_no_krr}(d) \end{tabular}} 
    & 5 & \(\text{NH2}_{\text{a}}\) & 5.74 $\pm$ 0.04 & 0.15 $\pm$ 0.03 & 0.01 $\pm$ 0.01 & 0.01 $\pm$ 0.01 & 0.00 $\pm$ 0.00 & 16.79 $\pm$ 0.15 & \\
    
    & 6 & \(\text{NH2}_{\text{b}}\) & 1.58 $\pm$ 0.02 & 0.00 $\pm$ 0.00 & 0.00 $\pm$ 0.00 & 0.00 $\pm$ 0.00 & 0.00 $\pm$ 0.00 & 5.27 $\pm$ 0.07 & \\
    
    & 3 & \(\text{NH2}_{\text{a}}\) & 5.06 $\pm$ 0.05 & 0.02 $\pm$ 0.02 & 0.00 $\pm$ 0.00 & 0.00 $\pm$ 0.00 & 0.00 $\pm$ 0.00 & 14.19 $\pm$ 0.11 & 10\% \\
    
    & 4 & \(\text{NH2}_{\text{b}}\) & 1.76 $\pm$ 0.01 & 0.00 $\pm$ 0.00 & 0.00 $\pm$ 0.00 & 0.00 $\pm$ 0.00 & 0.00 $\pm$ 0.00 & 5.83 $\pm$ 0.01 & \\
    
    & 2 & \(\text{NH2}_{\text{a}}\) & 5.57 $\pm$ 0.02 & 0.00 $\pm$ 0.00 & 0.00 $\pm$ 0.00 & 0.00 $\pm$ 0.00 & 0.00 $\pm$ 0.00 & 15.29 $\pm$ 0.02 & \\

    \bottomrule[1.5pt]
    \end{tabular}
    \end{adjustbox}
    \caption{True vs. predicted material parameters for the Cahn-Hilliard pattern (\autoref{hetero_pattern:type4}) without KRR denoising under high noise condition using the Bayesian-EUCLID framework.}
    \label{table:Type4_high_20k_noKRR}
\end{table}

\begin{figure}[H]
    \centering
    \begin{subfigure}[b]{0.475\textwidth}
        \includegraphics[width=\textwidth]{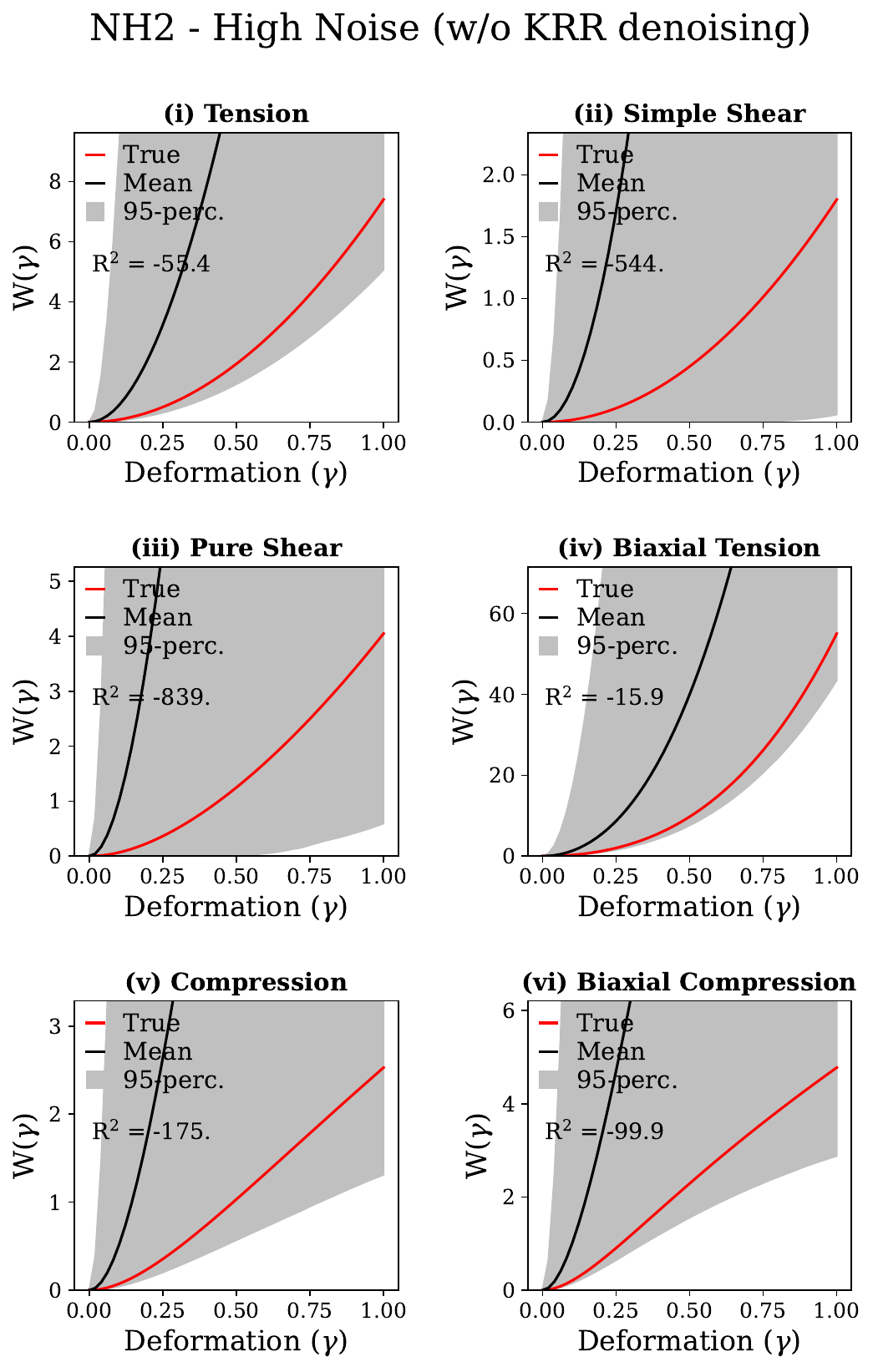}
        \caption{Material ID 2}
        \label{W_gamma_Type5_noKRR:ID2}
    \end{subfigure}
    \hfill
    \begin{subfigure}[b]{0.49\textwidth}
        \includegraphics[width=\textwidth]{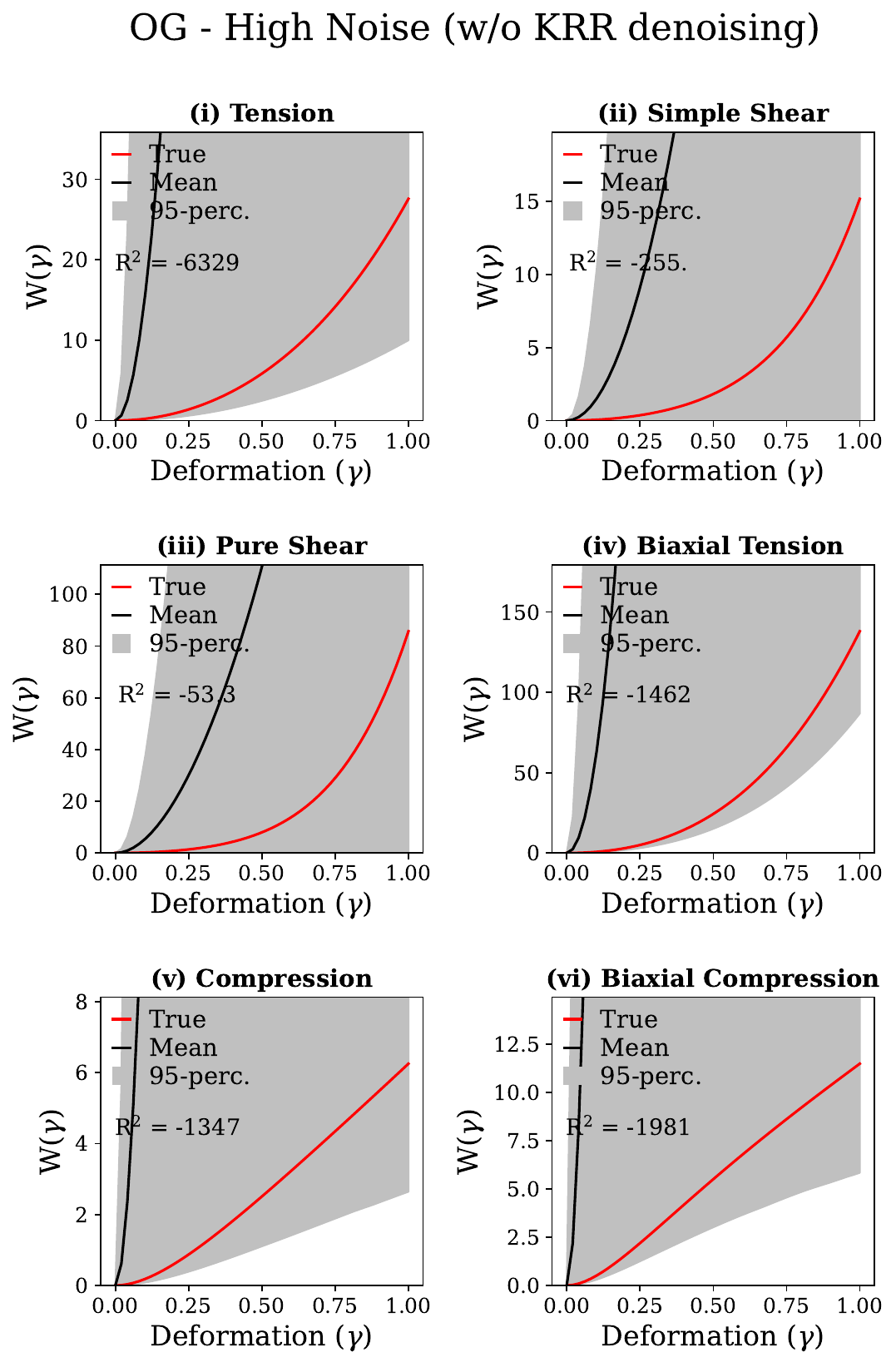}
        \caption{Material ID 3}
        \label{W_gamma_Type5_noKRR:ID3}
    \end{subfigure}
    \caption{Strain energy density plots for the circular inclusion-Ogden pattern (\autoref{hetero_pattern:type5}). Panels (a) and (b) correspond to the identified material sub-domains under high noise condition without KRR denoising. For each identified material segment ID (a) and (b), subplots (i)--(vi) show strain energy density plots \( W(\gamma) \) along various deformation paths (see supplementary Equation S1), providing a comparison between the discovered and true models.}
    \label{W_gamma_Type5_noKRR}
\end{figure}

\begin{figure}[H]
    \centering
    \begin{subfigure}[b]{0.49\textwidth}
        \includegraphics[width=\textwidth]{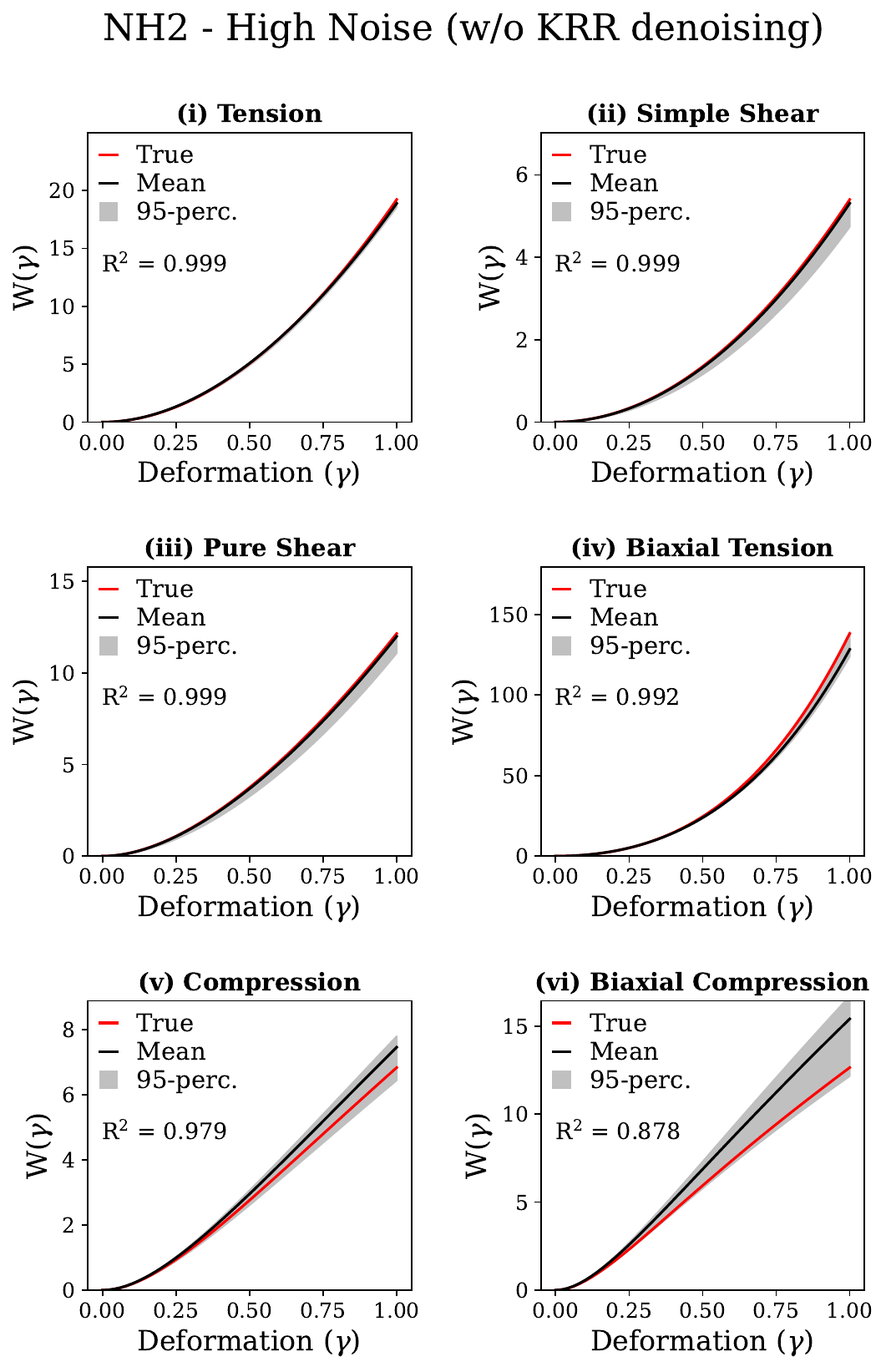}
        \caption{Material ID 1}
        \label{W_gamma_Type6_noKRR:ID1}
    \end{subfigure}
    \hfill
    \begin{subfigure}[b]{0.49\textwidth}
        \includegraphics[width=\textwidth]{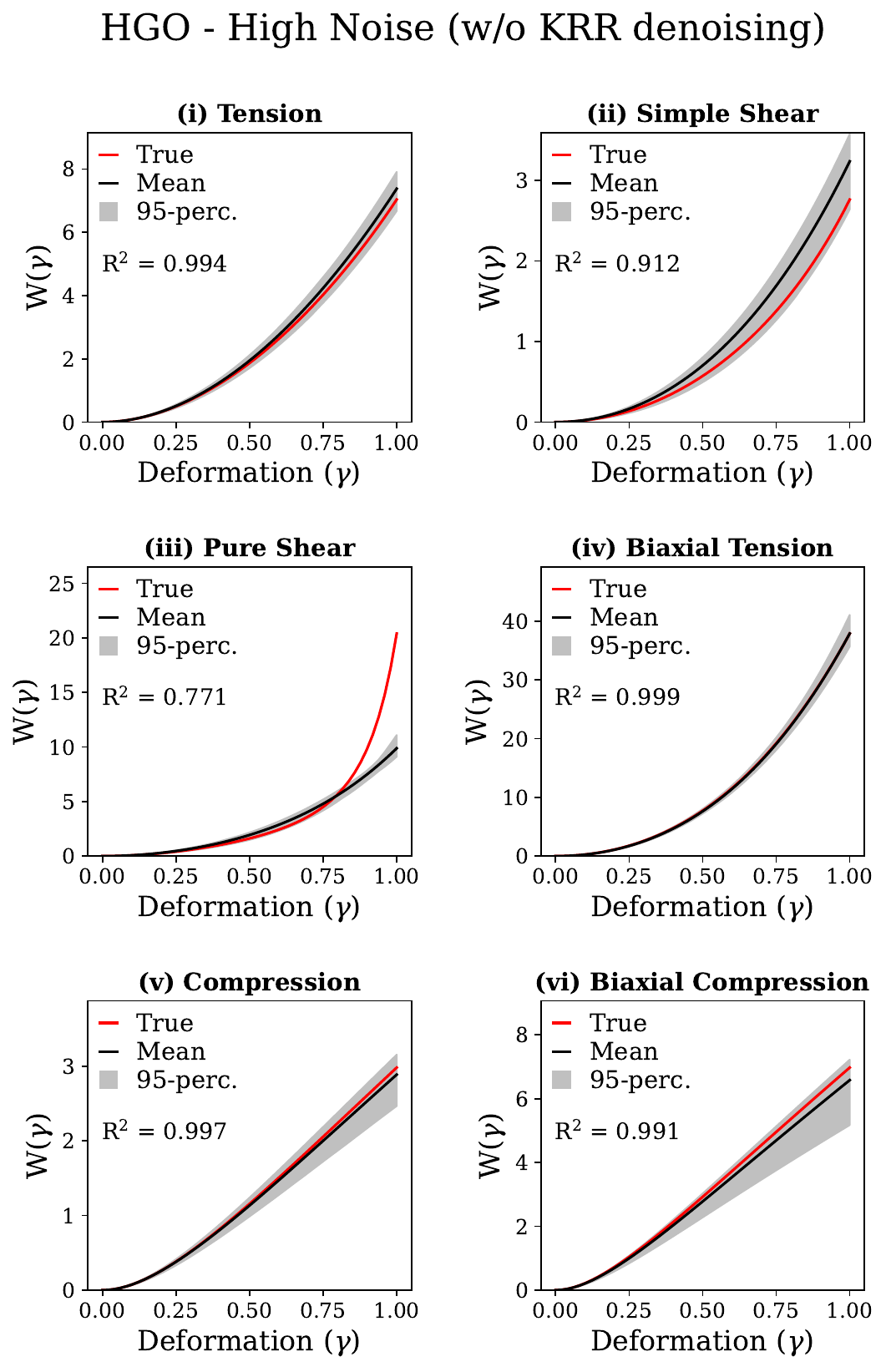}
        \caption{Material ID 2}
        \label{W_gamma_Type6_noKRR:ID2}
    \end{subfigure}
    \caption{Strain energy density plots for the circular inclusion with anisotropic fiber reinforcement pattern (\autoref{hetero_pattern:type6}). Panels (a) and (b) correspond to the identified material sub-domains under high noise condition without KRR denoising. For each identified material segment ID (a) and (b), subplots (i)--(vi) show strain energy density plots \( W(\gamma) \) along various deformation paths (see supplementary Equation S1), providing a comparison between the discovered and true models.}
    \label{W_gamma_Type6_noKRR}
\end{figure}

\autoref{table:Type1_high_10k_noKRR}, \ref{table:Type2_high_10k_noKRR}, \ref{table:Type3_high_10k_noKRR}, and \ref{table:Type4_high_20k_noKRR} 
summarize the recovered material parameters under high noise conditions for Type~1--4 cases, respectively. For the Type~5 and Type~6 configurations, 
corresponding to the Ogden and Holzapfel-Gasser-Ogden (HGO) models with feature exclusion, the strain energy density plots along six different 
deformation paths are shown in \autoref{W_gamma_Type5_noKRR} and \ref{W_gamma_Type6_noKRR}, respectively, in comparison to the ground truth. It can be observed that for Type~1, Type~4, and Type~6 configurations, the predicted material parameters listed in \autoref{table:Type1_high_10k_noKRR} and \ref{table:Type4_high_20k_noKRR} fall within the acceptable range, and the strain energy density plots in \autoref{W_gamma_Type6_noKRR} show close agreement with the corresponding ground truth. In contrast, for heterogeneity patterns Type~2, Type~3, and Type~5, which involve combinations of complex hyperelastic models such as Isihara (ISH), Haines--Wilson (HW), and Ogden (OG), the inverse parameter identification performs poorly, exhibiting large standard deviations and low \(R^2\) scores as evident from \autoref{table:Type2_high_10k_noKRR}, \ref{table:Type3_high_10k_noKRR}, and \autoref{W_gamma_Type5_noKRR}. These observations suggest that KRR-based denoising of the displacement field plays a crucial role in maintaining the reliability and applicability of the proposed framework across heterogeneous systems of varying complexity. \par

\section{Spatially correlated high noise study for Type 1 (Cross-shaped inclusion)}
\label{app:corr_noise}

In this study, we introduce spatially correlated Gaussian noise to the finite element displacement field (\(\boldsymbol{u}_{\text{FEM}}\)) for Type~1 (\autoref{hetero_pattern:type1}) cross-shaped inclusion pattern to investigate the performance of Hetero-EUCLID under correlated displacement noise. The noisy displacement field, \( \boldsymbol{u} = \{ u_{i}^{a} \; \forall \;  (a,i) \in \mathcal{D} : a = 1,\dots,n_n \, ; \, i = 1,2,3\} \) is modeled as
\begin{equation}
    \boldsymbol{u} = \boldsymbol{u}_{\text{FEM}} + \boldsymbol{u}_{\text{noise}} \quad \text{with} \quad \boldsymbol{u}_{\text{noise}} \sim \mathcal{N}(\boldsymbol{0}, \boldsymbol{\Sigma}) 
    \label{eq:spat_corr}
\end{equation}

where \(\boldsymbol{u}_{\text{noise}}\) is a zero--mean Gaussian
random field with covariance \(\boldsymbol{\Sigma}\). The covariance matrix \(\boldsymbol{\Sigma}\) induces spatial correlation based on the nodal coordinates, which is in contrast to the uncorrelated noise model (\autoref{eq26}), where noise at each node is independent. The entries of \(\boldsymbol{\Sigma}\) are defined through an exponential kernel so that nodes that are closer in space share more strongly correlated noise.
\begin{equation}
    \Sigma_{pq} = \exp\left(-\frac{d_{pq}}{\ell_c}\right),
    \qquad 
    d_{pq} = \|\boldsymbol{x}_p - \boldsymbol{x}_q\|_2 \quad \text{where}, \quad p,q \in \{1,\dots,n_n\}
    \label{eq:exp_kernel}
\end{equation}

 where \(d_{pq}\) represents the Euclidean distance between nodes $p$ and $q$, and \(\ell_c\)  denotes the correlation length.  For the \(50~\text{mm} \times 50~\text{mm} \times 1~\text{mm}\) specimen considered here, we choose \(\ell_c = 0.5~\text{mm}\) equivalent to 1\% of the specimen size of in--plane dimension. The noisy displacement field is then denoised using Kernel Ridge Regression (KRR) \citep{saunders1998ridge} with a Radial Basis Function (RBF) kernel.\par
\begin{figure}[H]
    \centering
    \includegraphics[width=\textwidth]{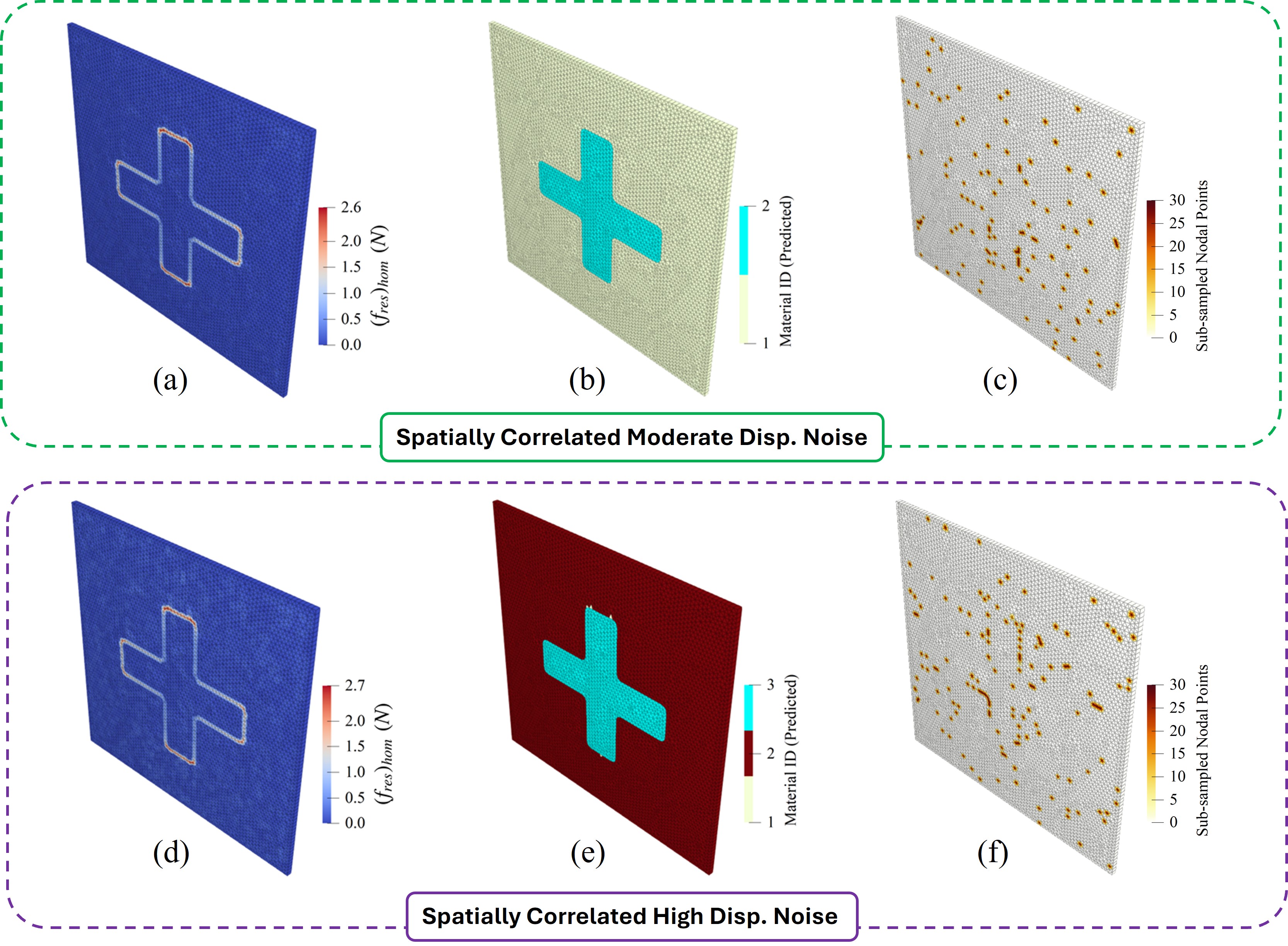}
    \caption{Model discovery for the cross-shaped inclusion pattern  (\autoref{hetero_pattern:type1}). For the spatially correlated \textit{moderate-noise} case, (a) depicts the residual force norm distribution (\autoref{eq19}) from the homogenized model, (b) illustrates the material segments identified through the segmentation algorithm, and (c) highlights the sub-sampled nodes from the \(\mathcal{D}^{\text{free}}\) set used to construct the heterogenized model. The spatially correlated \textit{high-noise} case follows a similar structure in panels (d)--(f), showing how the residual force field, segmentation, and sub-sampling behave under high noise levels. In both scenarios, the final heterogenized model is solved using the Bayesian framework to recover the region-wise constitutive parameters as listed in \autoref{table_Type1_10k_spatial_corr}.}
    \label{Type1_spatial_corr_mod_high_10k}
\end{figure}
\begin{table}[H]
    \centering
    \begin{adjustbox}{width=\textwidth}
    \begin{tabular}{@{} l c c c c c c c c c @{}}
    \toprule[1.5pt]
    \multicolumn{10}{l}{\textbf{Pattern Type 1: Cross-shaped inclusion} $\vert$ Mesh elements (C3D6): \textbf{10,442} $\vert$ Activated feature library terms, $\boldsymbol{n_{\mathit{f}} = 6}$} \\
    \midrule[1.5pt]
    \addlinespace
    \begin{tabular}{@{}l@{}}Case\end{tabular}
    & \begin{tabular}{@{}c@{}}Material\\ID\end{tabular} 
    & \begin{tabular}{@{}c@{}}Constitutive\\model\end{tabular} 
    & \begin{tabular}{@{}c@{}}$\theta_{1}$\\(MPa)\end{tabular} 
    & \begin{tabular}{@{}c@{}}$\theta_{2}$\\(MPa)\end{tabular} 
    & \begin{tabular}{@{}c@{}}$\theta_{3}$\\(MPa)\end{tabular} 
    & \begin{tabular}{@{}c@{}}$\theta_{4}$\\(MPa)\end{tabular} 
    & \begin{tabular}{@{}c@{}}$\theta_{5}$\\(MPa)\end{tabular} 
    & \begin{tabular}{@{}c@{}}$\theta_{19}$\\(MPa)\end{tabular} 
    & \begin{tabular}{@{}c@{}} $n_{\text{free}}$\\sampling\end{tabular} \\
    \midrule[1.0pt]
    \multirow{2}{*}{\begin{tabular}{@{}l@{}}Ground\\ Truth\end{tabular}} 
    & 1 & \(\text{NH2}_{\text{a}}\) & 1.80 & $-$ & $-$ & $-$ & $-$ & 6.00 & \multirow{2}{*}{$-$} \\
    
    & 2 & \(\text{NH2}_{\text{b}}\) & 5.40 & $-$ & $-$ & $-$ & $-$ & 15.00 & \\
    \midrule[1.0pt]
    
    \multirow{2}{*}{\begin{tabular}{@{}l@{}}Spatially Corr. Moderate Noise \\($\sigma_u = 2\times10^{-3}$) \textit{Refer} \ref{Type1_spatial_corr_mod_high_10k}(b) \end{tabular}} 
    & 1 & \(\text{NH2}_{\text{a}}\) & 1.81 $\pm$ 0.00 & 0.00 $\pm$ 0.00 & 0.00 $\pm$ 0.00 & 0.00 $\pm$ 0.00 & 0.00 $\pm$ 0.00 & 6.00 $\pm$ 0.00 & \multirow{2}{*}{2\%} \\
    & 2 & \(\text{NH2}_{\text{b}}\) & 5.41 $\pm$ 0.08 & 0.01 $\pm$ 0.03 & 0.00 $\pm$ 0.01 & 0.00 $\pm$ 0.01 & 0.00 $\pm$ 0.00 & 14.93 $\pm$ 0.15 & \\
    \midrule[1.0pt]

    \multirow{2}{*}{\begin{tabular}{@{}l@{}}Spatially Corr. High Noise \\($\sigma_u = 5\times10^{-3}$)  \textit{Refer} \ref{Type1_spatial_corr_mod_high_10k}(e)\end{tabular}} 
    & 2 & \(\text{NH2}_{\text{a}}\) & 1.81 $\pm$ 0.00 & 0.00 $\pm$ 0.00 & 0.00 $\pm$ 0.00 & 0.00 $\pm$ 0.00 & 0.00 $\pm$ 0.00 & 5.96 $\pm$ 0.00 & \multirow{2}{*}{2\%} \\
    & 3 & \(\text{NH2}_{\text{b}}\) & 5.18 $\pm$ 0.11 & 0.00 $\pm$ 0.01 & 0.00 $\pm$ 0.00 & 0.00 $\pm$ 0.01 & 0.00 $\pm$ 0.00 & 14.16 $\pm$ 0.26 & \\

    \bottomrule[1.5pt]
    \end{tabular}
    \end{adjustbox}
    \caption{True vs. predicted material parameters for cross-shaped inclusion pattern (\autoref{hetero_pattern:type1}) under spatially correlated moderate and high displacement noise conditions using Bayesian-EUCLID framework.}
    \label{table_Type1_10k_spatial_corr}
\end{table}

\autoref{Type1_spatial_corr_mod_high_10k} presents the segmentation outcomes for the Type~1 cross-shaped inclusion under spatially correlated displacement noise at moderate (\(\sigma_{u,\text{mod}} = 2\times10^{-3}\)) and high (\(\sigma_{u,\text{high}} = 5\times10^{-3}\)) noise levels. The corresponding Neo-Hookean (NH2) material parameters recovered by the Hetero-EUCLID framework are reported in \autoref{table_Type1_10k_spatial_corr}. The predicted parameters remain well within the acceptable tolerance with very low standard deviations when compared to the ground truth, demonstrating the reliability of the approach even under spatially correlated noise conditions. When compared with the uncorrelated noise case (\autoref{eq26}) reported in \autoref{table_Type1_none_high_10k}, the segmentation outcome and the recovered material parameters remain nearly identical, with no observable deterioration in accuracy. This demonstrates that introducing spatial structure into the noise field does not adversely affect either the interface identification or the inverse parameter estimation. Based on these observations for the Type~1 specimen, we expect a similar level of robustness for other heterogeneity patterns subjected to spatially correlated displacement noise. \par 

\section{Sensitivity analysis under low elastic contrast}\label{study_low_elastic}

In this section, we investigate the robustness of segmentation and material parameter prediction for the Type~1 cross-shaped heterogeneity pattern (\autoref{hetero_pattern:type1}), wherein the difference in elastic properties of the matrix (\(\text{NH2}_{\text{a}}\)) and inclusion (\(\text{NH2}_{\text{b}}\)) materials is significantly lesser compared to the difference in material properties for the study presented in \autoref{subsubsec:Crossinc}. Specifically, \autoref{table_Type1_10k_min_contrast} presents a case with approximately 25\% deviation in the material parameters \((\theta_{1}, \theta_{19})\) between the matrix (\(\text{NH2}_{\text{a}} : \{ \theta_{1} = 4.0, \, \theta_{19} = 11.0 \}\)) and the inclusion (\(\text{NH2}_{\text{b}} : \{ \theta_{1} = 5.4, \, \theta_{19} = 15.0 \}\)).\par

\begin{figure}[H]
    \centering
    \includegraphics[width=\textwidth]{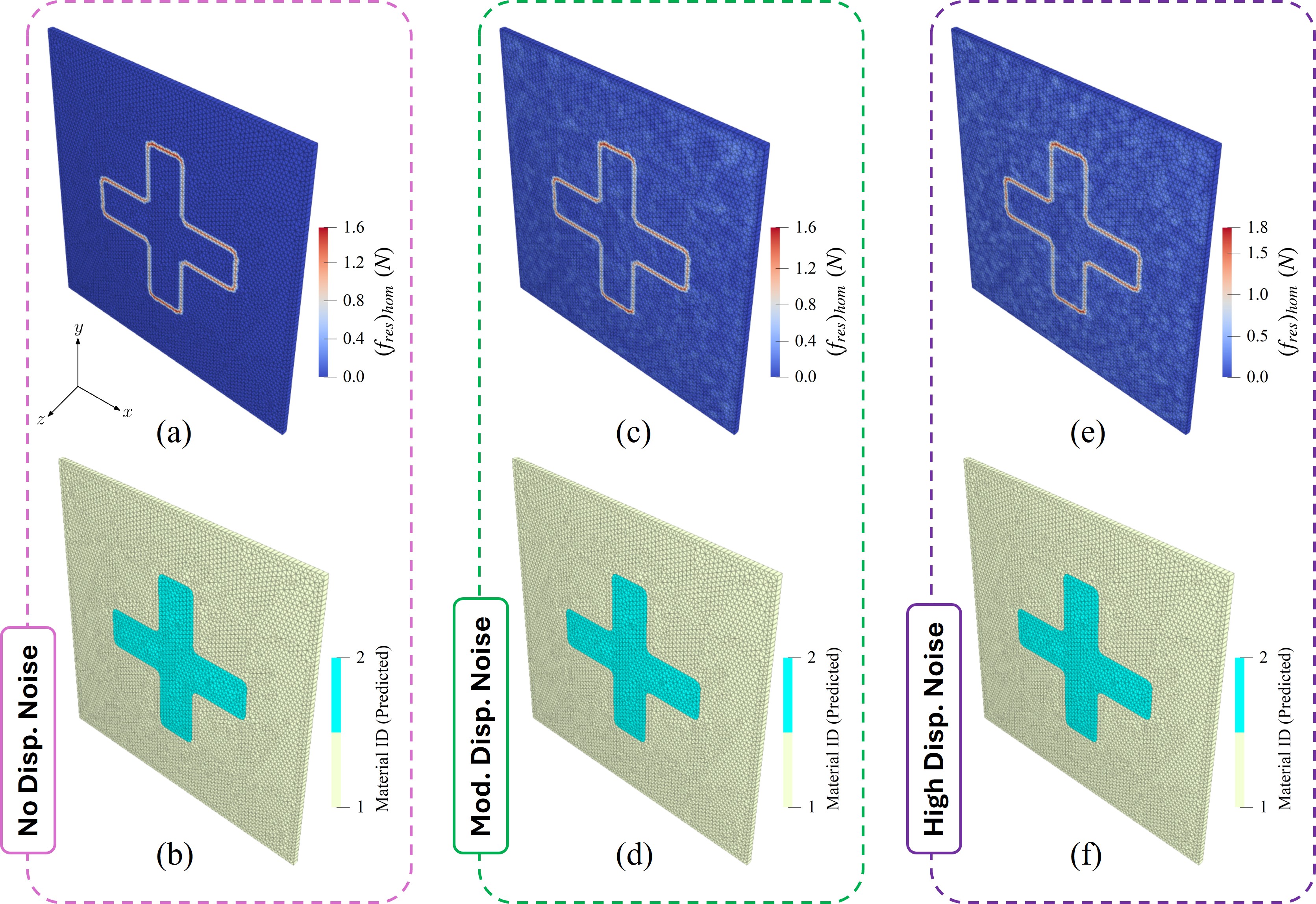}
    \caption{Model discovery for the cross-shaped inclusion pattern  (\autoref{hetero_pattern:type1}) under low elastic contrast. For the \textit{zero-noise} case, (a) depicts the residual force norm distribution (\autoref{eq19}) from the homogenized model, and (b) illustrates the material segments identified through the segmentation algorithm. The \textit{moderate} and \textit{high-noise} cases follow a similar structure in panels (c)--(d) and (e)--(f), respectively, showing how the residual force field and segmentation behave under moderate and high noise levels. In all three scenarios, the final heterogenized model is solved using the Bayesian framework to recover the region-wise constitutive parameters as listed in \autoref{table_Type1_10k_min_contrast}.}
    \label{Type1_allnoise_min_contrast}
\end{figure}

\autoref{Type1_allnoise_min_contrast} illustrates the segmentation outcomes and corresponding residual force distributions for noise-free, moderate, and high displacement noise conditions. The residual force norm \((f_{\text{res}})_{\text{hom}}\) computed from the homogenized model successfully captures the material interface even with minimal elastic contrast, as seen in \ref{Type1_allnoise_min_contrast}a, c, and e. The segmentation results, depicted in \ref{Type1_allnoise_min_contrast}b, d, and f, demonstrate that the seed-based growth algorithm can distinctly identify the inclusion region, highlighting the framework’s sensitivity in resolving low-contrast heterogeneities. The predicted material parameters are summarized in \autoref{table_Type1_10k_min_contrast}. For noise-free and moderate noise levels (\(\sigma_{u,\text{mod}} = 2 \times 10^{-3}\)), the predicted parameters \((\theta_{1}, \theta_{19})\) for both material regions closely match the ground truth with low standard deviations, reflecting a high degree of confidence in the prediction. Under high noise conditions (\(\sigma_{u,\text{high}} = 5 \times 10^{-3}\)), the deviation in parameter estimation increases significantly, particularly for \(\theta_{1}\). This would indicate that the Hetero-EUCLID framework performs accurate segmentation and characterization of materials with nearly similar mechanical properties for moderate displacement noises. However, the accuracy of the predicted material properties seems to deteriorate with increase in displacement noise if the constituent heterogeneities have similar material properties.

\begin{table}[H]
    \centering
    \begin{adjustbox}{width=\textwidth}
    \begin{tabular}{@{} l c c c c c c c c c @{}}
    \toprule[1.5pt]
    \multicolumn{10}{l}{\textbf{Pattern Type 1: Cross-shaped inclusion} $\vert$ Mesh elements (C3D6): \textbf{10,442} $\vert$ Activated feature library terms, $\boldsymbol{n_{\mathit{f}} = 6}$} \\
    \midrule[1.5pt]
    \addlinespace
    \begin{tabular}{@{}l@{}}Case\end{tabular}
    & \begin{tabular}{@{}c@{}}Material\\ID\end{tabular} 
    & \begin{tabular}{@{}c@{}}Constitutive\\model\end{tabular} 
    & \begin{tabular}{@{}c@{}}$\theta_{1}$\\(MPa)\end{tabular} 
    & \begin{tabular}{@{}c@{}}$\theta_{2}$\\(MPa)\end{tabular} 
    & \begin{tabular}{@{}c@{}}$\theta_{3}$\\(MPa)\end{tabular} 
    & \begin{tabular}{@{}c@{}}$\theta_{4}$\\(MPa)\end{tabular} 
    & \begin{tabular}{@{}c@{}}$\theta_{5}$\\(MPa)\end{tabular} 
    & \begin{tabular}{@{}c@{}}$\theta_{19}$\\(MPa)\end{tabular} 
    & \begin{tabular}{@{}c@{}}Free nodes\\sub-sampling\end{tabular} \\
    \midrule[1.0pt]
    \multirow{2}{*}{\begin{tabular}{@{}l@{}}Ground\\ Truth\end{tabular}} 
    & 1 & \(\text{NH2}_{\text{a}}\) & 4.00 & $-$ & $-$ & $-$ & $-$ & 11.00 & \multirow{2}{*}{$-$} \\
    
    & 2 & \(\text{NH2}_{\text{b}}\) & 5.40 & $-$ & $-$ & $-$ & $-$ & 15.00 & \\
    \midrule[1.0pt]
    
    \multirow{2}{*}{\begin{tabular}{@{}l@{}}No Noise, \ref{Type1_allnoise_min_contrast}(b) \\ ($\sigma_u = 0$) \end{tabular}} 
    & 1 & \(\text{NH2}_{\text{a}}\) & 4.00 $\pm$ 0.00 & 0.00 $\pm$ 0.00 & 0.00 $\pm$ 0.00 & 0.00 $\pm$ 0.02 & 0.00 $\pm$ 0.00 & 11.00 $\pm$ 0.00 & \multirow{2}{*}{5\%} \\
    
    & 2 & \(\text{NH2}_{\text{b}}\) & 5.40 $\pm$ 0.01 & 0.00 $\pm$ 0.00 & 0.00 $\pm$ 0.00 & 0.00 $\pm$ 0.00 & 0.00 $\pm$ 0.00 & 15.00 $\pm$ 0.02 & \\
    \midrule[1.0pt]
    
    \multirow{2}{*}{\begin{tabular}{@{}l@{}}Mod. Noise, \ref{Type1_allnoise_min_contrast}(d) \\($\sigma_u = 2\times10^{-3}$) \end{tabular}} 
    & 1 & \(\text{NH2}_{\text{a}}\) & 3.98 $\pm$ 0.23 & 0.00 $\pm$ 0.00 & 0.00 $\pm$ 0.03 & 0.00 $\pm$ 0.01 & 0.00 $\pm$ 0.00 & 11.02 $\pm$ 0.02 & \multirow{2}{*}{5\%} \\
    
    & 2 & \(\text{NH2}_{\text{b}}\) & 4.50 $\pm$ 1.71 & 0.18 $\pm$ 0.39 & 0.07 $\pm$ 0.14 & 0.01 $\pm$ 0.01 & 0.00 $\pm$ 0.00 & 14.32 $\pm$ 0.97 & \\
    \midrule[1.0pt]
    
    \multirow{2}{*}{\begin{tabular}{@{}l@{}}High Noise, \ref{Type1_allnoise_min_contrast}(f) \\($\sigma_u = 5\times10^{-3}$) \end{tabular}} 
    & 1 & \(\text{NH2}_{\text{a}}\) & 2.42 $\pm$ 1.95 & 0.01 $\pm$ 0.02 & 0.55 $\pm$ 0.52 & 0.00 $\pm$ 0.00 & 0.01 $\pm$ 0.01 & 7.97 $\pm$ 0.02 & \multirow{2}{*}{5\%} \\
    
    & 2 & \(\text{NH2}_{\text{b}}\) & 4.81 $\pm$ 0.77 & 0.04 $\pm$ 0.16 & 0.12 $\pm$ 0.14 & 0.05 $\pm$ 0.07 & 0.03 $\pm$ 0.02 & 11.27 $\pm$ 0.35 & \\

    \bottomrule[1.5pt]
    \end{tabular}
    \end{adjustbox}
    \caption{True vs. predicted material parameters for cross-shaped inclusion pattern (\autoref{hetero_pattern:type1}) with a low elastic contrast between the matrix and inclusion materials, evaluated across different noise levels using Hetero-EUCLID framework.}
    \label{table_Type1_10k_min_contrast}
\end{table}

\section{Pseudo-code for heterogeneous hyperelastic material model discovery}
\label{pseudocode_heteroEUCLID}
    \begin{algorithm}[H]
        \caption{Hetero-EUCLID algorithm}
        \label{alg:heterogeneous_discovery}
        \begin{algorithmic}[1]
        \Statex \textbf{Input:} Full-field displacement data $\mathbf{u}$, boundary reaction forces $\mathbf{R}$
        
        \State $\mathcal{S} \gets$ homogeneous material model \Comment{modeling assumption}
            \For{$t = 1, \dots ,n_t$} \Comment{$n_t$: number of load steps}
                \State ${\boldsymbol{\theta}^\text{ols}} \gets \mathtt{ols}(\boldsymbol{A}, \boldsymbol{b})$ \Comment{Solve $\min_{\theta} \|\boldsymbol{A\theta} - \boldsymbol{b}\|^2$}
                \State $(\boldsymbol{f}_{\mathrm{res}})_{\mathrm{hom}} \gets \boldsymbol{A}_{\mathrm{free}} {\boldsymbol{\theta}^\text{ols}} - \boldsymbol{b}_{\mathrm{free}}$ 
                       \Comment{Vector of $(f_x^a, f_y^a, f_z^a)$ for all free nodes}
                \For{each free node $a \in \mathcal{D}^{\mathrm{free}}$}
                    \State $f_{\mathrm{res}}^{a} \gets \sqrt{(f_x^a)^2 + (f_y^a)^2}$ \Comment{Euclidean Norm}
                \EndFor
            \EndFor
        \State \{$\mathcal{S}_k: k \in {1 \dots (i-1)}\} \gets \mathtt{getMaterialSegments}$({$\boldsymbol{N}_\text{flag}$}, Nodal positions, Connectivity)  \Comment{\autoref{alg:segm}}
        \State \textbf{Bayesian EUCLID:} \hfill {\footnotesize (refer \ref{BayesianEUCLID})}
        \State $n_{\text{free}} \gets \mathrm{SampleFreeNodes}(\mathcal{D}^{\mathrm{free}}, 0.02\text{--}0.10)$
        \State $n_{\text{flag}} \gets \mathrm{SampleInterfaceNodes}(\mathcal{D}^{\mathrm{interface}}, 0.20\text{--}0.40, (f_{\mathrm{res}}^{a})_{\mathrm{het}})$ \hfill {\footnotesize (refer \ref{flagged-subsample})}
        
        \State Posterior inference performed via Markov Chain Monte Carlo (MCMC)
        \For{$k = 1, \dots ,N_{\mathrm{chains}}$}
            \State Allocate chain storage $C_k$ of size $(N_{\mathrm{burn}} + N_G)$
            \State Initialize $\theta_{r} \gets$ model coefficients
            \State Initialize $\sigma^2 \gets$ noise variance
            \State Initialize $v_s \gets$ scaling hyperparameter
            \State Initialize $p_0 \gets$ prior weight
            \State Initialize $z \gets$ latent node assignments
            \For{$q = 1, \ldots, (N_{\mathrm{burn}} + N_G)$}
                \State Sample $\theta \sim p(\theta \mid \cdot)$
                \State Sample $\sigma^2, v_s, p_0 \sim p(\sigma^2, v_s, p_0 \mid \cdot)$
                \State order $\gets \mathrm{Perm}(1, \ldots, n_f)$ \Comment{Random permutation of node indices}
                \For{$j = 1, \dots ,n_f$}
                    \State $i \gets \mathrm{order}[j]$
                    \State Sample $z_i \sim p(z_i \mid \theta, \sigma^2, v_s, p_0)$
                \EndFor
                \State Record $C_k[q] \gets \{\theta, \sigma^2, v_s, p_0, z\}$
            \EndFor
        \EndFor
        \State Discard first $N_{\mathrm{burn}}$ samples from each chain
        \State Retain last $N_G$ samples as approximate posterior draws
        \State \textbf{Output:} Material parameters ($\boldsymbol{\theta}^{(1)},...,\boldsymbol{\theta}^{(k)}$) for each segment, with standard deviations and mean
        \end{algorithmic}
    \end{algorithm}

\bibliographystyle{elsarticle-num-names} 
\bibliography{references}

\end{document}